\documentclass[aps,prd,superscriptaddress,onecolumn,a4paper,floatfix,nofootinbib]{revtex4}

\usepackage{graphicx}
\usepackage[]{amssymb}
\usepackage[]{amsmath}
\usepackage{units}
\usepackage{xspace}
\usepackage{rotating}
\usepackage[]{amssymb}
\usepackage[]{amsmath}
\usepackage[]{dcolumn}
\usepackage[format=hang,justification=raggedright]{caption,subfig}
\usepackage[usenames]{color}
\ifthenelse{
  \isundefined{\beginnumbering}
}{
  \newcommand{\pstart}{}
  \newcommand{\pend}{}
}{
\firstlinenum{1}
\linenumincrement{1}
\setlength{\linenumsep}{3pc}
}

\newcommand{\numu}{\ensuremath{\nu_{\mu}}}

\newcommand{\numubar}{\ensuremath{\overline{\nu}_{\mu}}}
\newcommand{\nue}{\ensuremath{\nu_{e}}}
\newcommand{\nuebar}{\ensuremath{\overline{\nu}_{e}}}
\newcommand{\dm}{\ensuremath{\Delta m^2}}
\newcommand{\st}{\ensuremath{\sin^{2}2{\theta}}}

\newcommand{\pt}{\ensuremath{p_{T}}}
\newcommand{\pz}{\ensuremath{p_{z}}}

\newcommand{\pzpt}{\ensuremath{\pz{},\pt{}}}
\newcommand{\D}{\ensuremath{\mathrm{d}}}
\newcommand{\hpxsec}{\ensuremath{\D^{2}N/\D p_{z}\D p_{T}}}
\newcommand{\dedx}{\ensuremath{\D E /\D x}}

\newcommand{\dmsq}{\ensuremath{|\Delta m^{2}|}}
\newcommand{\sintwo}{\ensuremath{\sin^{2}2{\theta}}}

\newcommand{\chisq}{\ensuremath{\chi^{2}}}

\newcommand{\numucc}{\ensuremath{\nu_{\mu}} charged-current}

\newcommand{\nuecc}{\ensuremath{\nu_{\mathrm{e}}} charged-current}
\newcommand{\nc}{neutral-current}

\newcommand{\enu}{\ensuremath{E_{\nu}}}

\newcommand{\gevc}{\ensuremath{\mathrm{GeV}/\mathit{c}}}
\newcommand{\mevc}{\ensuremath{\mathrm{MeV}/\mathit{c}}}
\newcommand{\gevcsq}{\ensuremath{\mathrm{GeV}/\mathit{c}^{2}}}

\newcommand{\evmass}{\ensuremath{\mathrm{eV}^{2}/\mathit{c}^{4}}}

\newcommand{\gcalor}{{\tt GCALOR}}
\newcommand{\geant}{{\tt GEANT}}
\newcommand{\geantth}{{\tt GEANT3}}
\newcommand{\gfluka}{{\tt GEANT-FLUKA}}

\newcommand{\flukafive}{{\tt FLUKA05}}

\newcommand{\minos}{MINOS}
\newcommand{\numi}{NuMI}

\newcommand{\ngen}{{\tt NEUGEN}}

\newcommand{\ngthree}{{\tt NEUGEN-v3}}

\newcommand{\xsec}{cross-section}
\newcommand{\xsecs}{cross-sections}
\newcommand{\unc}[1]{#1}

\newcommand{\pcom}[1]{}

\newcommand{\nddist}{\unit[1040]{m}}
\newcommand{\dpipe}{\unit[675]{m}}

\begin{document}

\title{A Study of Muon Neutrino Disappearance Using the Fermilab Main Injector Neutrino Beam}
\preprint{FERMILAB-PUB-07-577-E}
\newcommand{\Cambridge}{Cavendish Laboratory, University of Cambridge, Cambridge CB3 0HE, United Kingdom}
\newcommand{\FNAL}{Fermi National Accelerator Laboratory, Batavia, Illinois 60510, USA}
\newcommand{\RAL}{Rutherford Appleton Laboratory, Chilton, Didcot, Oxfordshire, OX11 0QX, United Kingdom}
\newcommand{\UCL}{Department of Physics and Astronomy, University College London, Gower Street, London WC1E 6BT, United Kingdom}
\newcommand{\Caltech}{Lauritsen Laboratory, California Institute of Technology, Pasadena, California 91125, USA}
\newcommand{\ANL}{Argonne National Laboratory, Argonne, Illinois 60439, USA}
\newcommand{\Athens}{Department of Physics, University of Athens, GR-15771 Athens, Greece}
\newcommand{\NTUAthens}{Department of Physics, National Tech. University of Athens, GR-15780 Athens, Greece}
\newcommand{\Benedictine}{Physics Department, Benedictine University, Lisle, Illinois 60532, USA}
\newcommand{\BNL}{Brookhaven National Laboratory, Upton, New York 11973, USA}
\newcommand{\CdF}{APC -- Universit\'{e} Paris 7 Denis Diderot, Paris Cedex 13, France}
\newcommand{\Cleveland}{Cleveland Clinic, Cleveland, Ohio 44195, USA}
\newcommand{\Delhi}{Dept. of Physics and Astrophysics, University of Delhi, Delhi 110007, India}
\newcommand{\GEHealth}{GE Healthcare, Florence South Carolina 29501, USA}
\newcommand{\Harvard}{Department of Physics, Harvard University, Cambridge, Massachusetts 02138, USA}
\newcommand{\HolyCross}{Holy Cross College, Notre Dame, Indiana 46556, USA}
\newcommand{\IIT}{Physics Division, Illinois Institute of Technology, Chicago, Illinois 60616, USA}
\newcommand{\Indiana}{Indiana University, Bloomington, Indiana 47405, USA}
\newcommand{\ITEP}{High Energy Experimental Physics Department, Institute of Theoretical and Experimental Physics, 
  B. Cheremushkinskaya, 25, 117218 Moscow, Russia}
\newcommand{\JMU}{Physics Department, James Madison University, Harrisonburg, Virginia 22807, USA}
\newcommand{\LASL}{Nuclear Nonproliferation Division, Threat Reduction Directorate, Los Alamos National Laboratory, Los Alamos, New Mexico 87545, USA}
\newcommand{\Lebedev}{Nuclear Physics Department, Lebedev Physical Institute, Leninsky Prospect 53, 117924 Moscow, Russia}
\newcommand{\LLL}{Lawrence Livermore National Laboratory, Livermore, California 94550, USA}
\newcommand{\MIT}{Lincoln Laboratory, Massachusetts Institute of Technology, Lexington, Massachusetts 02420, USA}
\newcommand{\Minnesota}{University of Minnesota, Minneapolis, Minnesota 55455, USA}
\newcommand{\Crookston}{Math, Science and Technology Department, University of Minnesota -- Crookston, Crookston, Minnesota 56716, USA}
\newcommand{\Duluth}{Department of Physics, University of Minnesota -- Duluth, Duluth, Minnesota 55812, USA}
\newcommand{\Oxford}{Subdepartment of Particle Physics, University of Oxford,  Denys Wilkinson Building, Keble Road, Oxford OX1 3RH, United Kingdom}
\newcommand{\Pittsburgh}{Department of Physics and Astronomy, University of Pittsburgh, Pittsburgh, Pennsylvania 15260, USA}
\newcommand{\IHEP}{Institute for High Energy Physics, Protvino, Moscow Region RU-140284, Russia}
\newcommand{\RoyalH}{Physics Department, Royal Holloway, University of London, Egham, Surrey, TW20 0EX, United Kingdom}
\newcommand{\Carolina}{Department of Physics and Astronomy, University of South Carolina, Columbia, South Carolina 29208, USA}
\newcommand{\SLAC}{Stanford Linear Accelerator Center, Stanford, California 94309, USA}
\newcommand{\Stanford}{Department of Physics, Stanford University, Stanford, California 94305, USA}
\newcommand{\StJohnFisher}{Physics Department, St. John Fisher College, Rochester, New York 14618 USA}
\newcommand{\Sussex}{Department of Physics and Astronomy, University of Sussex, Falmer, Brighton BN1 9QH, United Kingdom}
\newcommand{\TexasAM}{Physics Department, Texas A\&M University, College Station, Texas 77843, USA}
\newcommand{\Texas}{Department of Physics, University of Texas, Austin, Texas 78712, USA}
\newcommand{\TechX}{Tech-X Corporation, Boulder, Colorado 80303, USA}
\newcommand{\Tufts}{Physics Department, Tufts University, Medford, Massachusetts 02155, USA}
\newcommand{\UNICAMP}{Universidade Estadual de Campinas, IF-UNICAMP, CP 6165, 13083-970, Campinas, SP, Brazil}
\newcommand{\USP}{Instituto de F\'{i}sica, Universidade de S\~{a}o Paulo,  CP 66318, 05315-970, S\~{a}o Paulo, SP, Brazil}
\newcommand{\Washington}{Physics Department, Western Washington University, Bellingham, Washington 98225, USA}
\newcommand{\WandM}{Department of Physics, College of William \& Mary, Williamsburg, Virginia 23187, USA}
\newcommand{\Wisconsin}{Physics Department, University of Wisconsin, Madison, Wisconsin 53706, USA}
\newcommand{\deceased}{Deceased.}

\affiliation{\ANL}
\affiliation{\Athens}
\affiliation{\Benedictine}
\affiliation{\BNL}
\affiliation{\Caltech}
\affiliation{\Cambridge}
\affiliation{\UNICAMP}
\affiliation{\CdF}
\affiliation{\FNAL}
\affiliation{\Harvard}
\affiliation{\IIT}
\affiliation{\Indiana}
\affiliation{\IHEP}
\affiliation{\ITEP}
\affiliation{\JMU}
\affiliation{\Lebedev}
\affiliation{\LLL}
\affiliation{\UCL}
\affiliation{\Minnesota}
\affiliation{\Duluth}
\affiliation{\Oxford}
\affiliation{\Pittsburgh}
\affiliation{\RAL}
\affiliation{\USP}
\affiliation{\Carolina}
\affiliation{\Stanford}
\affiliation{\Sussex}
\affiliation{\TexasAM}
\affiliation{\Texas}
\affiliation{\Tufts}
\affiliation{\Washington}
\affiliation{\WandM}
\affiliation{\Wisconsin}

\author{P.~Adamson}
\affiliation{\FNAL}
\affiliation{\UCL}

\author{C.~Andreopoulos}
\affiliation{\RAL}

\author{K.~E.~Arms}
\affiliation{\Minnesota}

\author{R.~Armstrong}
\affiliation{\Indiana}

\author{D.~J.~Auty}
\affiliation{\Sussex}

\author{S.~Avvakumov}
\affiliation{\Stanford}

\author{D.~S.~Ayres}
\affiliation{\ANL}

\author{B.~Baller}
\affiliation{\FNAL}

\author{B.~Barish}
\affiliation{\Caltech}

\author{P.~D.~Barnes~Jr.}
\affiliation{\LLL}

% ...

\collaboration{The MINOS Collaboration}
\noaffiliation

\date{\today}

\begin{abstract}

We report the results of a search for \numu{} disappearance by the Main Injector Neutrino Oscillation Search~\cite{prl}. The experiment uses two detectors separated by \unit[734]{km} to observe a beam of neutrinos created by the Neutrinos at the Main Injector facility at Fermi National Accelerator Laboratory. The data were collected in the first \unit[282]{days} of beam operations and correspond to an exposure of $\unit[1.27\times 10^{20}]{}$ protons on target. Based on measurements in the Near Detector, in the absence of neutrino oscillations we expected $336\pm 14$ \numu{} charged-current interactions at the Far Detector but observed 215. This deficit of events corresponds to a significance of 5.2 standard deviations.  The deficit is energy dependent and is consistent with two-flavor neutrino oscillations according to $\dmsq=\unit[2.74^{+0.44}_{-0.26}\times 10^{-3}]{\evmass{}}$ and $\st{}>0.87$ at 68\% confidence level. 


\end{abstract}


\maketitle


\section{Introduction}

\pstart
There is now strong evidence that \nue{} produced in the Sun~\cite{Aharmim:2005gt,Hosaka:2005um} and \nuebar{} produced in nuclear reactors~\cite{Araki:2004mb} change flavor.  There is additional and compelling evidence that \numu{}/\numubar{} produced in the atmosphere~\cite{Ashie:2005ik,Ambrosio:2004ig,Allison:2005dt} and more recently by accelerators~\cite{Ahn:2006zz,prl} disappear while propagating through both the atmosphere and the Earth over distances of \unit[250-13000]{km}. This behavior is consistent with three-flavor neutrino oscillations induced by mixing between non-degenerate mass eigenstates $\nu_{1},\nu_{2},\nu_{3}$ and the states $\nu_{e},\nu_{\mu},\nu_{\tau}$ created in weak interactions.  The flavor and mass eigenstates are related by a unitary matrix $U_{\mathrm{PMNS}}$~\cite{Maki:1962mu,Pontecorvo:1967fh} which is typically expressed in terms of three angles $\theta_{12},\theta_{23},\theta_{13}$ and a CP-violating phase $\delta$. The energy and distance scales at which oscillations occur are determined by the difference in squared masses, $\Delta m^{2}_{ij}=m^{2}_{i}-m^{2}_{j}$, of the $\nu_{1},\nu_{2},\nu_{3}$. 
\pend

\pstart
The data suggest that the phenomenology of solar and reactor neutrinos is driven by a squared-mass splitting $7.6\leq \dm_{\odot} \leq \unit[8.6 \times 10^{-5}]{\evmass{}}$ ( 68\% confidence level -- C.L.)~\cite{Aharmim:2005gt} whereas the behavior of atmospheric and accelerator produced \numu{} is determined by a much larger splitting $1.9 \leq |\dm_{\mathrm{atm}}| \leq \unit[3.0 \times 10^{-3}]{\evmass{}}$ (90\% C.L.)~\cite{Ashie:2005ik}. In this case muon-neutrino disappearance may be described as two-flavor neutrino oscillations:
\pend
\begin{equation}
P\left(\nu_{\mu}\rightarrow\nu_{\mu}\right) = 1 - \st{}\sin^{2} \left(1.27\frac{\dm [\mathrm{eV}^{2}] L [\mathrm{km}]}{E[\mathrm{GeV}]}\right) \label{eq:oscprob}
\end{equation}
\pstart
where \dm{} is the effective difference of squared masses, $\theta$ is a mixing angle between the mass and weak eigenstates, $E$ is the neutrino energy and $L$ is the distance from the neutrino production point to the observation point. The analysis reported here is conducted in the two-neutrino framework, and we can tentatively identify $\dm{} \approx \dm_{\mathrm{atm}} \approx \dm_{32}$ and $\st{} = \sin^{2}(2\theta_{23})$ in the limit $P\left(\numu{} \rightarrow \nue{}\right) = 0$. More precisely, the experiment measures $\dm = \rm{sin}^{2}\theta_{12}\Delta m^{2}_{31}+{\rm cos}^{2}\theta_{12}\Delta m^{2}_{32}$ where effects of order ${\rm sin}^2(\theta_{13})$ have been neglected~\cite{Nunokawa:2005nx}.
\pend



\pstart
The Main Injector Neutrino Oscillation Search (\minos{}) is a long-baseline, two-detector neutrino oscillation experiment that uses a muon-neutrino beam produced by the Neutrinos at the Main Injector~\cite{numitdr,Kopp:2005bt} (\numi{}) facility at Fermi National Accelerator Laboratory (FNAL). \numi{} is able to provide a range of neutrino energies but has mostly been operated in the ``low-energy'' (LE10/185kA of Tab.~\ref{tab:beam-config}) configuration, which maximizes the neutrino flux at $E\approx\unit[3]{GeV}$. Neutrinos are observed by two functionally identical detectors, located at two sites, the Near Detector (ND) at FNAL and the Far Detector (FD) in the Soudan Underground Laboratory in Minnesota.  The detectors are separated by \unit[734]{km} and are designed to detect muon-neutrino charged-current interactions for $\unit[1 < E < 100]{GeV}$.  The characteristic $L/E\approx\unit[245]{km/GeV}$ allows \minos{} to rigorously test the oscillation hypothesis and make precision measurements of the \dm{} and \st{} mixing parameters governing muon-neutrino disappearance at the atmospheric neutrino mass-scale.
\pend

\pstart
\minos{} employs two detectors to significantly reduce the effect that systematic uncertainties associated with the neutrino flux, \xsecs{} and detector efficiency have upon the \numu{} disappearance measurement. Data collected by the Near Detector in several different configurations of the \numi{} beam (Tab.~\ref{tab:beam-config} and Fig.~\ref{fig:targ-horns}) were used to constrain the simulation of the neutrino flux and the detector response to neutrino interactions.  Four independent analyses were used to predict the energy spectrum expected at the Far Detector for the case that \numu{} do not disappear.  The analyses were fully developed using Near Detector data only.  After we were satisfied with the procedure, the energy spectrum measured at the Far Detector was inspected and then compared to the predictions; oscillations cause an energy dependent suppression of the \numucc{} rate. A fit to the Far Detector data was performed by incorporating Eq.~\ref{eq:oscprob} into the Far Detector prediction. The fit also included the most significant sources of systematic uncertainty and resulted in a measurement of the oscillation parameters. 
\pend


\pstart
\minos{} began collecting \numi{} beam data in March 2005. The Far Detector dataset used in the analysis reported here was recorded  between May 20, 2005 and February 25, 2006. The total exposure is $\unit[1.27\times 10^{20}]{}$ protons on target (POT) and only includes data collected while \numi{} was operating in the low-energy beam configuration. This exposure allows \minos{} to measure \dm{} and \st{} with a precision that is comparable to the best existing measurements~\cite{Ashie:2005ik,Ahn:2006zz}. The results reported here use the same dataset and are identical to those reported in~\cite{prl} but include a more detailed description of the experiment, analysis and results. Since the publication of~\cite{prl}, \minos{} has accumulated a total of $\unit[3.5\times10^{20}]{POT}$ through July 2007 and preliminary results based on $\unit[2.5\times10^{20}]{POT}$ have been presented~\cite{lp07}. Analysis of the full $\unit[3.5\times10^{20}]{POT}$ dataset is ongoing, and accumulation of further data is foreseen.
\pend


\pstart
We begin in Sec.~\ref{sec:beam} by describing the neutrino beam-line design, operation and simulation. The neutrino detectors are described in Sec.~\ref{sec:det} along with the calibration procedure and the simulation of neutrino interactions.  We also discuss neutrino data collection and event reconstruction in Sec.~\ref{sec:det} and the selection of \numucc{} events in Sec.~\ref{sec:sandb}. The manner in which Near Detector neutrino data was used to constrain the neutrino flux calculation is presented in Section~\ref{sec:skzp}.  Four methods for predicting the Far Detector \numucc{} spectrum in the absence of oscillations are discussed in Sec.~\ref{sec:extrap}. We conclude in Sec.~\ref{sec:osc} with a description of the oscillation analysis and measurement of the parameters \dm{} and \st{}. 
\pend



\section{The Neutrino Beam}
\label{sec:beam}
\begin{figure*}
  \centering
  \includegraphics[width=\textwidth,keepaspectratio,clip]{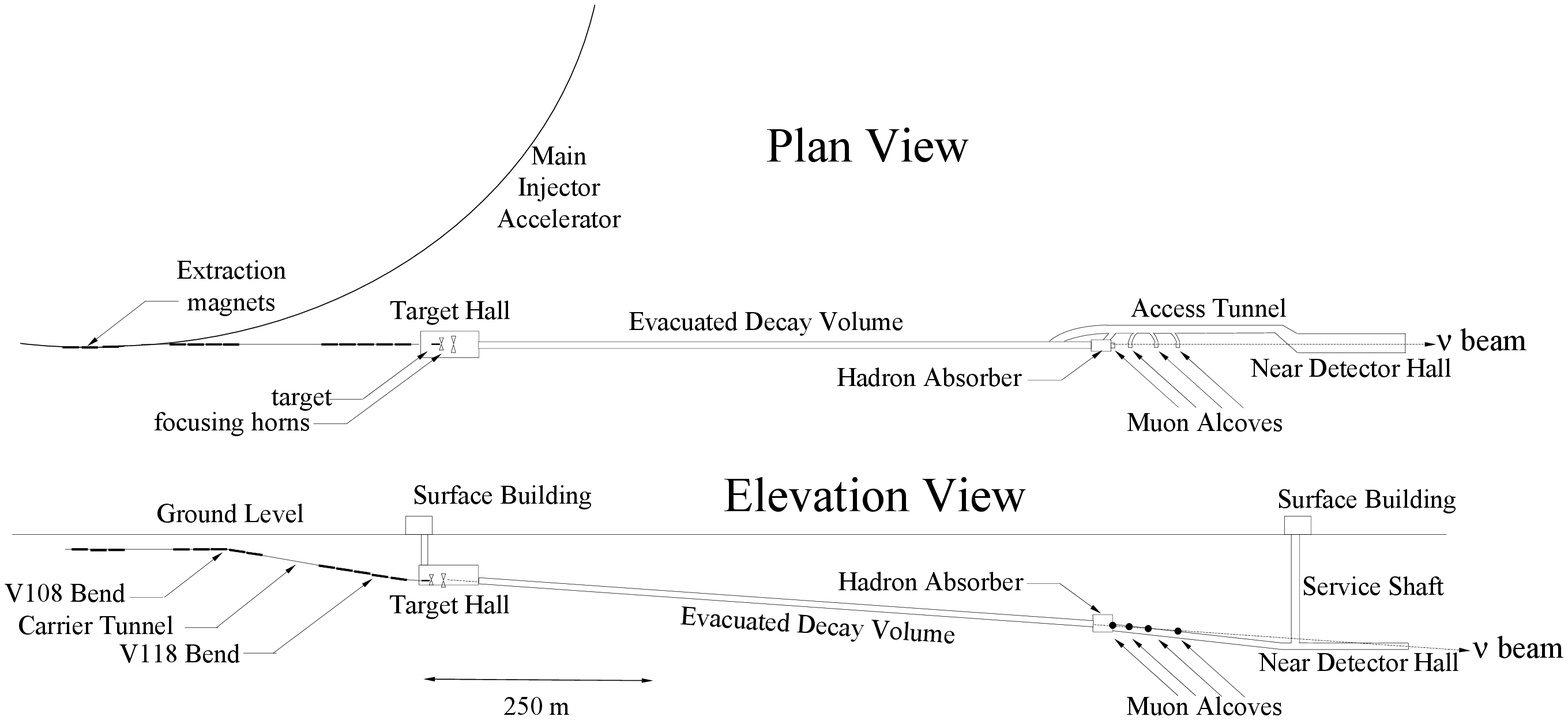}
  \caption{ \label{fig:numi} Plan and elevation views of the \numi{} beam facility.  A proton beam is directed onto a target, where the secondary pions and kaons are focused into an evacuated decay volume via magnetic horns.  Ionization chambers at the end of the beam-line measure the secondary hadron beam and tertiary muon beam.}
\end{figure*}

\pstart
The \numi{} Neutrino Beam~\cite{numitdr,Kopp:2005bt} generates neutrinos mainly from the decays of pions and kaons produced in the \numi{} target, with a smaller contribution from muon decays. A schematic diagram of the \numi{} beam-line is shown in Fig.~\ref{fig:numi}.  Protons of \unit[120]{\gevc{}} momentum are extracted from the Main Injector (MI) accelerator in a \unit[10]{$\mu$s} spill, bent downward by \unit[58]{mrad} to point at the Far Detector, and impinged upon the \numi{} hadron production target.  Positioned downstream of the target, two toroidal magnets called ``horns'' sign-select and focus the secondary mesons from the target, as shown in Fig.~\ref{fig:targ-horns}.  The mesons are directed into a \dpipe{} long evacuated volume, where they may decay to muons and neutrinos.  At the end of the decay volume, a beam absorber stops the remnant hadrons. The absorber is followed by approximately \unit[240]{m} of un-excavated rock which stops the tertiary muons, leaving only neutrinos. Figure~\ref{fig:numi-spectra} shows \numucc{} spectra from three configurations of the \numi{} beam.  The following section provides a brief overview of the beam-line and instrumentation.
\pend

\subsection{Beam Hardware and Performance}
\subsubsection{Primary Proton Beam}

\pstart
The Main Injector accepts batches of protons from the \unit[8]{\gevc{}} Booster accelerator, which are then accelerated to \unit[120]{\gevc{}}. In most MI cycles seven batches were accelerated. Protons were removed from the MI ring in two increments, with a pair of batches being sent to the Antiproton Accumulator for the Tevatron program, while the remaining five batches were directed into the \numi{} primary-proton line and transported \unit[350]{m} to the \numi{} target.  The \numi{} extractions typically contained a total of $\unit[2.1\times10^{13}]{protons}$ with a cycle time of $2.2-\unit[2.4]{s}$. Extractions of up to six batches and \unit[$2.5\times10^{13}$]{protons} were achieved during the first year of operations.  The proton beam centroid at the target was stable to within $\unit[\pm 0.1]{mm}$ and the area of the beam-spot varied within the range $\unit[3.3-4.5]{mm^{2}}$.  Further details of the primary proton beam delivery system and its performance are given in Ref.~\cite{bishai2006}.
\pend

\subsubsection{Target and Horns}

\begin{figure}
  \centering
  \includegraphics[width=\columnwidth]{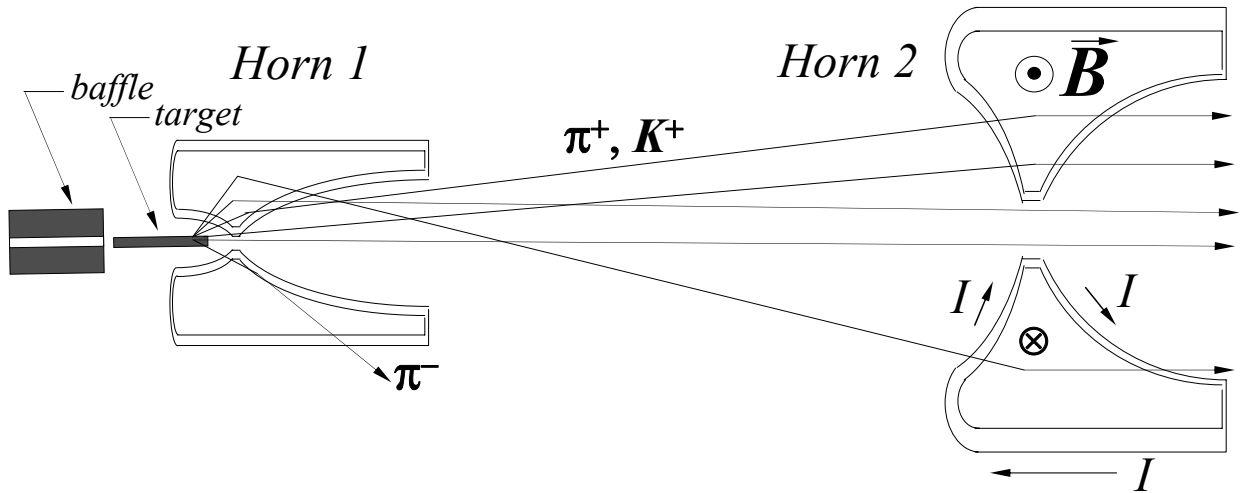}
  \caption{\label{fig:targ-horns} In the \numi{} neutrino beam,  horns 1 and 2 are separated by \unit[10]{m}.  A collimating baffle upstream of the target protects the horns from direct exposure to misdirected proton beam pulses.  The target and baffle system can be moved further upstream of the horns to produce higher energy neutrino beams~\cite{kostin2002}.  The vertical scale is four times that of the horizontal (beam axis) scale.}
\end{figure}


\begin{figure*}
  \centering
\includegraphics[width=\columnwidth]{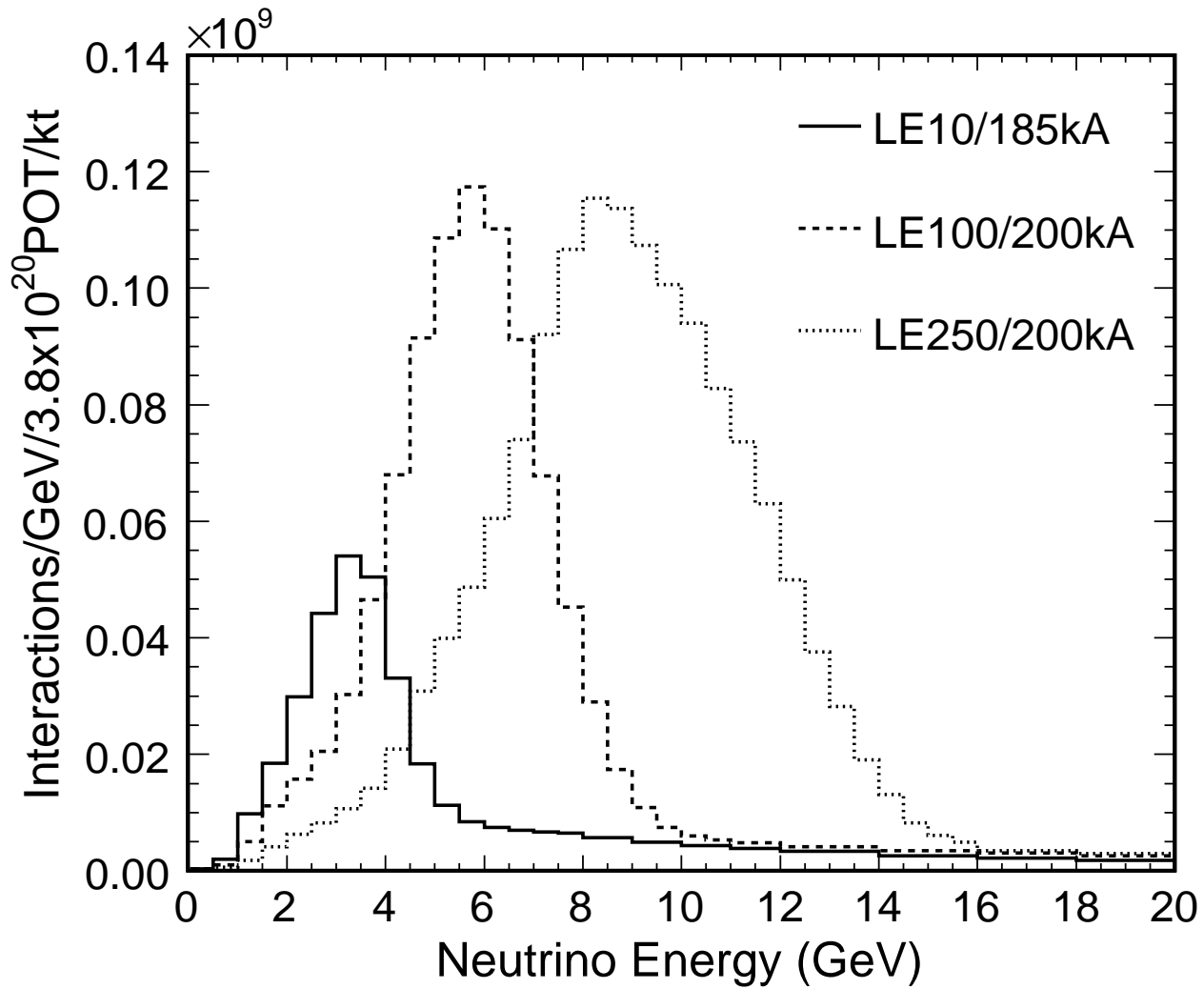}
  \caption{\label{fig:numi-spectra} Calculated rate of \numucc{} interactions in the MINOS Near Detector, located \nddist{} from the \numi{} target. The figures were made by combining the flux calculation discussed in \ref{sec:beam_sim} with the cross-sections discussed in Sec.~\ref{sec:mc}. Three spectra are shown, corresponding to the LE10/185kA, LE100/200kA, and LE250/200kA configurations of Tab.~\ref{tab:beam-config}.}
\end{figure*} 

\pstart
The production target is a rectangular graphite rod, segmented longitudinally into 47 segments, or fins. The target dimensions are \unit[6.4]{mm} in width, \unit[15]{mm} in height and \unit[940]{mm} in length. The beam spot size at the target is \unit[1.2-1.5]{mm}~\cite{bishai2006}.  A collimating baffle upstream of the target provides protection for the target, its cooling lines, as well as downstream beam components.  The baffle is a \unit[1.5]{m} long graphite rod with an \unit[11]{mm} diameter inner bore. 
\pend



\pstart
The particles produced in the target are focused by two magnetic horns~\cite{Abramov2002}, shown schematically in Fig.~\ref{fig:targ-horns}.  The \unit[200]{kA} pulsed current produces a maximum \unit[30]{kG} toroidal field which selects particles produced at the target by charge-sign and momentum.  Measurements of the horns show the expected $1/r$ fall-off in the magnetic field to within a percent and the field at distances from the beam axis smaller than the inner conductors is observed to be less than \unit[20]{G}.  The absolute value of the current flowing through the horns was calibrated to within $\pm 0.5\%$ and was observed to vary less than 0.2\% over the course of the data collection period.  The horns have parabolic-shaped inner conductors~\cite{Abramov2002}, for which the focal length for point-to-parallel focusing of particles from the target is proportional to momentum~\cite{budker1961,Baratov1977d}. The alignment of the target and horn system relative to the beam axis was checked using the proton beam itself~\cite{zwaska2006}.
\pend

\pstart
The relative longitudinal positions of the two horns and the target optimizes the momentum focus for pions and therefore the typical neutrino energy.  To fine-tune the beam energy, the target is mounted on a rail-drive system with \unit[2.5]{m} of longitudinal travel, permitting remote change of the beam energy without directly accessing the horns and target~\cite{kostin2002}.  In its furthest downstream location, the target is cantilevered approximately \unit[65]{cm} into the first parabolic horn.  Characteristics of the predicted neutrino spectra from several target position and horn current settings are listed in Tab.~\ref{tab:beam-config} and \numucc{} energy distributions are shown in Fig.~\ref{fig:numi-spectra}.  Data from the LE10/185kA configuration were used in the oscillation search. Data from the other beam configurations were used to constrain the beam Monte Carlo (see Sec.~\ref{sec:skzp}).  
\pend

\begin{table*}
\begin{ruledtabular}
\begin{tabular}{lcccd}
Beam		&	Target          &       Horn  & Peak  & \mbox{Exposure}  \\
Configuration	&	Position (cm) 	&Current (kA) & $\enu{}\pm \mathrm{r.m.s.}$ (GeV)	& \unit[10^{18}]{POT}\\ \hline\hline
LE10/0kA	&	10		&	0	& $7.4 \pm 4.1$\footnote{The \unit[0]{kA} ``horn-off'' beam is unfocused and has a broad energy distribution} &  2.69 \\
LE10/170kA	&	10		&	170	& $3.1\pm 1.1$ &  1.34 \\
LE10/185kA	&	10		&	185	& $3.3\pm 1.1$ & 127. \\
LE10/200kA	&	10		&	200	& $3.5\pm 1.1$ & 1.26 \\
LE100/200kA	&	100		&	200	& $5.6\pm 1.5$ & 1.11 \\
LE250/200kA	&	250		&	200	& $8.6\pm 2.7$ & 1.55 \\

\end{tabular}
\end{ruledtabular}
    \caption{Beam configurations and data sets used in this publication. The target position refers to the distance the target was displaced upstream of its default position inside the first focusing horn.  The peak (i.e., most probable) neutrino energy \enu{} is determined after multiplying the muon-neutrino flux predicted by the beam Monte Carlo simulation by charged-current \xsec{}. The r.m.s.\ refers to the root mean square of the peak of the neutrino energy distribution. \label{tab:beam-config}}
\end{table*}


\subsubsection{Decay Volume and Absorber}
\pstart
Particles are focused by the horns into a \dpipe{} long, \unit[2]{m} diameter steel pipe, evacuated to \unit[0.5]{Torr} to reduce meson absorption and scattering.  This length is approximately the decay length of a \unit[10]{GeV} pion. The entrance to the decay pipe is sealed by a two-piece aluminum-steel window. The central ($\mbox{radius} < \unit[50]{cm}$) portion of the window is made of \unit[1]{mm} thick aluminum and is strengthened by an outer ($\mbox{radius} > \unit[50]{cm}$) section made of \unit[1.8]{cm} thick steel. The design reduces scattering in the window while maintaining vacuum integrity. The decay volume is surrounded by \unit[2.5-3.5]{m} of concrete shielding.  At the end of the decay volume is a beam absorber consisting of a water-cooled aluminum and steel core followed by steel and concrete blocks.
\pend


\subsubsection{Instrumentation}

\pstart
The primary proton beam position is monitored along the transport line by 24 capacitive beam position monitors (BPMs), the beam intensity by two toroidal beam current transformers and 44 loss monitors, and the beam position and spot size by ten retractable segmented foil secondary emission monitors (SEMs)~\cite{kopp2006a}.  During normal operations only the last SEM upstream of the target is inserted in the beam.  The absolute toroid uncertainty  was determined to be $\pm1.0\%$  by precision current pulses and monitored for drift throughout the run by comparison of the toroids in the \numi{} line with monitors in the Main Injector.  The beam position and size at the target are measured to within \unit[$\pm50$]{$\mu$m}~\cite{bishai2006}.  
\pend

\pstart
In the target hall, the current flowing in the horns was monitored spill-to-spill, as was the temperature of the current-delivering strip-line.  Both were seen to change by approximately $0.2\%$ due to thermal variations in the target hall.  The temperature of the upstream collimating baffle was continuously monitored, as this was observed to be a good measure of the proton beam halo scraping on the baffle. During \numi{} operations it was found that approximately 0.3\% of the beam was obstructed by the baffle.
\pend

\begin{figure*}[h]
  \centering
\includegraphics[width=0.8\columnwidth]{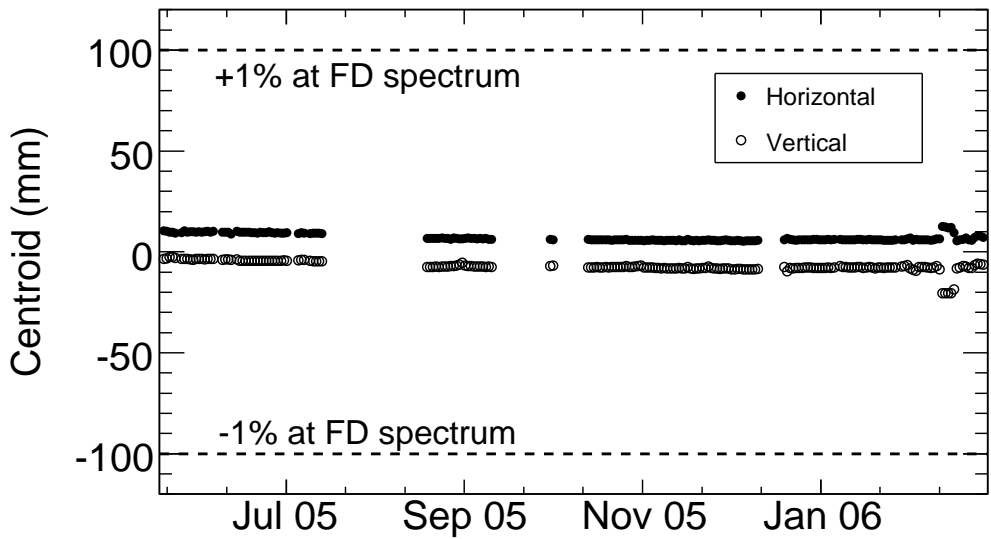}
  \caption{\label{fig:mu_mon} Horizontal and vertical beam centroid measured by the first muon station. The observed variations in the centroid correspond to \unit[20]{$\mu$rad} variations in the beam angle. Such deviations are expected to have less than 0.2\% effect on the \numu{} energy spectrum at the Far Detector . }
\end{figure*} 


\pstart
Ionization chambers are used to monitor the secondary and tertiary particle beams~\cite{kopp2006b}.  An array is located immediately upstream of the beam absorber to monitor the remnant hadrons at the end of the decay pipe. There are also three muon monitoring stations, one downstream of the absorber, one after \unit[12]{m} of rock, and a third after an additional \unit[18]{m} of rock.  The muon stations confirmed the relative alignment of the neutrino beam direction to approximately \unit[20]{$\mu$rad} throughout the run as shown in Fig.~\ref{fig:mu_mon}. This is expected to limit spectral variations at the Far Detector to less than 0.2\%.  In addition, the charge (per spill per proton on target) deposited in the muon monitors varied by only $\pm 2\%$, indicating a similar level of stability in the neutrino flux. 
\pend

\pstart
The relative position of a point on the surface above the Far Detector and the \numi{} target were measured by GPS survey with an accuracy of \unit[1]{cm} in the horizontal plane, and \unit[6]{cm} in the vertical~\cite{Bocean:2006mh,numitdr}. Translating the surface coordinates to the underground Far Detector hall introduces an additional uncertainty of \unit[70]{cm} in the vertical. The positions of beam-line components with respect to the target were surveyed and are known with an accuracy of \unit[0.5]{mm} (95\% C.L.).   These data indicate that the beam angle with respect to the ideal Near to Far Detector vector is less than \unit[15]{$\mu$rad} (68\%~C.L.). This introduces a negligible uncertainty into the calculation of the neutrino flux at the Far Detector. 
\pend

\subsection{Simulation of the Neutrino Beam}
\label{sec:beam_sim}
\begin{figure}[h]
  \centering
  \subfloat[\label{fig:near_spec}]{\includegraphics[width=0.55\columnwidth]{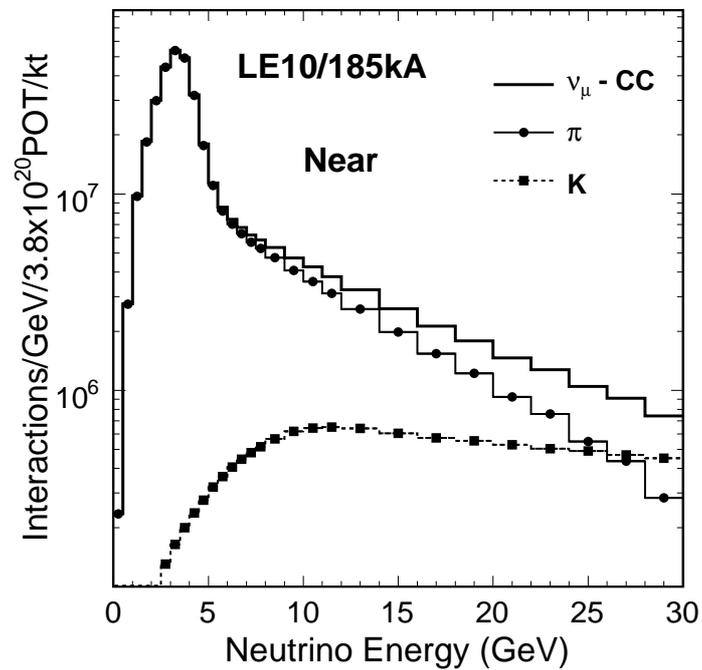}}\\
  \subfloat[\label{fig:near_far_overlay}]{\includegraphics[width=0.55\columnwidth]{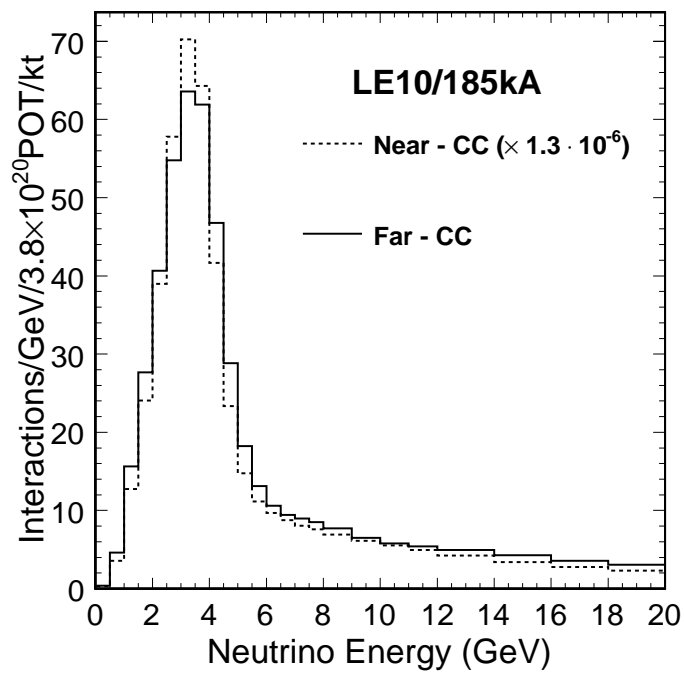}}
  \caption{
  \protect\subref{fig:near_spec} Expected energy distribution of \numu{} charged-current interactions in the Near Detector. The contributions from $\pi$ and $K$ decays are shown separately. \protect\subref{fig:near_far_overlay} Comparison of the neutrino energy spectrum at the Near and the Far Detectors.  The two are not identical, due to solid angle differences between the two detectors (see Fig.~\ref{fig:near-far-diagram}).  \label{fig:near-far}}

\end{figure} 

\begin{figure*}[h]
  \centering
  \includegraphics[width=1.0\textwidth]{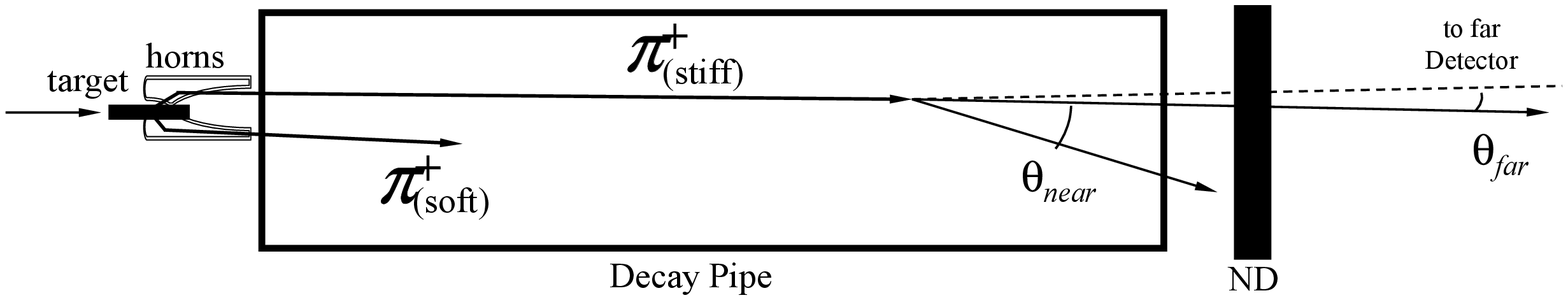}
  \caption{
The differences in the Near Detector and Far Detector spectra are due to the different angular acceptance of the two detectors and the proximity of the Near Detector to the beam-line \label{fig:near-far-diagram}}
\end{figure*} 

\pstart
The neutrino beam-line is modeled in three stages: (1) a simulation of the hadrons produced by \unit[120]{\gevc{}} protons incident on the \numi{} target, (2) the propagation of those hadrons and their progeny through the magnetic focusing elements, along the \dpipe{} decay pipe and into the primary beam absorber allowing for decay of unstable particles and (3) the calculation of the probability that any produced neutrino traverses the Near and Far detectors.
\pend


\pstart
The production of secondary mesons in the \numi{} target was calculated using the \flukafive{}~\cite{fluka4} Monte Carlo.  Particles exiting the target are recorded and later propagated in a \geantth{}~\cite{g3man} simulation of the \numi{} beam-line. The simulation describes the magnetic focusing horns, surrounding shielding and decay pipe. The magnetic field inside the horns was modeled as a perfect $1/r$ toroidal field.  The magnetic field in the inner conductor of the horn was calculated assuming that the current was uniformly distributed throughout the \unit[3]{mm} thickness of the aluminum conductor.  This assumption was motivated by the large skin depth $\delta\approx\unit[7.7]{mm}$ for the approximately $\unit[1]{ms}$ current pulse.  The \unit[10-30]{G} magnetic field in the inner apertures of the horns was neglected, as was the approximately $\unit[1]{G}$ field in the decay pipe due to residual magnetization of the iron vessel. The \gfluka{} code is used to describe hadronic interactions in the beam-line and the associated production of secondary particles as well as the full particle decay chains. Decays in which a neutrino is produced are saved and later used as input for neutrino event simulation in the Near and Far Detectors.
\pend

\pstart
The neutrino event simulation uses each decay recorded by the \numi{} simulation with a probability and neutrino energy determined by the decay kinematics and the (randomly chosen) trajectory through the Near or Far Detector. For two-body $\pi/K\rightarrow\mu\nu$ decays of relativistic mesons, the neutrino energy is given by 
\pend
\begin{equation}
E_\nu\approx
\frac{(1-\frac{m_\mu^2}{M^2}) E}{1+\gamma^2\tan^2\theta_\nu} \quad .
\label{eq:enu-eqn}
\end{equation}
\pstart
\noindent where $m_\mu$ and $M$ are the muon and parent hadron masses, $E$ the parent hadron energy, $\gamma=E/M$ is the parent's Lorentz boost, and $\theta_\nu$ is the angle (in the lab) between the neutrino and parent hadron directions.  The neutrino is forced to pass through either the Near or Far MINOS Detector, with probability for the particular meson decay given by
\pend
\begin{equation}
\frac{dP}{d\Omega_\nu}\approx\frac{1}{4\pi}\frac{4\gamma^2(1+\tan^2\theta_\nu)^{3/2}}{(1+\gamma^2\tan^2\theta_\nu)^2} \quad ,
\label{eq:angle}
\end{equation}
\pstart
\noindent In deriving both expressions, $\beta=\sqrt{1-(1/\gamma^2)}\approx1$ is assumed. We have also accounted for the effect of $\mu$ polarization (e.g., induced in $\pi\rightarrow\mu\nu$) on the neutrino flux from $\mu$ decays.  Three-body kaon decays are also included, but contribute $<0.1\%$ to the \numu{} event rate.
\pend

\subsection{Expected Neutrino Energy Spectra}
\label{sec:expected_spectra}
\pstart
Figure~\ref{fig:numi-spectra} shows the energy spectra of \numucc{} interactions in the MINOS Near Detector in three of the beam configurations of Tab.~\ref{tab:beam-config}. In the LE10/185kA beam configuration, the neutrino in 87\% of \numucc{} interactions was produced by $\pi^+\rightarrow\mu^+\nu_\mu$, with $K^+$ decays contributing the additional 13\%. $K^0$ and $\mu$ decays contribute $<0.1\%$ to \numu{} event rate, though they do contribute significantly to the \nue{} rate. The contributions of $\pi$ and $K$ to the neutrino energy spectrum is shown in Fig.~\ref{fig:near-far}.  The predicted spectrum for the LE10/185kA beam is composed of 92.9\% \numu{}, 5.8\% \numubar{} and 1.3\% \nue{}+\nuebar{}.
\pend

\pstart
The uncertainties on the neutrino flux due to focusing effects are expected to be largest at the edges of apertures of the horns.  The uncertainties are shown in Fig~\ref{fig:miserr_n}-\ref{fig:miserr_fon} and described in more detail in Sec~\ref{sec:flux_uncertainties}.  Uncertainties at higher neutrino energies are smaller because higher energy neutrinos arise from high energy $\pi$ and $K$ mesons which pass through the field-free apertures of the horns without focusing. 
\pend

\pstart
The neutrino energy spectra at the MINOS Near and Far Detectors are not identical, due to the different solid angles subtended by the two detectors and due to the energy-dependence of the meson decay locations.  As indicated in Fig.~\ref{fig:near-far-diagram}, higher-energy pions decay further down the length of the decay volume, and hence a variety of decay angles $\theta_\nu$ result in neutrinos which strike the Near Detector.  The Far Detector, by contrast, can only be struck for very small-angle decays.  By Eq.~\ref{eq:enu-eqn}, the different allowed decay angles, as well as the close proximity of fast pion decays to the Near Detector, results in different energy spectra at the two detectors. These differences are conveniently characterized in terms of the Far/Near spectral ratio shown in Fig.~\ref{fig:fn_before_after}.  Thus, while the Near Detector in principle measures the neutrino energy spectrum in the absence of oscillations, corrections must be applied to this spectrum to derive a predicted spectrum at the Far Detector, as is discussed in Sec.~\ref{sec:extrap}. 
\pend

\subsection{Uncertainties In The Neutrino Flux}
\label{sec:flux_uncertainties}

\begin{figure*}[h]
  \centering
 \subfloat[\label{fig:miserr_n}]{\includegraphics[width=0.45\textwidth]{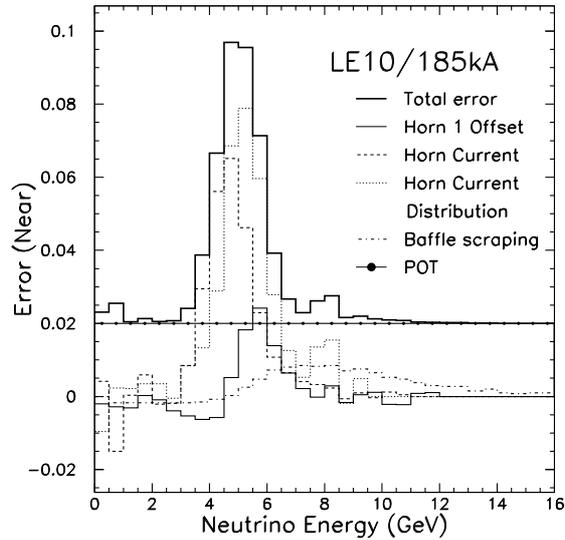} }\\
 \subfloat[\label{fig:miserr_fon}]{\includegraphics[width=0.45\textwidth]{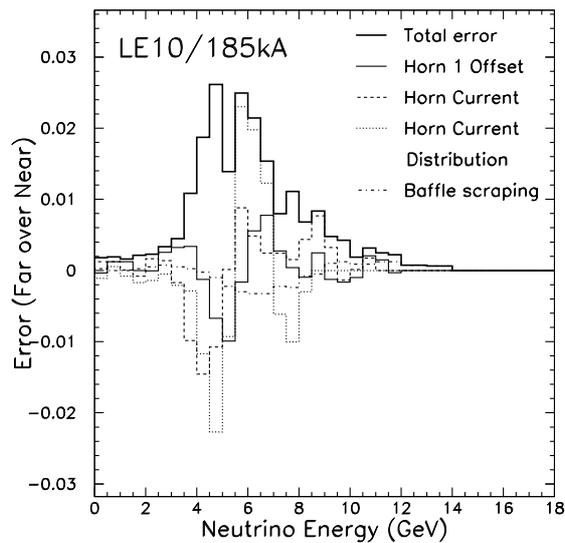} }
  \caption{
The effect that beam focusing uncertainties have on \protect\subref{fig:miserr_n} the Near Detector \numucc{} spectrum and \protect\subref{fig:miserr_fon} on the ratio (Far/Near) of \numucc{} spectra at the two detectors.  The lines show the fractional change in the number of events \protect\subref{fig:miserr_n} or event ratio \protect\subref{fig:miserr_fon} expected in each energy bin due to a one standard deviation shift in various beam parameters. These include the horn and target alignment, proton beam halo scraping on the collimator baffle, knowledge of the horn current, and the modeling of the current distribution in the horn conductor.  The total error is the sum in quadrature of the individual uncertainties. \label{fig:miserr}}
\end{figure*}

\pstart
The dominant contribution to the uncertainty in the flux at both detector sites is caused by uncertainty in the yield of hadrons off the target as a function of \pz{} and \pt{}, the components of the hadron momentum along and transverse to the beam-line. The magnitude of the uncertainty on the energy spectra is difficult to estimate given the lack of hadron production data.  As described in Sec.~\ref{sec:skzp}, the data in the Near Detector can be used to constrain the beam flux calculation, and in particular the yield of hadrons off the target. The constrained calculation improves agreement with the Near Detector data and reduces uncertainties in the prediction of the Far Detector flux. The procedure used to apply this constraint is described in Sec.~\ref{sec:skzp}. Unless otherwise noted the Monte Carlo simulation will use the flux calculation constrained by the Near Detector data.
\pend



\pstart
Systematic uncertainties on the predicted neutrino flux from beam focusing effects are readily calculable, and are constrained using data from the instrumentation in the primary, secondary, and tertiary beams. Figure~\ref{fig:miserr_n} shows the expected uncertainty on the neutrino energy spectrum at the Near Detector. Focusing uncertainties produce spectral distortions in both detectors that occur predominantly in the portions of the neutrino energy spectrum which correspond to parent hadrons which cross the edges of the horn apertures (see Fig.~\ref{fig:targ-horns}).  The largest beam focusing uncertainties include our knowledge of the absolute current flowing through the horns, the alignment of the horns and target to the rest of the beam-line~\cite{zwaska2006}, the shielding geometry, the uncertainty in the fraction of the proton beam scraping on the upstream collimating baffle (which acts as a target further upstream of the horns), and the modeling of the current distribution  in the horns' inner conductors (e.g., skin-depth effect). The expected distortion in the ratio of the Far and Near Detector spectra are smaller (see Fig.~\ref{fig:miserr_fon}) than the individual Near or Far Detector uncertainties (Fig.~\ref{fig:miserr_n}).  Hence, use of the Near Detector to measure the spectrum results in a prediction for the spectrum at the Far Detector with smaller uncertainty.
\pend



\section{The Neutrino Detectors}

\label{sec:det}

\begin{figure*}
\centering
  \subfloat[\label{fig:nd}]{\includegraphics[width=0.5\textwidth]{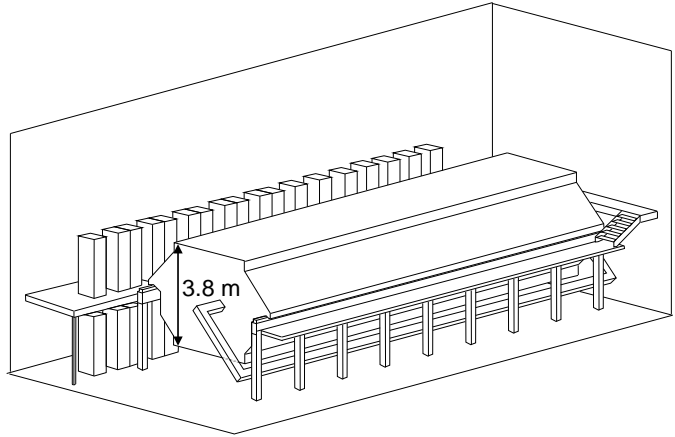}}
\\
  \subfloat[\label{fig:fd}]{\includegraphics[width=0.5\textwidth]{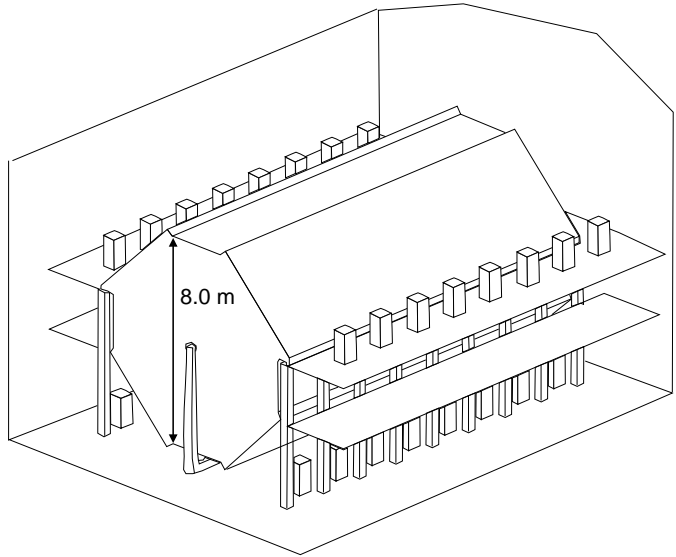}}
  \caption{The MINOS Near \protect\subref{fig:nd} and Far \protect\subref{fig:fd} Detectors. The Far Detector consists of two functionally identical modules, only one of which is shown in the figure.
   \label{fig:detectors} }
\end{figure*}


\begin{figure}
\begin{center}
  \includegraphics[width=\columnwidth]{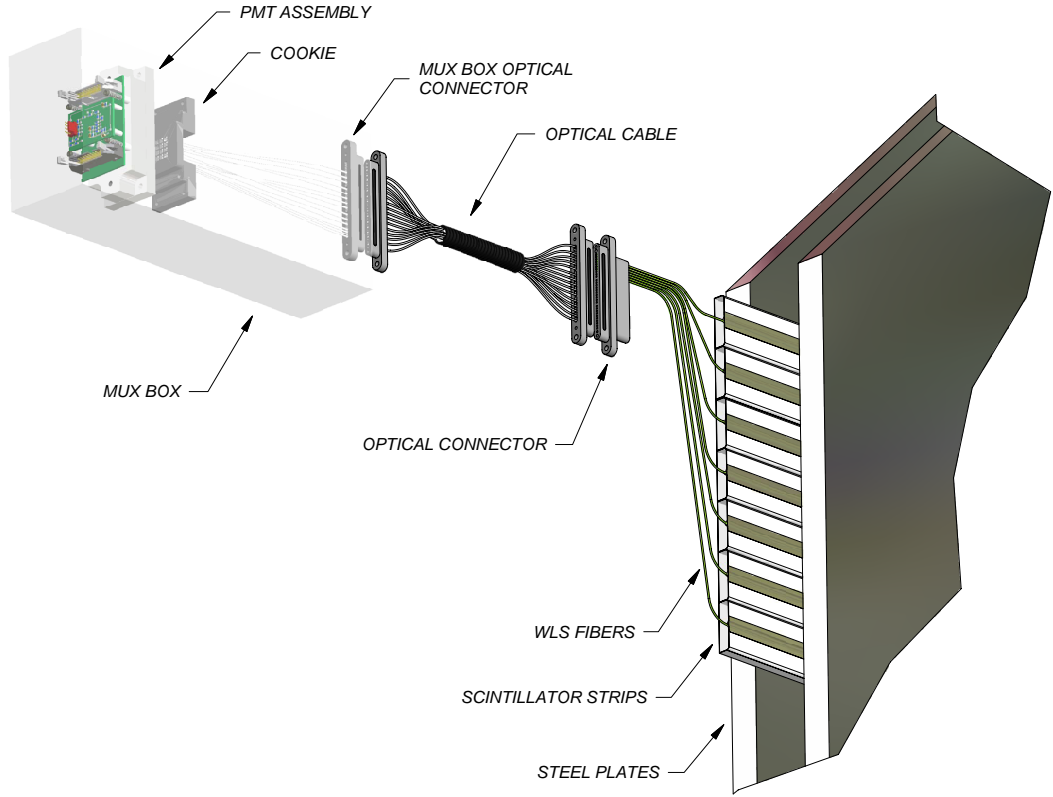}
  \caption{Optical readout of the MINOS detectors.  Scintillation light is captured by wavelength-shifting fibers embedded in the scintillator and then transferred to multi-anode photomultipliers via clear optical fibers. The Far Detector uses 16-anode phototubes and the Near Detector uses 64-anode phototubes. In the Far Detector eight fibers are optically summed on a single pixel in a manner which permits the eight-fold ambiguity to be resolved during the event reconstruction. \label{fig:optical_readout}}
\end{center}
\end{figure}

\pstart 
\minos{} consists of two neutrino detectors (shown in Fig.~\ref{fig:detectors}) separated by a long baseline. The Near Detector resides on the Fermilab site, \unit[104]{m} underground and \nddist{} downstream of the \numi{} target.  The Far Detector is located in the Soudan Underground Laboratory at a distance of \unit[735]{km} from the \numi{} target and \unit[705]{m} beneath the surface. The detectors are capable of observing muon-neutrino and electron-neutrino charged-current and neutral-current interactions having a visible energy larger than about \unit[$500$]{MeV}. Each detector has a toroidal magnetic field which is used to measure muon momenta via curvature and discriminate between \numu{} and \numubar{}. The two detectors were designed to respond to neutrino interactions in the same way so as to reduce the impact that uncertainties in the neutrino flux, cross-sections and detector acceptance have on oscillation measurements. In this section we will first describe the common features of both detectors, then those specific to the Far and Near Detectors, followed by a description of the data collection, event reconstruction, calibration and simulation.  A more comprehensive description of the MINOS detectors is presented in ~\cite{det_nim}. 
\pend

\subsection{Detector Technology}

\pstart
The \minos{} detectors are magnetized steel/scintillator tracking/sampling calorimeters designed to measure muon-neutrinos produced by the \numi{} beam.  The active medium comprises \unit[4.1]{cm}-wide, \unit[1.0]{cm}-thick plastic scintillator strips arranged side and encased within aluminum sheets to form light-tight modules of 20 or 28 strips. Modules are combined to form scintillator planes which are then mounted on a \unit[2.54]{cm}-thick steel absorber plate. The detectors are composed of a series of these steel/scintillator planes hung vertically at a \unit[5.94]{cm} pitch. The scintillator strips in successive planes are rotated by $90^{\circ}$ to measure the three dimensional event topology.   The strips were produced in an extrusion process during which a \unit[2.0]{mm}-wide, \unit[2.0]{mm}-deep groove was driven along each strip. The entire strip, apart from the groove, was co-extruded with a \unit[0.25]{mm}-thick $\mathrm{TiO_{2}}$ doped polystyrene reflective layer.  Scintillation light is collected by a \unit[1.2]{mm}-diameter wavelength-shifting fiber that was glued into the groove with optical epoxy. The fiber transports the light to an optical connector located on the edge of the scintillator layer where it is transferred to a clear polystyrene fiber and routed to multi-anode photomultipliers, as shown in Fig.~\ref{fig:optical_readout}. Light emitting diodes are used to produce controlled pulses of light during the data-taking to track gain changes in and characterize the response of the photomultipliers and electronics~\cite{Adamson:2002ze}. The data acquisition (DAQ)~\cite{Belias:2004bj} and timing system synchronize and continuously read out the front-end electronics. Software triggering in the DAQ provides flexible event selection and data processing. GPS timestamps allow data from the two detectors to be synchronized with the beam pulses. The two detectors have different front-end electronics due to the disparate rates of neutrino interactions and cosmic-ray crossings at the two sites. 
\pend

\subsection{The Far Detector}
\pstart
The Far Detector has a mass of \unit[5400]{metric tons} and consists of 486 steel plates arranged in two modules separated by a \unit[1.1]{m} gap. The plates are \unit[2.54]{cm} thick, \unit[8]{m} wide octagons. A \unit[15.2]{kA} magnet coil is routed through a hole in the center of each plane and induces an average field of \unc{\unit[1.27]{T}} in the steel. Each scintillator plane covers the \unit[8]{m} octagon and consists of 192 scintillator strips read out from both ends by Hamamatsu 16-anode photomultipliers. The signals from eight strip-ends, separated by approximately $\unit[1]{m}$ within a single plane, are optically summed onto a single photomultiplier channel. The optical summing pattern is different for the two sides of the detector, allowing for resolution of the eight-fold ambiguity associated with each hit. 
\pend

\pstart
The front-end electronics~\cite{Oliver:2004ek} digitize signals from the photomultipliers with a 14 bit ADC (\unit[2]{fC} precision) when the common dynode signal exceeds approximately $\unit[0.25]{photoelectrons}$ and time-stamps them with a \unit[1.5625]{ns} least significant bit.  To reduce the electronics dead-time, hits are only digitized if more than one of the photomultipliers serving a contiguous group of 20 or 24 planes on one side of the detector are above threshold in coincidence. Data selected for further processing are pedestal suppressed and sent to a trigger farm where software triggering and data processing are performed and GPS timestamps are applied. 
\pend

\pstart
The DAQ has data buffering that is large enough to allow it to wait for the GPS time-stamp of the spill to arrive from the Near Detector over the Internet. This is used to form a bias-free beam trigger by recording all hits in the detector within a $\unit[100]{\mu s}$ window around the time of the spill. To avoid splitting events, the window size is extended for each spill to ensure that the entire window is bounded on both sides by an activity free period of at least \unit[156]{ns}. Finally, all hits in the $\unit[30]{\mu s}$ prior to each trigger are added to the event to provide a mapping of channels that were engaged in digitization at the time of the trigger. Fake spill triggers are also generated to monitor backgrounds. When spill information is not available at the Far Detector site, and for all out-of-spill triggering, candidate events are formed from time sequential blocks of hits bounded on each side by at least \unit[156]{ns} in which no detector activity occurred; trigger algorithms based on the spatial and energy clustering of hits in the detector are then applied to select events of potential interest. The integrated trigger rate in the Far Detector is typically \unit[4]{Hz} and is dominated by cosmic-rays and single photoelectron noise. 
\pend

\subsection{The Near Detector}
\pstart
The \unit[980]{metric ton} Near Detector consists of 282 steel/scintillator planes arranged in a single magnetized module.  Each steel plate has a ``squashed-octagon'' shape \unit[6.2]{m} wide and \unit[3.8]{m} high with a $\unit[30\times 30]{cm^{2}}$ hole offset \unit[56]{cm} from the horizontal center of the detector to accommodate a magnet coil.  The \unit[40]{kA} carried by the coil induces a \unc{\unit[1.17]{T}} field at the neutrino beam center, located \unit[1.49]{m} to the left of the coil. The coil and detector geometry was designed to provide a magnetic field in the region around the beam center that is similar to the field in the Far Detector. The Near Detector has two different types of scintillator planes: partially instrumented and fully instrumented. The planes are smaller than those in the Far Detector and are read out on only one side using Hamamatsu 64 anode photomultipliers. Fully instrumented Near Detector planes have 96 strips and cover a $\unit[13.2]{m^{2}}$ area. These planes are attached to one in every five steel plates along the entire length of the detector. Partially instrumented planes are attached to four out of five plates in the upstream portion of the detector. These planes are comprised of 64 strips covering a $\unit[6.0]{m^{2}}$ area.
\pend

\pstart
The Near Detector is organized in two sections. The upstream 121 steel plates form the calorimeter, which is used to define the interaction vertex, find the upstream portion of muon tracks, and measure the energy of the neutrino induced hadronic shower. In the calorimeter each plate is instrumented with scintillator and the signals from each strip are read out independently. The downstream 161 plates are used as a muon spectrometer and only one in every five is instrumented. Furthermore, in the spectrometer, the signals from four strips are summed onto one electronics channel. The four-fold ambiguity is resolved in the event reconstruction program by extrapolating the muon track found in the upstream portion of the detector.
\pend

%

\pstart
At the typical beam intensity of $\unit[2.2\times 10^{13}]{POT/spill}$, an average of 16 (44) neutrino interactions~\footnote{This includes interactions occurring throughout the detector, many of which cannot be fully reconstructed but which nevertheless induce some activity in the detector.}  occurred in the Near Detector during each $\unit[10]{\mu{}s}$ spill in the LE10/185kA (LE250/200kA) beam configuration. About half of these events occur in the calorimeter region and may be fully reconstructed. The Near Detector readout electronics~\cite{Cundiff:2006yz} is designed to measure neutrino interactions throughout the spill, without dead-time, and with a timing resolution that allows efficient separation of time-adjacent events.  Analog to digital conversion has a floating precision with a minimum least significant bit of \unit[1.4]{fC} and occurs in contiguous \unit[18.8]{ns} time-intervals corresponding to the \unit[53]{MHz} RF of the Main Injector. The electronics has two primary operating modes, ``spill-gate'' and ``dynode'', which are switched in real time.  In spill-gate mode the output from every photomultiplier pixel is digitized continuously in a $\unit[13]{\mu{}s}$ period starting about $\unit[1.5]{\mu{}s}$ before the arrival of neutrinos at the Near Detector. In dynode mode, used for out-of-spill acquisition of cosmic-rays, continuous digitization for a period of \unit[150]{ns} is initiated independently for each photomultiplier when the dynode signal exceeds a programmable threshold. The integrated trigger rate is typically \unit[30]{Hz}.
\pend





\subsection{Event Reconstruction}
\label{sec:reco}

\begin{figure*}
\begin{center}
\includegraphics[width=\textwidth,bb=0 320 389 468,clip]{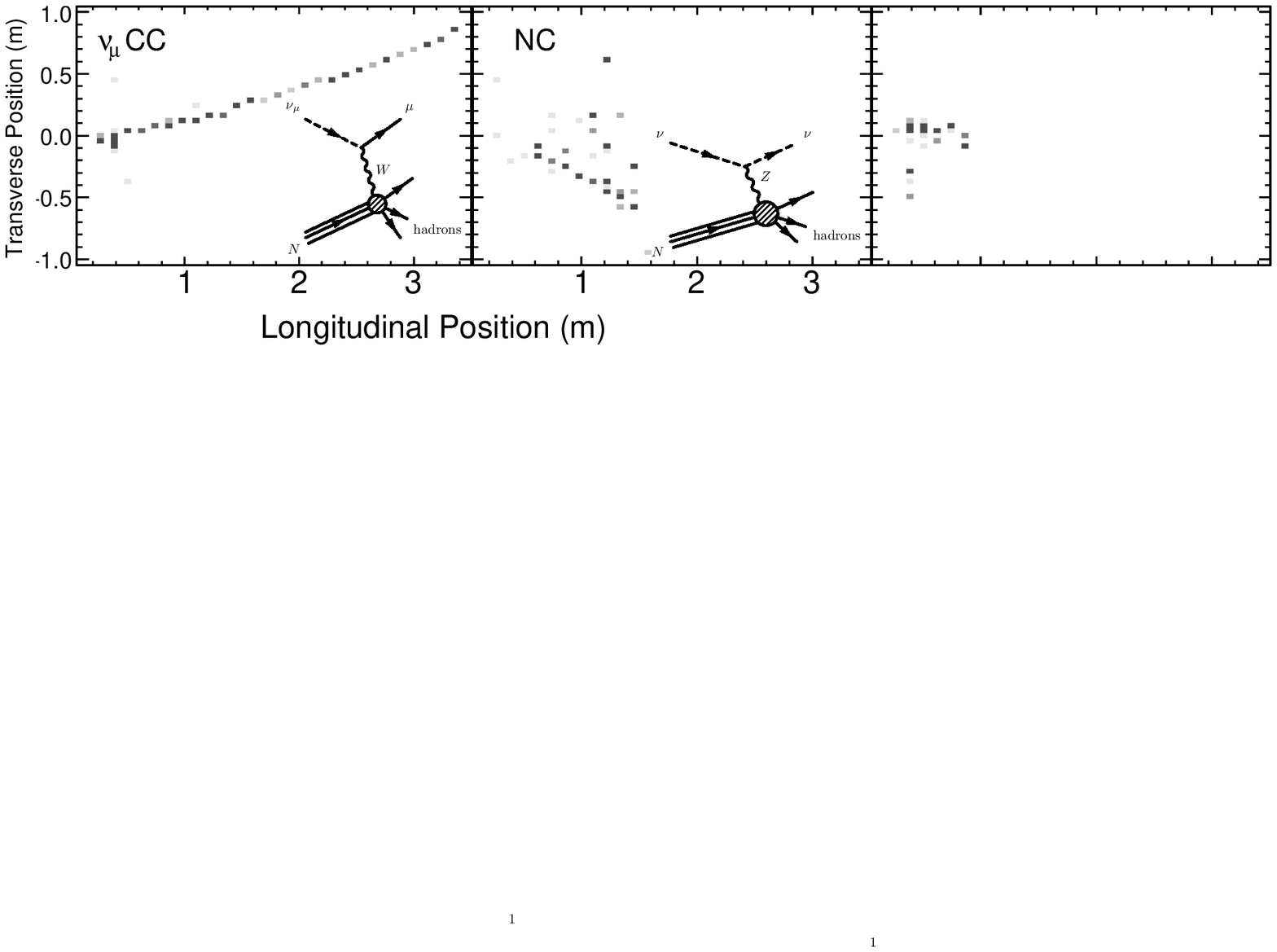}
\caption{\label{fig:cc_nc_events} Muon-neutrino charged-current (\numucc{}) and neutral-current (NC) events simulated in the MINOS detector. Shaded rectangles indicate energy depositions (hits) in the detector's strips. The strips are organized in planes transverse to the beam axis and the strips in successive planes are rotated by $\pm 90^{\circ}$ to provide two orthogonal views of the event. This figure shows the data from one view.}
\end{center}
\end{figure*}

\pstart
The reconstruction procedure uses the topology and timing of hits to identify neutrino interactions inside the detector as well as through-going muon tracks from cosmic-rays or neutrino interactions in the surrounding rock.  Two contained vertex interactions are shown in Fig.~\ref{fig:cc_nc_events}. In \numucc{} events a $W$ boson is exchanged between the neutrino and the target. The final state consists of a muon and hadrons from the recoil system. Neutral-current (NC) events are mediated by $Z$ exchange and only the fraction of energy carried by the recoil system is visible in the final state. The chief goal of the reconstruction procedure is to estimate the visible energy of \numucc{}, \nuecc{}, and \nc{} interactions while also providing a distilled set of quantities describing the event in order to discriminate between the three processes.  As shown in Fig.~\ref{fig:cc_nc_events}, the strongest evidence of a \numucc{} event comes from the presence of a track that penetrates through several detector planes and is sufficiently distinguishable from any additional hits around the track starting point. These additional hits are associated with the hadronic shower created by the recoil system and their pulse-height may be used to estimate the shower energy. Tracks are occasionally found in \nc{} events and additional event-topological quantities are used to discriminate such events from actual \numucc{} events (see Sec.~\ref{sec:evtsel}).
\pend

\begin{figure*}
\centering
\subfloat[\label{fig:event_vz}]{\includegraphics[width=0.65\textwidth]{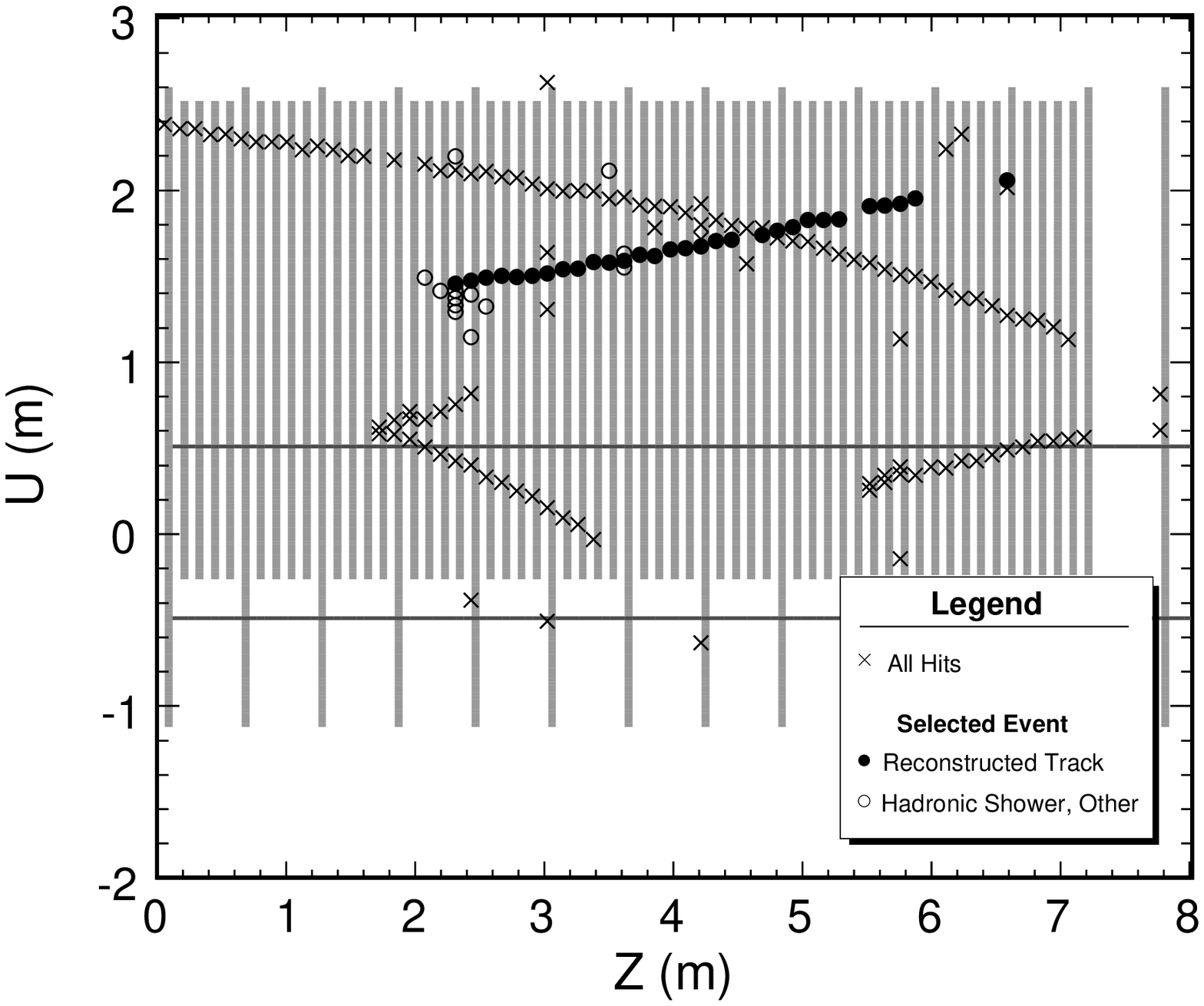} } \\
\subfloat[\label{fig:event_xy}]{\includegraphics[width=0.325\textwidth]{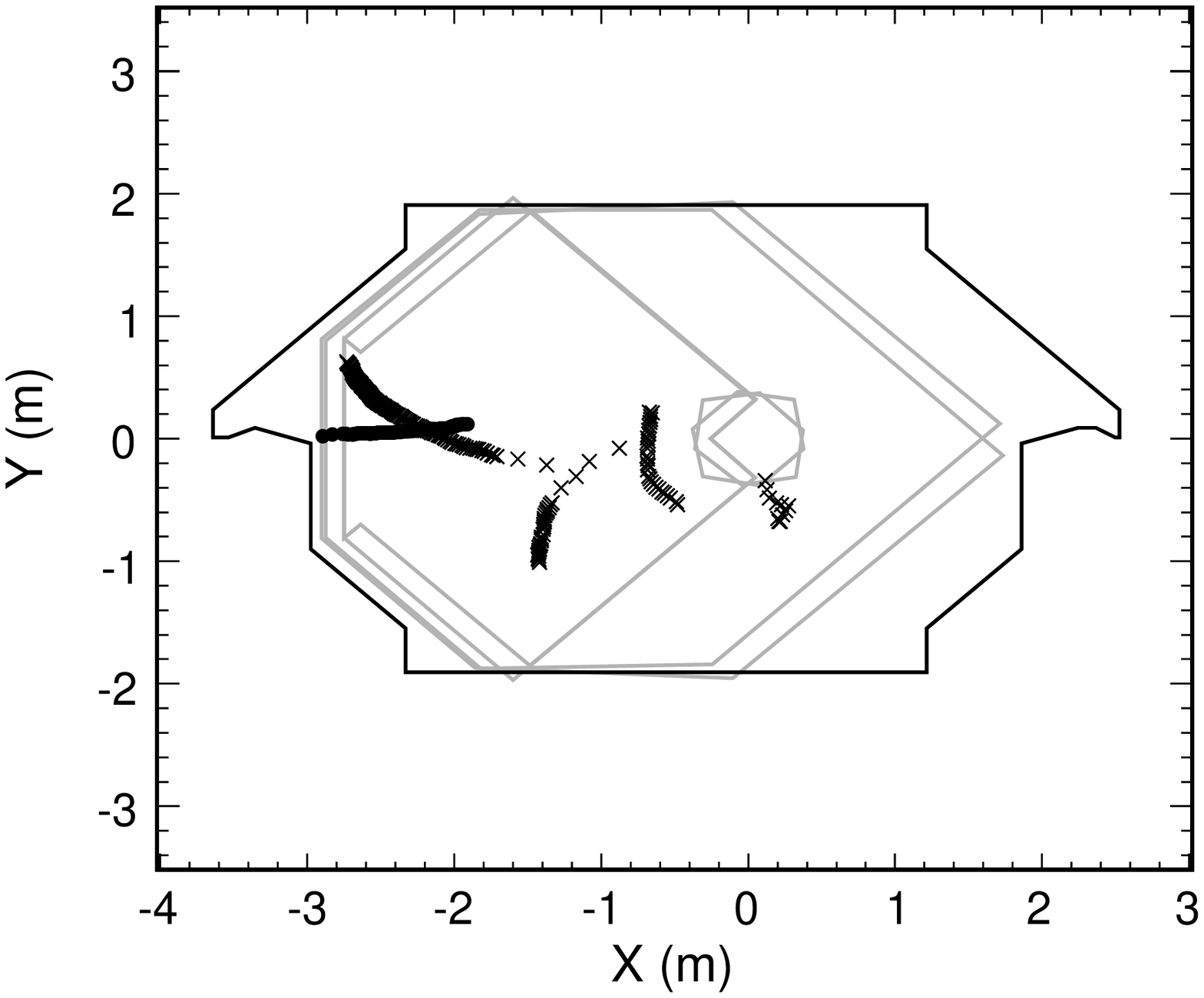}}
\subfloat[\label{fig:event_t}]{\includegraphics[width=0.325\textwidth]{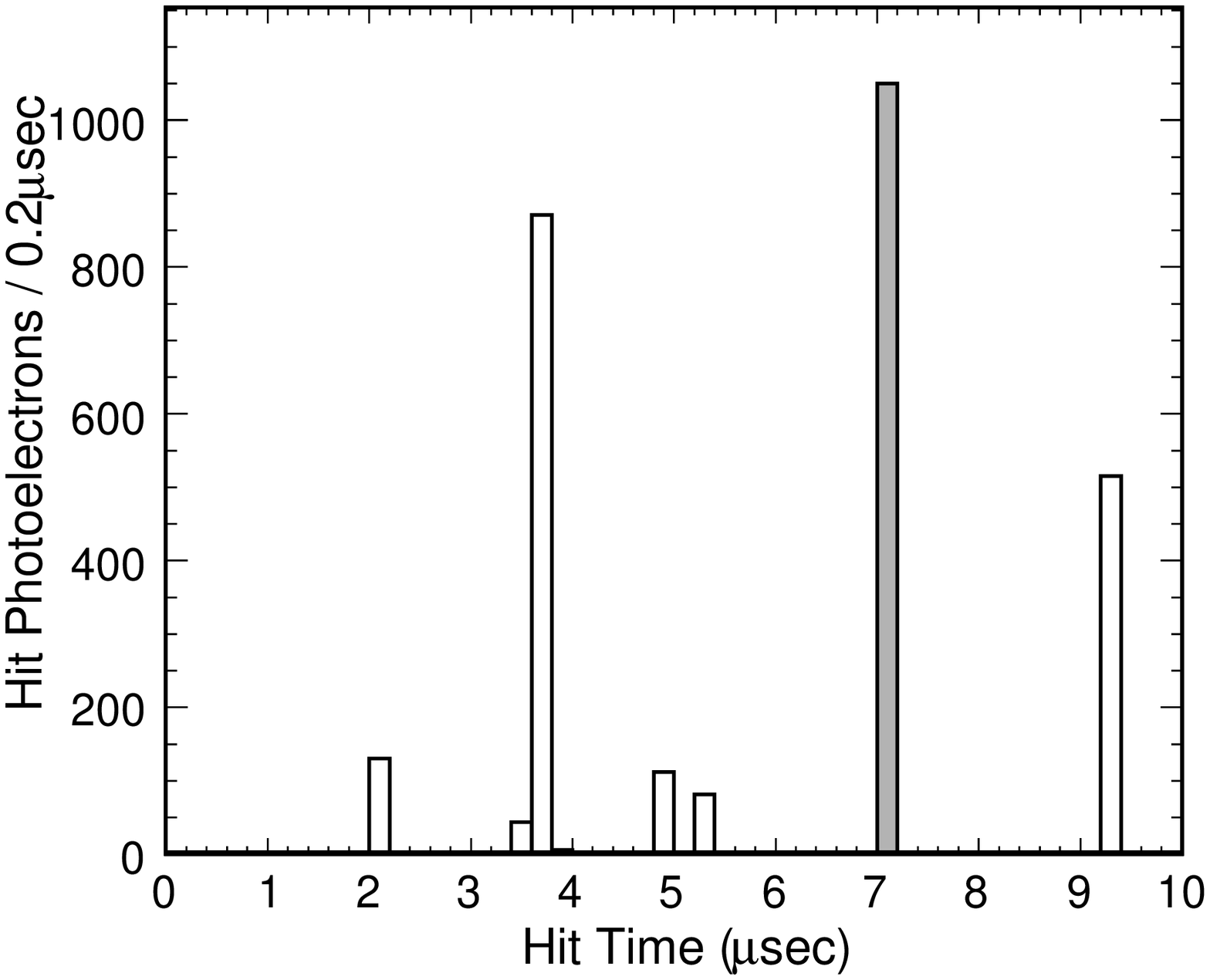}}
\caption{\label{fig:nd_event} One beam spill observed in the Near Detector. For clarity of presentation a spill containing a smaller than average number neutrino events was chosen. Data from the one of the two strip orientation is shown in \protect\subref{fig:event_vz}. Neutrinos are incident from the left and only the upstream section of the detector is shown. Grey vertical bars indicate the scintillator coverage. The timing and topological pattern of hits in the detector has been used to reconstruct and select an event containing a \unit[5.6]{GeV} muon and a \unit[2.6]{GeV} hadronic shower. A beam's eye view of the detector is shown in \protect\subref{fig:event_xy}, along with the reconstructed horizontal and vertical positions of track hits in each detector plane. Figure \protect\subref{fig:event_t} shows the detector signal as a function of time, with signals from the selected event shaded. The bin width is about ten times larger than the detector's timing resolution. }
\end{figure*}


\pstart
The high intensity of the \numi{} beam leads to multiple neutrino interactions inside the Near Detector in each beam spill. The first stage in the reconstruction procedure divides the activity in the detector into one or more events, each of which contains hits that are localized in space and time. Figure~\ref{fig:nd_event} shows the activity recorded in the Near Detector for a single LE010/185kA beam spill. The plot also indicates how hits from a single interaction can be identified using timing and spatial information. In the Far Detector, where the rate is much lower, there is rarely more than one event per beam spill, and most spills actually contain no neutrino interactions.
\pend

\pstart
A track-finding algorithm is then applied to each event. This algorithm uses a Hough Transform~\cite{hough} to find track segments, which are then chained together (taking into account timing and spatial correlations) to form longer tracks. The track momentum is estimated from range~\footnote{When reconstructing the momentum by range, the track is assumed to be due to a muon and the range tables of \cite{Groom:2001kq} are employed.} if the track stops within the detector, or from a measurement of its curvature in the toroidal magnetic field if it exits. The curvature measurement is obtained from fitting the trajectory of the track using a Kalman Filter technique, which takes into account bending of the track from both multiple Coulomb scattering and the magnetic field. This procedure also gives an indication of the charge of the reconstructed track.  For typical muon tracks produced by beam \numucc{} interactions, the momentum resolution is approximately \unc{5\%} for range, and \unc{10\%} for curvature measurements.
\pend

\pstart
Showers are constructed from clusters of strips that are localized in space and time. The energy of a shower is computed from the summed pulse-height of the individual hits, where the pulse-height contribution of any reconstructed tracks that share the same hit is subtracted. The shower energy calibration is discussed in Sec.~\ref{sec:shwr_calib}. The energy resolution for neutrino induced hadronic showers is approximately 59\% at \unit[1]{GeV} (32\% at \unit[3]{GeV}).
\pend



\pstart
The total reconstructed energy of each event is estimated by summing the energy of the most energetic track with the energy of any shower present at the upstream end of the track. We select \numucc{} interactions by requiring that the event has at least one well reconstructed track with a starting-point, interpreted as the neutrino interaction point (vertex), in the fiducial volume. In the Near Detector, the fiducial volume is a cylinder of radius \unit[1]{m} from the beam center~\footnote{The radius is calculated from the beam center in the first detector plane. The cylinders follow the z axis of the detectors which differs from the beam axis by \unit[58]{mrad}.} and length \unit[4]{m} beginning \unit[1]{m} downstream of the front face of the detector. The Far Detector fiducial volume is a cylinder of radius \unit[3.7]{m} from the detector center. Vertices are required to be at least \unit[50]{cm} from the front and rear planes of the two detector modules and, to assure a track that is long enough to be analyzed, greater than \unit[2]{m} upstream of the last plane of module two. Tracks are required to have a negative charge to suppress \numubar{} and only events with neutrino energy less than \unit[30]{GeV} are used so as to preferentially select neutrinos from  $\pi$, rather than $K$, decays.  A final set of criteria are applied to remove rare periods in which the magnetic field coil was not energized or in which the high voltage was not on in some portion of the detector. 
\pend

\pstart
The resulting sample in the Near Detector is 89\% pure \numucc{} (91\% in the Far Detector, assuming no neutrino oscillations) with the dominant background coming from neutral-current events in which a (usually short) track was reconstructed. Section~\ref{sec:evtsel} describes the way in which event-topological quantities were used to improve the sample purity. In the following subsections we describe some characteristics of the dataset before purification.
\pend


\subsubsection{Near Detector Data}
\label{sec:nd_dist}

\pstart
We have reconstructed more than $2.6\times10^{6}$ neutrino events in the Near Detector in several different beam configurations (see Tab.~\ref{tab:beam-config}). These data have been used to verify and improve the quality of our neutrino interaction simulation, detector calibration, event reconstruction, neutral current rejection procedure (described in Sec.~\ref{sec:evtsel}), and beam flux calculation (Sec.~\ref{sec:skzp}). Since we are ultimately interested in predicting the Far Detector energy spectrum, it is not necessary that the Monte Carlo simulation reproduce the Near Detector data exactly. Instead we must have confidence that any discrepancies have causes which are common to both detectors such that the Near Detector data may be used to improve the Far Detector prediction. For example, uncertainties in the neutrino flux, neutrino \xsecs{} and energy resolutions (though not the absolute energy scale) may be mitigated by using Near Detector measurements. 
\pend


\begin{figure}
\centering
\subfloat[]{\includegraphics[width=0.4\textwidth,clip]{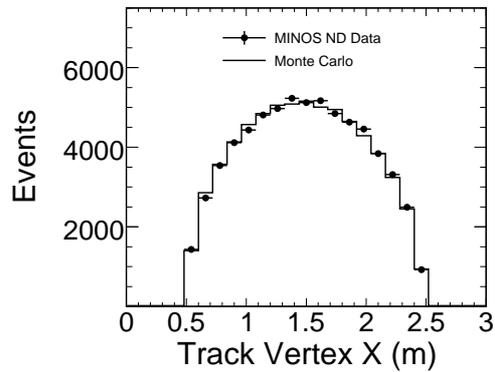}}
\\
\subfloat[]{\includegraphics[width=0.4\textwidth,clip]{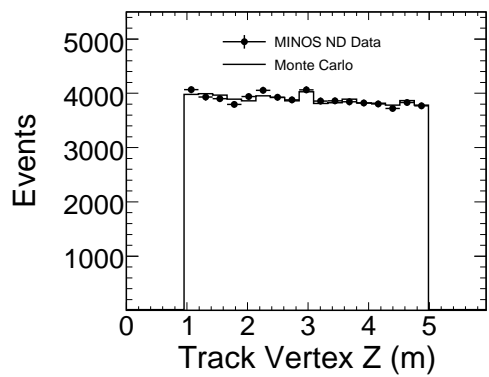}}
\\
\subfloat[]{\includegraphics[width=0.4\textwidth,clip]{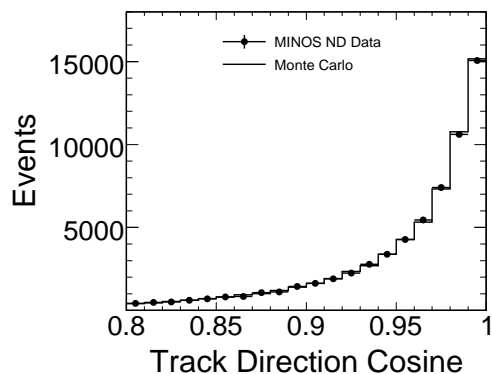}}
 \caption{Distributions of basic reconstructed quantities in the Near Detector for events reconstructed and selected using the procedure described in the text.  The distributions are normalized to the same numbers of events. The left-hand and center plot show the distribution of reconstructed track vertices for $X$ (perpendicular to detector axis) and $Z$ (parallel to detector axis) detector co-ordinates, and the right-hand plot shows the distribution of the reconstructed track direction relative to the incident neutrino beam. \label{fig:nd_lowlevel}}
\end{figure}

\begin{figure}[h]
\centering
\includegraphics[width=\columnwidth]{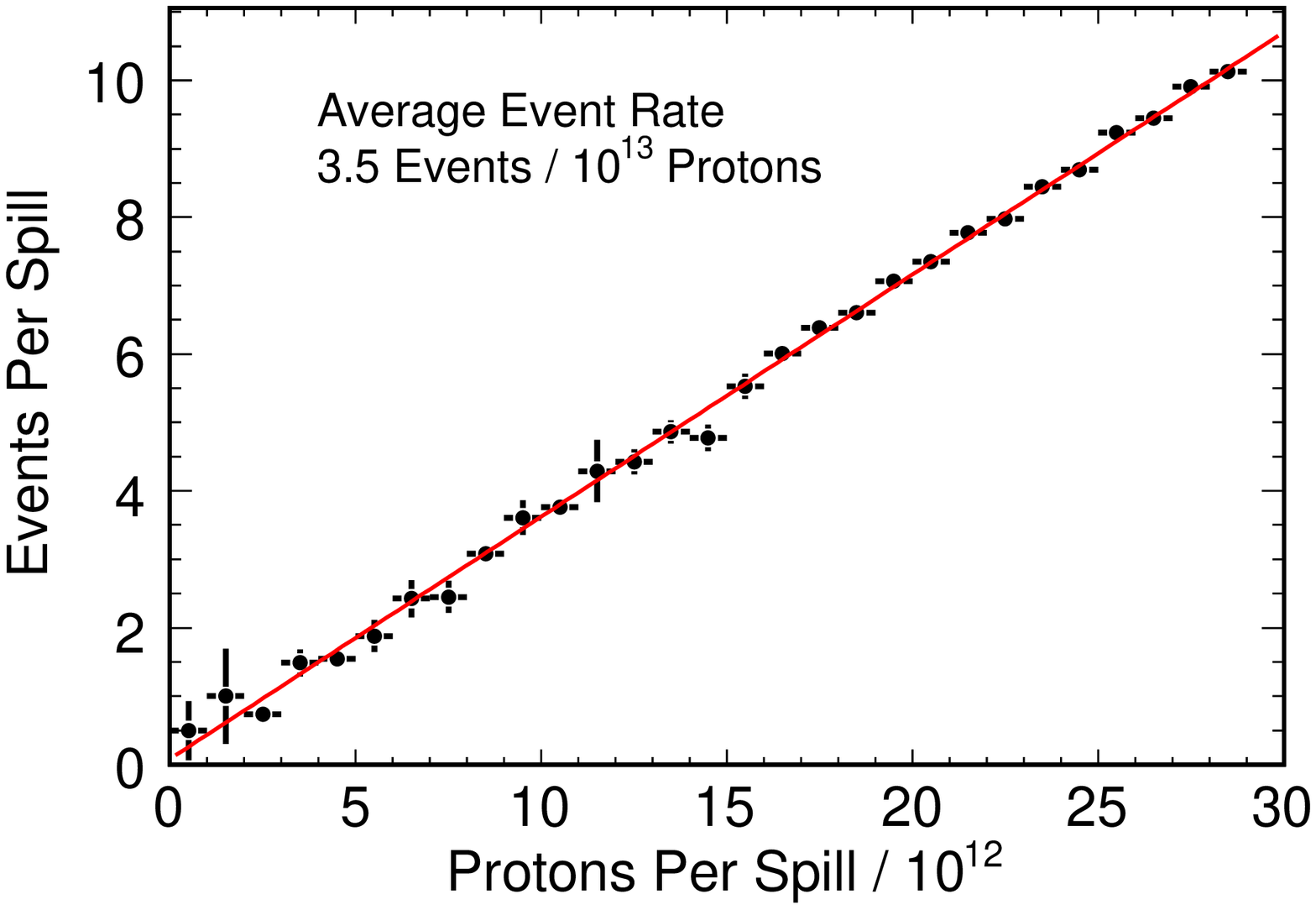} 
\caption{\label{fig:events_per_ppp} Mean number of reconstructed events per Near Detector spill as a function of spill intensity.} 
\end{figure}

\begin{figure*}[h]
\centering
\includegraphics[width=\textwidth]{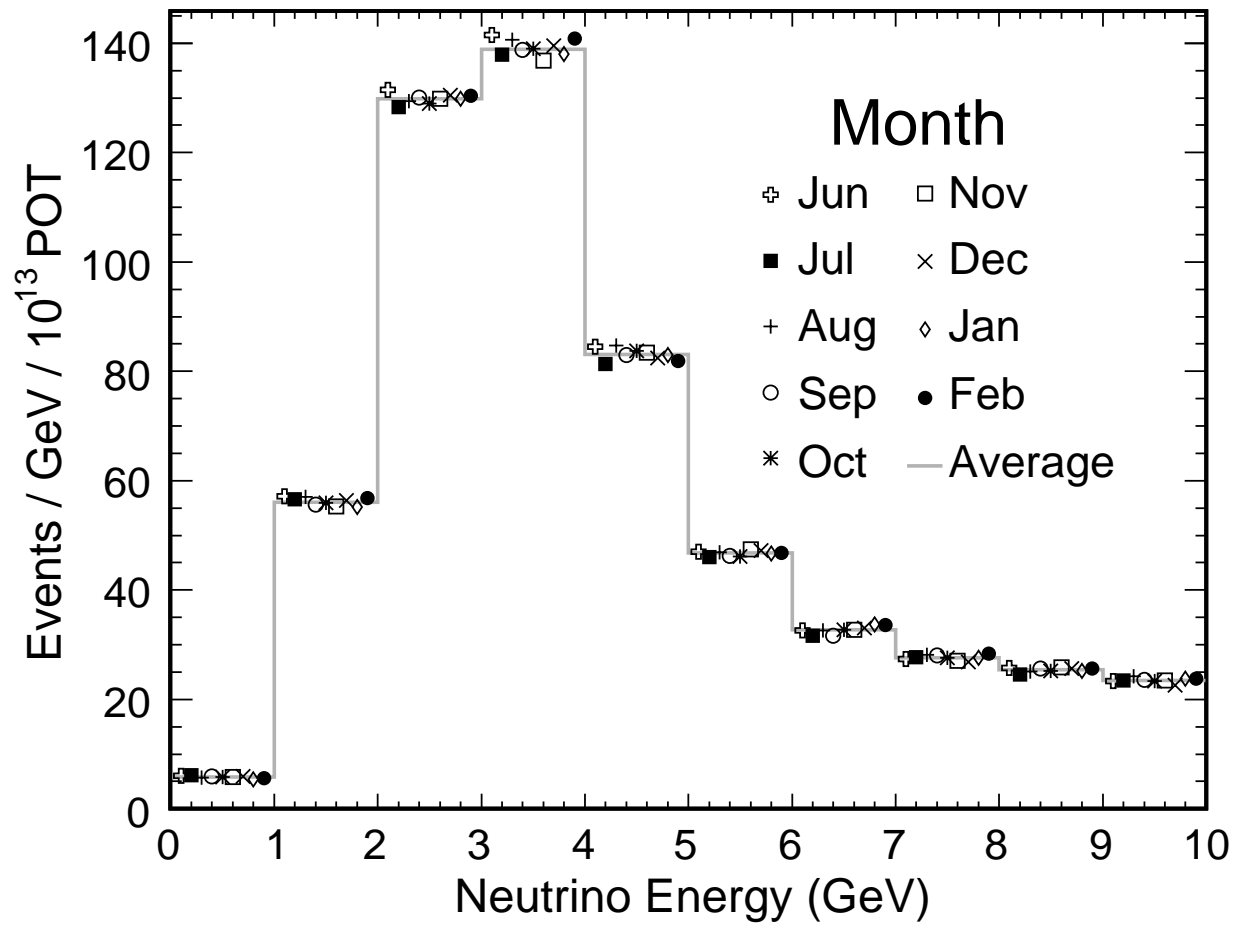} 
\caption{\label{fig:enu_by_month} Reconstructed neutrino energy distributions by calendar month. The distributions are normalized to protons on target and only data obtained in the LE10/185kA configuration is included. Markers representing individual months are offset to clarify the presentation and the line represents the average over all months. } 
\end{figure*}

\pstart
The Monte Carlo simulation reproduces many reconstructed quantities in the Near Detector, including the vertex distributions and track angular distribution shown in Fig.~\ref{fig:nd_lowlevel}. Figure~\ref{fig:events_per_ppp} shows the mean number of reconstructed events as a function of beam intensity ranging between $5\times 10^{11}$ and \unit[$2.7\times 10^{13}$]{POT/spill}. The linearity of the curve indicates that the Near Detector is able to measure individual neutrino interactions with little dependence on the interaction rate. The behavior of the curve near the origin suggests a negligible background from non-beam related events. The reconstructed neutrino energy spectrum measured during several months is overlaid in Fig.~\ref{fig:enu_by_month}. The average spill intensity in June 2005 is \unit[$1.65\times 10^{13}$]{POT/spill} and rises to a maximum of \unit[$2.4\times 10^{13}$]{POT/spill} in November 2005. The distributions are consistent within statistical errors and show no rate dependence. We therefore conclude that the Near Detector event reconstruction is stable over this range of spill intensities, and does not introduce any observable biases in the reconstructed neutrino energy spectrum. Consequently, we do not assign any intensity based uncertainty in the detector efficiency.
\pend

\subsubsection{Far Detector Data}
\label{sec:fd_dist}

\pstart
Several criteria were applied to assure the integrity of the Far Detector data. These explicitly exclude data taken during magnet coil and high voltage power failures and periods where the GPS timing system was not operational. The POT-weighted live-time of the Far Detector was 98.9\% during the period May 20, 2005 -- Feb 25, 2006 in which the oscillation dataset was accumulated. Two additional conditions are applied to Far Detector data in order to reduce contamination from cosmic ray events to a negligible level. First, the direction cosine of the primary reconstructed track with respect to the beam direction must be greater than 0.6 in order to reject the high angle tracks typical of cosmic rays. Furthermore, the event time must lie within a $\unit[50]{\mu s}$ window around the predicted time of the beam spill at the Far Detector site.
\pend

\begin{figure}
\begin{center}
  \includegraphics[width=\columnwidth]{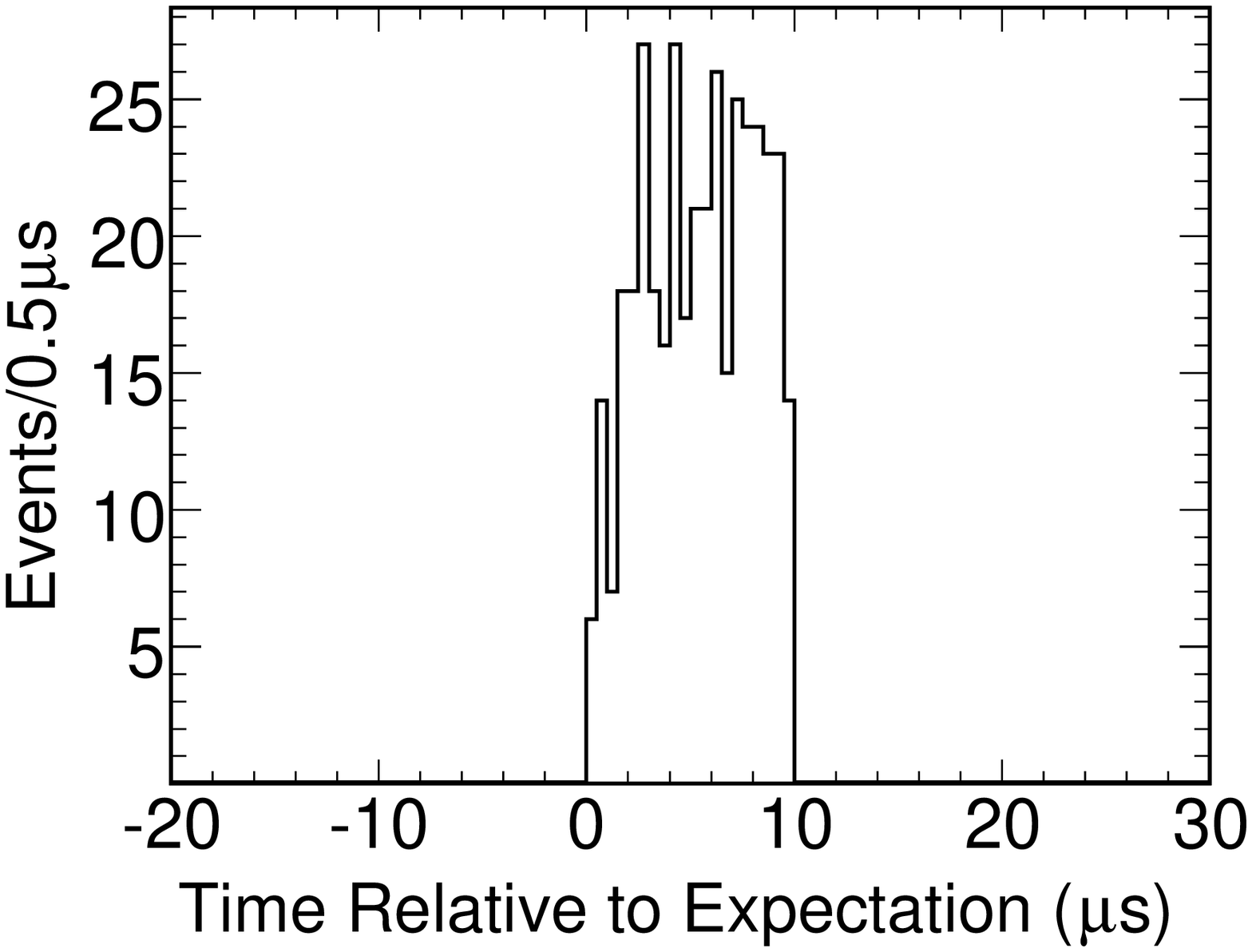}
  \caption{Distribution of the event time of the 384 Far Detector neutrino events relative to the time of the nearest beam spill~\cite{tofpaper}.}
   \label{fig:fdtiming}
\end{center}
\end{figure}

\begin{figure}[h]
\centering
\subfloat[\label{fig:fd_pot}]{\includegraphics[width=0.7\textwidth]{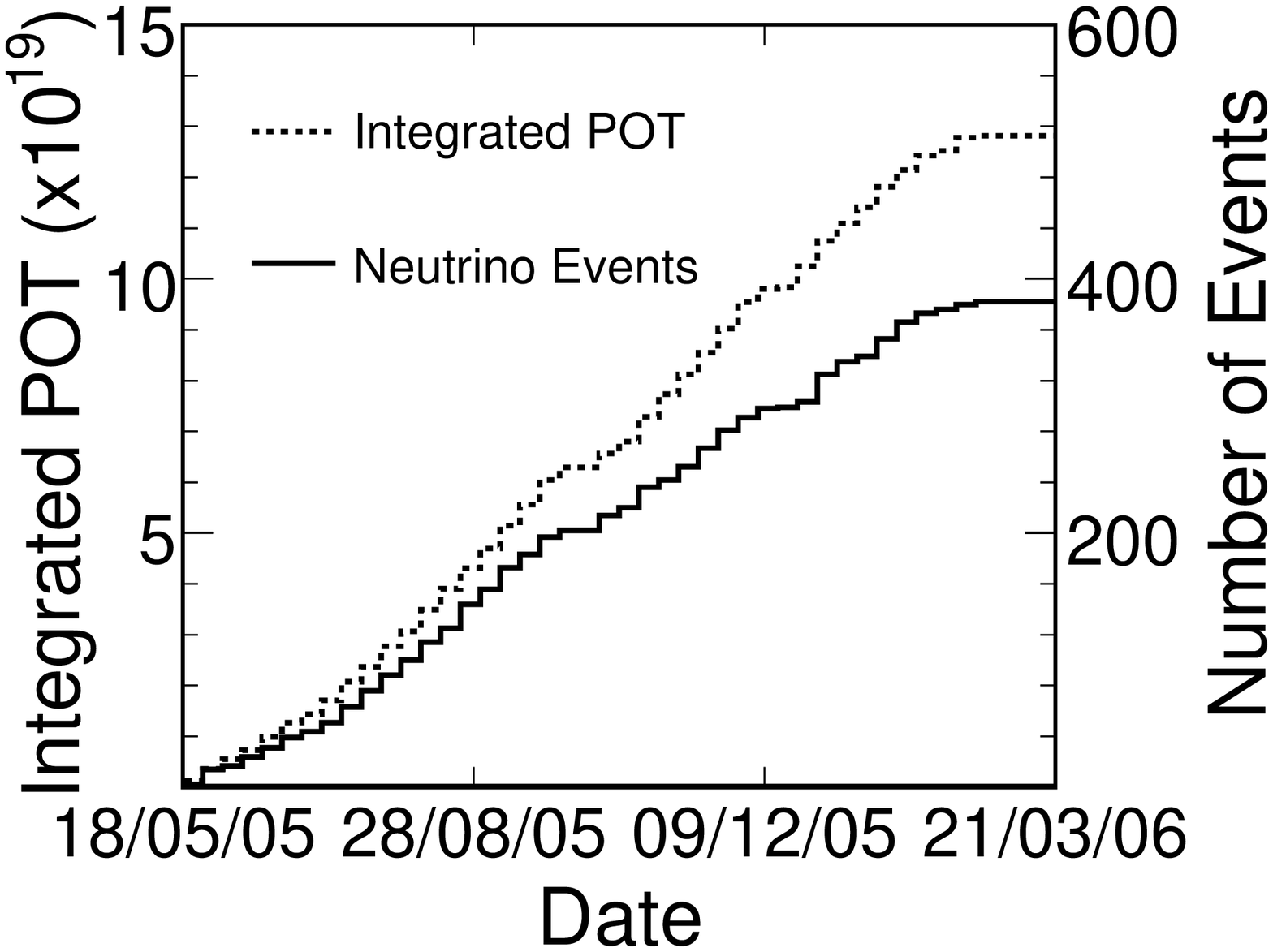} }
\\
\subfloat[\label{fig:nu_per_pot}]{\includegraphics[width=0.7\textwidth]{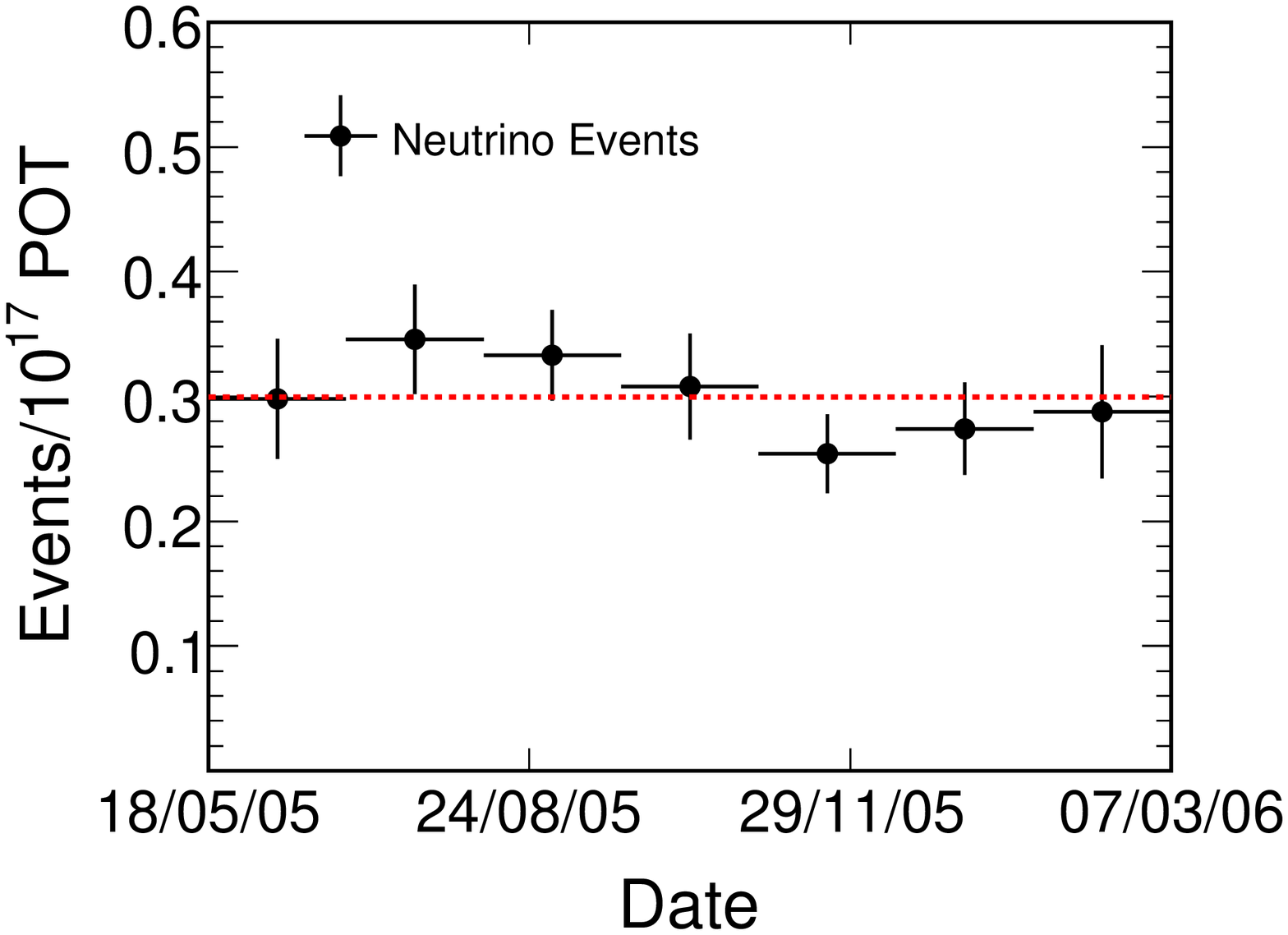}}
 \caption{\label{fig:ccpot} \protect\subref{fig:fd_pot} Cumulative distributions of the total number of protons on target and the number of Far Detector neutrino events observed as a function of time. \protect\subref{fig:nu_per_pot} Number of observed Far Detector neutrino interactions per $10^{17}$ protons on target as a function of time.}
\end{figure}

\pstart
Figure~\ref{fig:fdtiming} shows the time (relative to the spill time) of the 384 selected neutrino events that satisfy the criteria described above. The width of this distribution is consistent with the Main Injector spill length, and there is no evidence of background contamination within the $\unit[50]{\mu s}$ window.  Figure~\ref{fig:fd_pot} shows the cumulative distributions of the number of total protons on target and the number of Far Detector neutrino events as a function of time. The two distributions follow each other closely. Figure~\ref{fig:nu_per_pot} shows the number of neutrinos per $10^{17}$ protons on target and exhibits no significant time dependence.  
\pend

\begin{figure}
\centering
\subfloat[]{\includegraphics[width=0.4\textwidth,clip]{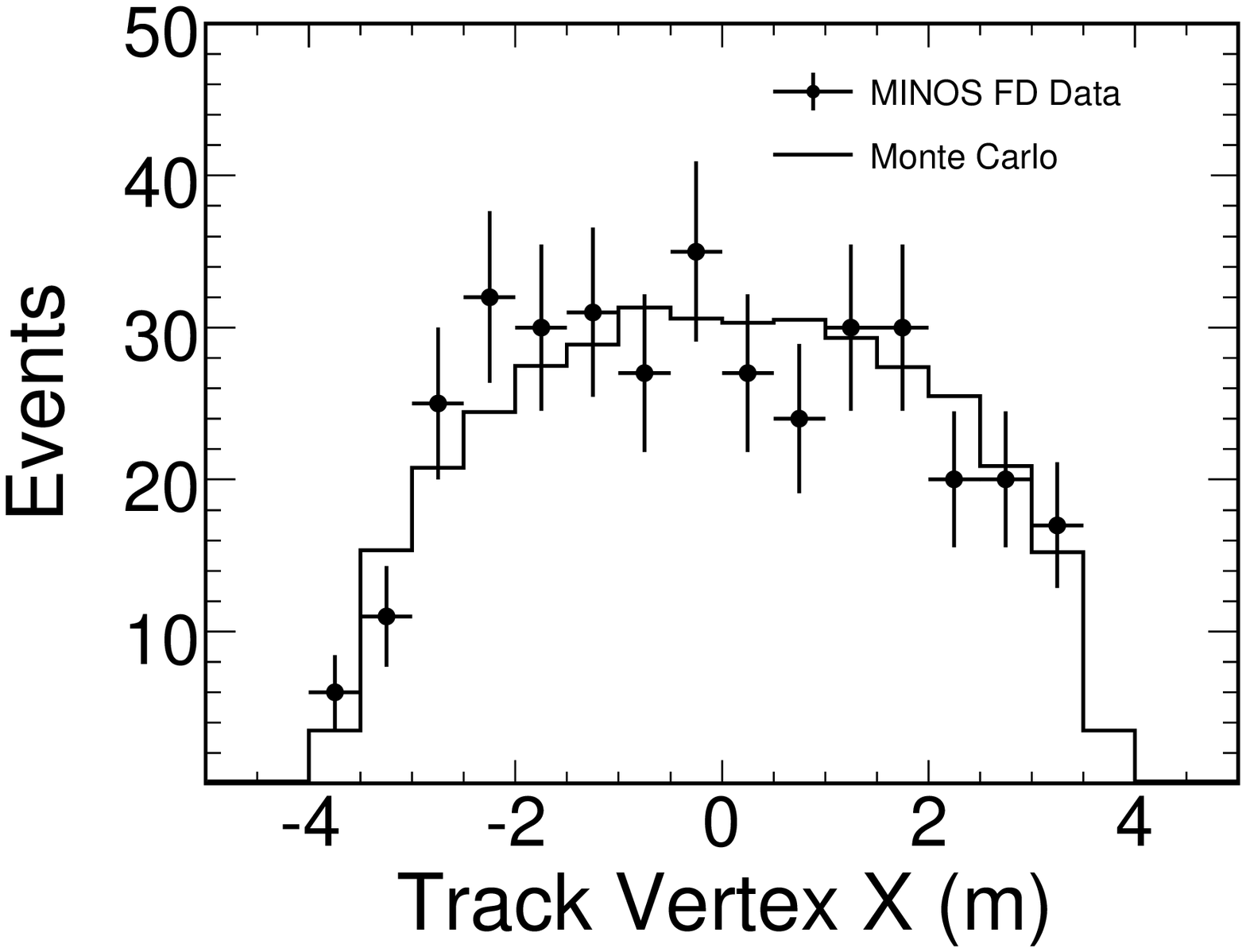}}
\\
\subfloat[]{\includegraphics[width=0.4\textwidth,clip]{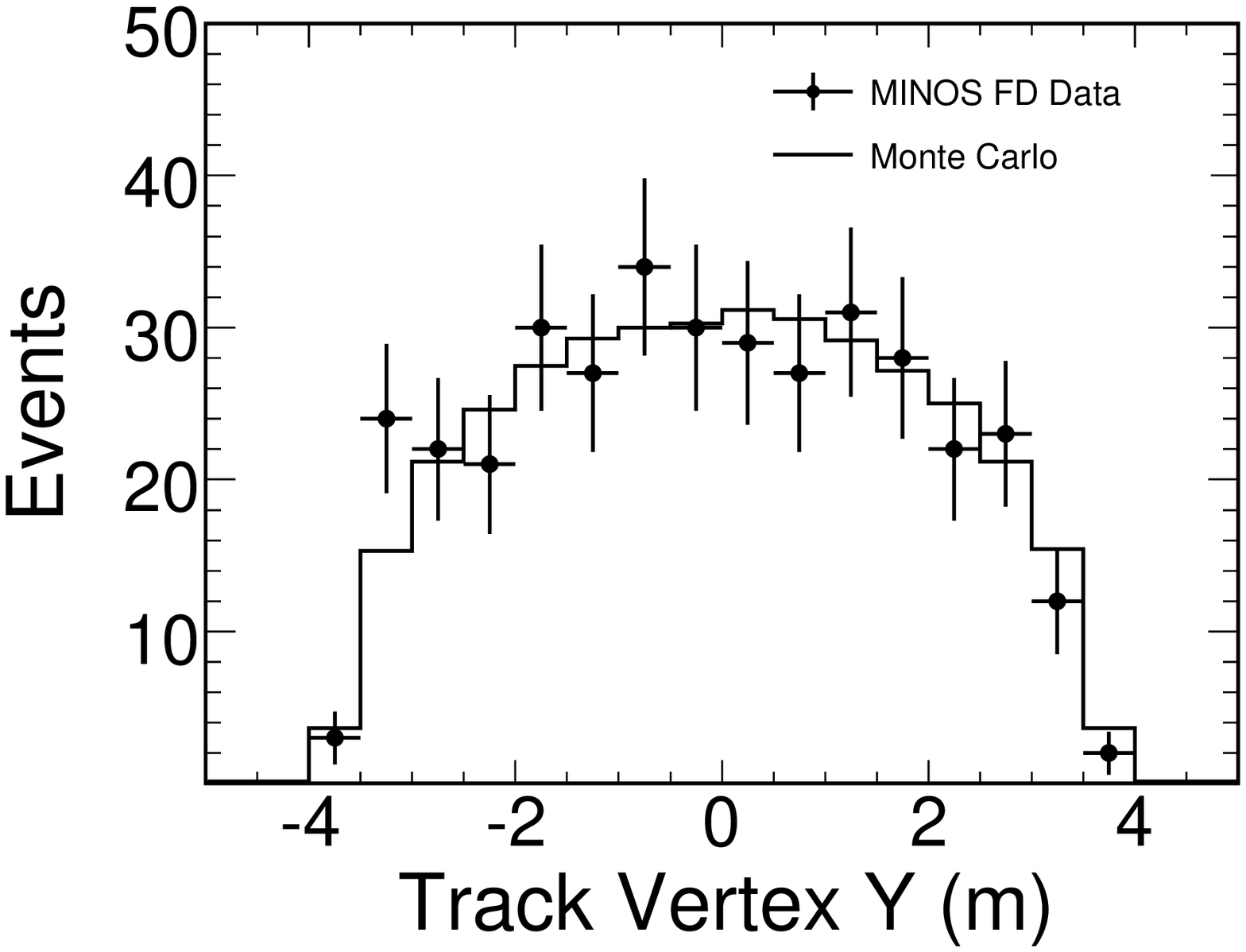}}
\\
\subfloat[]{\includegraphics[width=0.4\textwidth,clip]{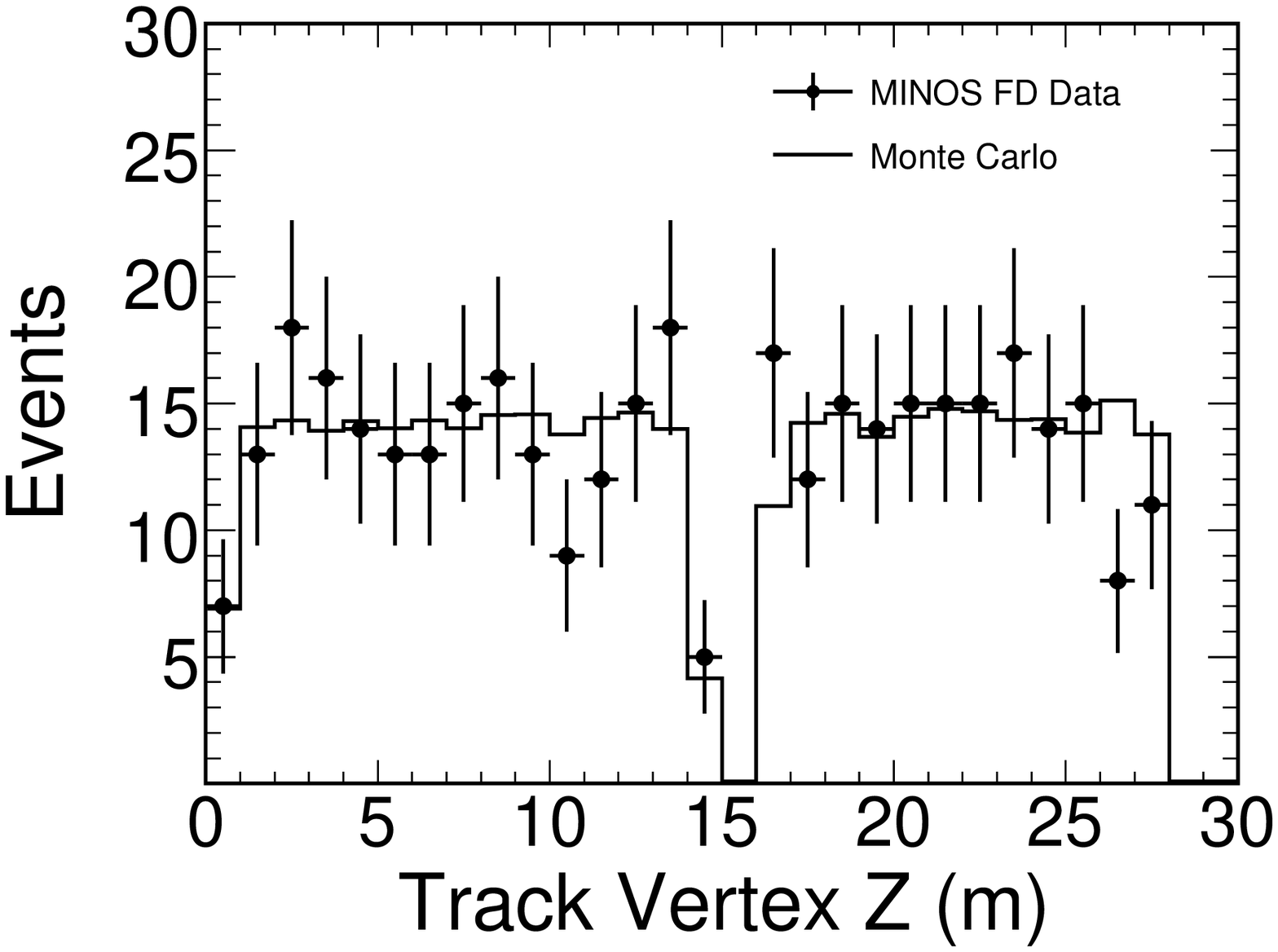}}
\caption{Distributions of track vertices for Far Detector data and Monte Carlo simulation. The Monte Carlo distributions are normalized to the same number of events as the data. \label{fig:fd_vtx}}
\end{figure}

\pstart
A visual scan of the events was carried out in order to ensure that there was no background contamination in the selected event sample and that the events were well reconstructed. Figure~\ref{fig:fd_vtx} shows the reconstructed vertex distributions of the Far Detector events.  The Monte Carlo distributions are normalized to the same number of events as the data to account for the possible effects of oscillations and are in agreement with the data. Additional comparisons between the data and simulation are shown in the sections that follow.
\pend


\clearpage

\subsection{Modeling Neutrino Interactions}
\label{sec:mc}

\begin{figure}
\centering
\includegraphics[width=\columnwidth]{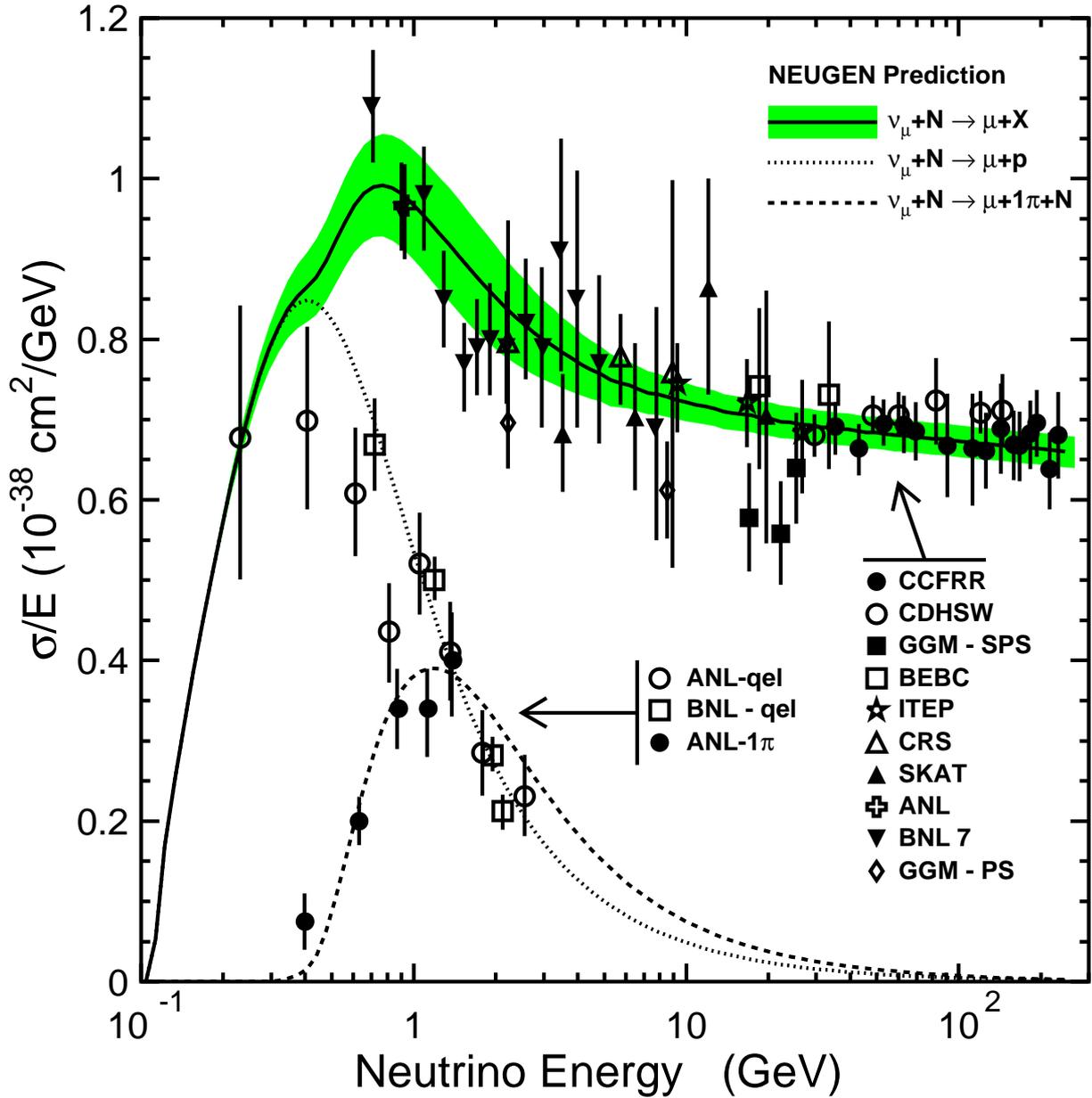}
\caption{\label{fig:ng_xsec}\ngthree{} calculation of  muon-neutrino charged-current \xsecs{} per nucleon on an isoscalar target.  The \xsec{} per GeV is shown as a function of the neutrino energy for inclusive scattering, quasi-elastic scattering and single pion production.  The calculation is compared with experimental data tabulated by~\cite{durham}.  The shaded band corresponds to the \xsec{} uncertainties described in the text.}
\end{figure}
 
\pstart
Neutrino interactions are modeled by the \ngthree{}~\cite{Gallagher:2002sf} program. \ngen{} simulates both quasi-elastic and inelastic neutrino scattering. The latter includes a Rein-Sehgal~\cite{Rein:1980wg} based treatment of neutrino induced resonance production, charged- and neutral-current coherent pion production and a modified leading order deep-inelastic scattering (DIS) model~\cite{Bodek:2004ea} extended to improve the treatment in the transition region between DIS and resonant production. KNO scaling~\cite{Koba:1972ng} is used to calculate the final state multiplicity in the DIS regime. Hadrons produced in the neutrino scattering are allowed to interact while exiting the target nucleus (``final state interactions''). The final state interaction calculation incorporates pion elastic and inelastic scattering, single charge exchange and absorption~\cite{Ransome:2005vb}.  The calculation is benchmarked by a comparison of final states in $\nu + d$ and $\nu + \mathrm{Ne}$ interactions as measured in the BEBC and ANL--\unit[12]{ft} bubble chambers~\cite{Merenyi:1992gf}.
\pend

\pstart
Figure~\ref{fig:ng_xsec} shows the \numucc{} \xsec{} as a function of neutrino energy in the laboratory frame. Based on a comparison of the model predictions to independent data, some of which~\cite{durham} is shown in Fig.~\ref{fig:ng_xsec}, we assign a systematic uncertainty of 3\% on the normalization of the DIS ($W>\unit[1.7]{\gevcsq{}}$) \xsec{}, and a 10\% uncertainty in the normalization of the single-pion and quasi-elastic \xsecs{}. We estimate a 20\% uncertainty in the relative contribution of non-resonant states to the 1$\pi$ and 2$\pi$ production \xsecs{} for $W<\unit[1.7]{\gevcsq{}}$. This uncertainty was determined from the parameter uncertainties and variations observed in fits to both inclusive and exclusive channel data, and in fits to data in different invariant mass regions. Final state interactions are expected to have a significant effect on the visible energy of the hadronic final state~\cite{Kordosky:2006gt}.  In particular there are significant uncertainties in the rate of pion absorption, the mechanism for transferring the pion's energy to a nucleon cluster, and the amount of energy eventually visible to the detector. We account for these uncertainties by studying the shift in the reconstructed shower energy when we turn the simulation of final-state interactions off, and when we modify the simulation so that all of an absorbed pion's energy is lost. We find that the predicted response to hadronic showers changes by approximately 10\%~\cite{Kordosky:2006gt} in these two extreme cases and use this as a conservative estimate of the uncertainty on the absolute hadronic energy scale.
\pend

\pstart
 The \minos{} detector simulation is based on \geantth{}~\cite{g3man} and is used to generate raw energy depositions (\geant{} hits) which serve as the input to our detector response model. The simulation randomly samples neutrinos from the flux predicted by the beam simulation (Sec.~\ref{sec:beam_sim}) and traces them through the Near and Far Detector halls. Events are generated inside the detectors as well as in the surrounding support structure and rock. The simulation includes a detailed geometric model of the detector which describes the material crossed by neutrinos and neutrino induced tracks to within \unc{1\%} plane-to-plane and \unc{0.3\%} averaged over the detector. The position of individual scintillator strips was determined with a precision of approximately \unit[1]{mm} using cosmic-ray tracks. The magnetic field is modeled via finite element analysis driven by bench measurements of the steel B--H curve. Since performing this analysis we have recalibrated our field and found that it increased by 12.3\% and 9.2\% averaged over the fiducial volumes in the Near and Far Detectors, respectively. This recalibrated field shifts the momentum scale for muons exiting the Near Detector by approximately 6.2\% (4.6\% in the Far Detector) but does not significantly affect the scale for muons which stop in the detector. 
\pend



\pstart
The detector response simulation has been tuned by directly incorporating the channel-by-channel calibration constants determined by the detector calibration procedure (see Sec.~\ref{sec:calib} below). Additional data, measured in pre-installation tests, were used to constrain details of the scintillator and photomultiplier response models. The simulation has been benchmarked against through-going and stopping cosmic rays in the Near and Far Detectors, beam neutrino events in the Near Detector, as well as a series of test-beam measurements~\cite{mike_thesis,tricia_thesis,jeff_thesis,anatael_thesis} collected with a scaled down version of the \minos{} detectors in the CERN PS East Hall. The test-beam measurements are used to fix the energy scale and validate the simulation of electromagnetic and hadronic showers. The hadronic shower code \gcalor{}~\cite{Zeitnitz:1994bs} was found to be in relatively good agreement with the data (see Sec.~\ref{sec:shwr_calib}) and is used in our simulations.  
\pend





\subsection{Detector Calibration}
\label{sec:calib}

\begin{figure}
\begin{center}
\includegraphics[width=\columnwidth]{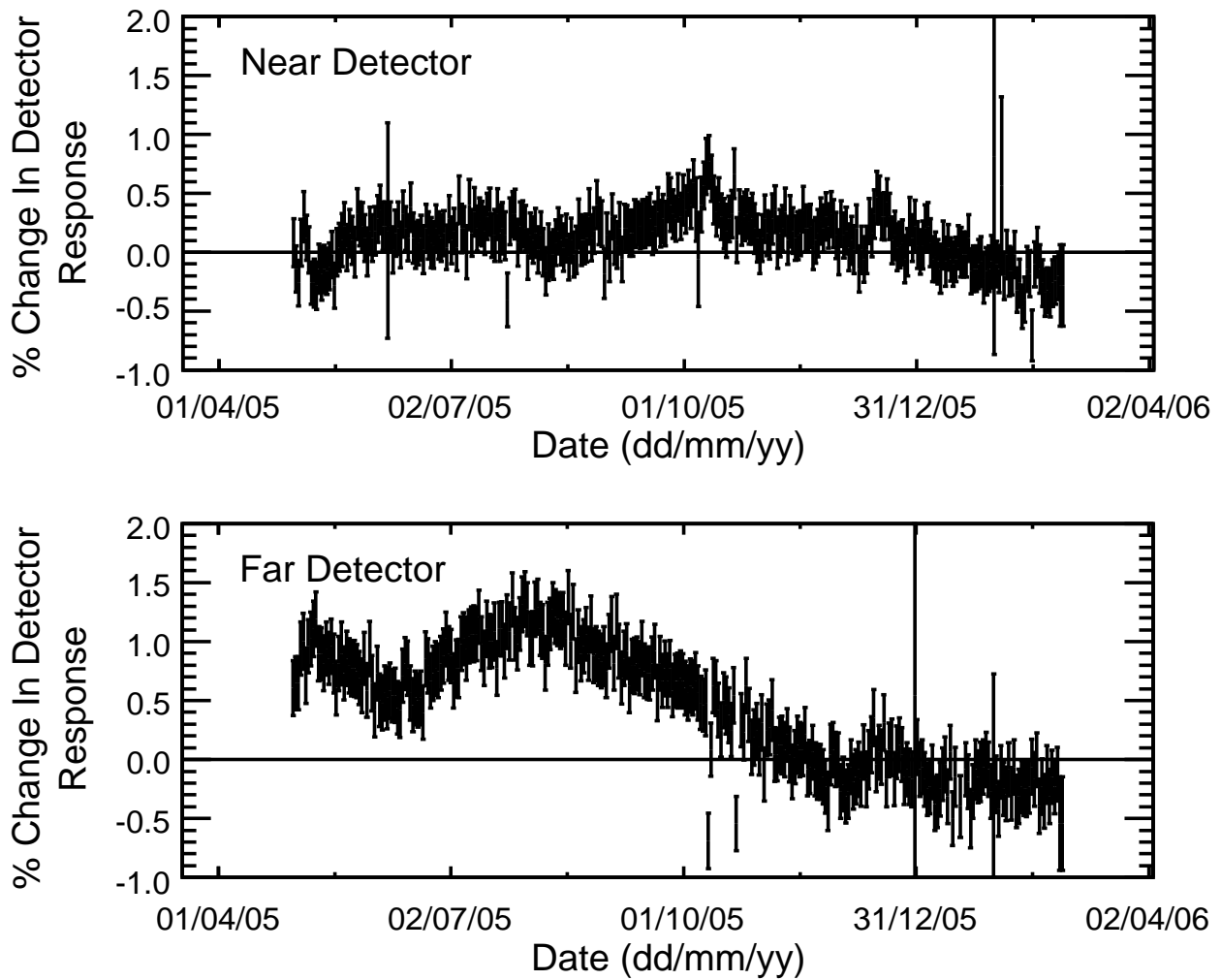}
\caption{\label{fig:drift} Variations in the median signal per plane deposited by through-going cosmic-ray muons observed during the data-taking period covered by this paper. The time dependence is largely due to variations in the environmental conditions in the Near and Far Detector halls and aging of the scintillator.  The zero point on the ordinate is arbitrary.}
\end{center}
\end{figure}

\begin{figure}
\begin{center}
\includegraphics[width=\columnwidth]{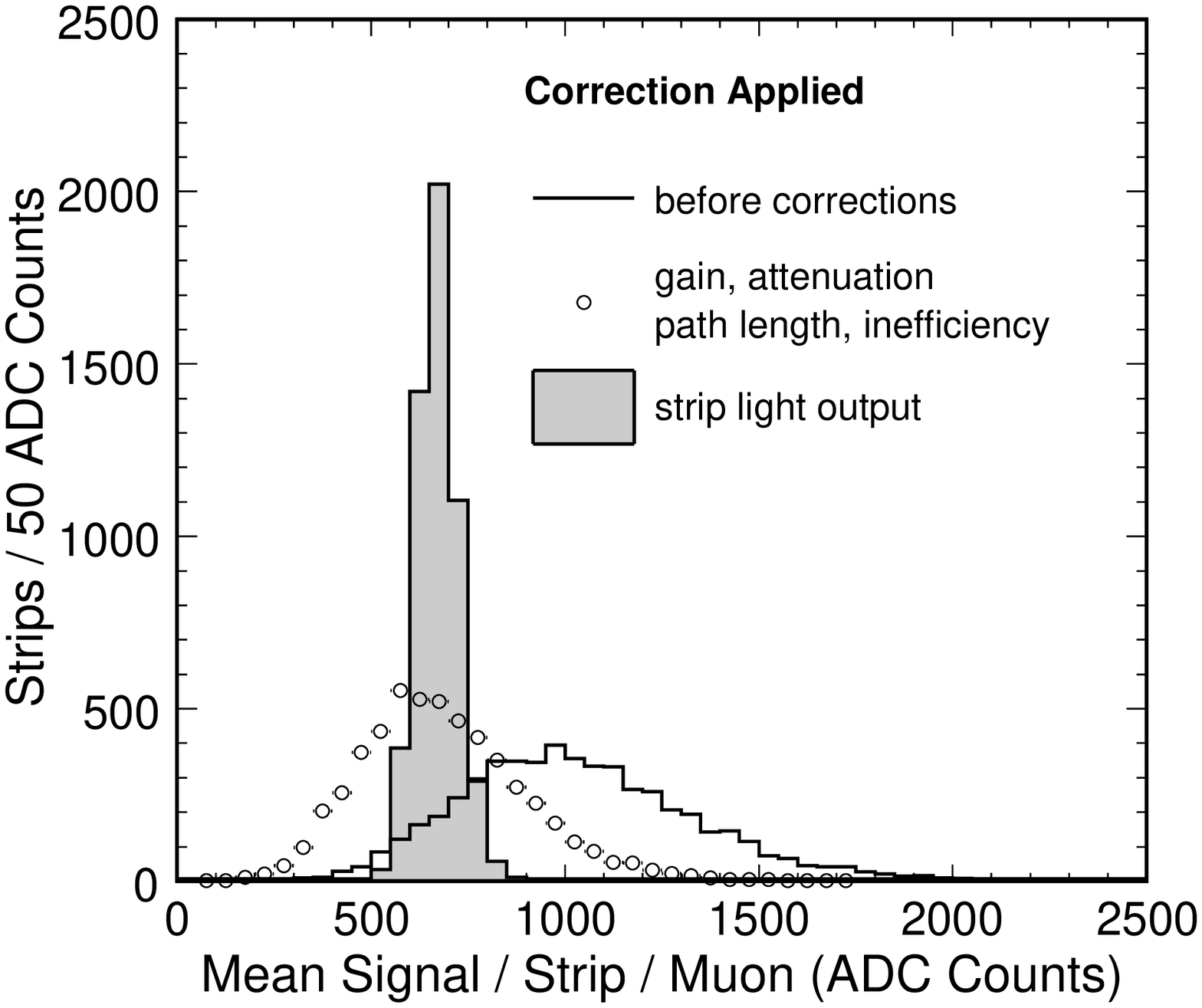}
\caption{\label{fig:muon_dedx}  Mean signal per scintillator strip induced by beam muons. Corrections are made with cosmic-ray muons and account for non-uniformities in channel gain $D(t)$, attenuation in optical fibers $A(i,x)$, and strip light output $U(i)$. }
\end{center}
\end{figure}


\pstart
The principal tools for calibrating the detector response are an LED based light-injection (LI) system, a test-bench scan of the scintillator modules with a radioactive source and cosmic-ray muons.  The detector is calibrated in a multi-stage procedure that converts the raw photomultiplier signal $Q_{raw}\left(i,t,x\right)$ measured by channel $i$ at time $t$ for an energy deposition at position $x$ into a fully corrected signal $Q_{cor}$. Each calibration stage produces a ``calibration constant''.  The fully corrected $Q_{cor}$ is defined as the product of $Q_{raw}\left(i,t,x\right)$ and the calibration constant from each stage:
\pend
\begin{equation*}
Q_{cor} = Q_{raw}\left(i,t,x\right) \times D\left(t\right)  \times U\left(i\right) \times A\left(i,x\right)  \times E
\end{equation*}
\pstart
\noindent where $D,U,A$ and $E$ refer to:
\pend
\begin{description}
\pstart
\item[Drift correction $D\left(t\right)$:] The channel gains and their variation over time are measured with the LED based light-injection system, demonstrating that short-term ($<\unit[24]{h}$) gain variations are small and occurred mostly due to environmental changes in the detector halls. Light-injection data were eventually superseded by measurements of the mean signal per plane induced by through-going cosmic-ray muons since the muon data were also able to correct for variations in the scintillator light-output. The detector response varies by $<2\%$ over the data-taking period as shown in Fig.~\ref{fig:drift}. The decreasing response in the Fear Detector is likely due to aging of the scintillator.
\pend

\pstart
\item[Uniformity correction $U\left(i\right)$:] Through-going cosmic-ray muons are used to account for differences in light output between individual strips as well as attenuation in the optical fibers and connections.  Event-by-event corrections are applied to account for the muon track angle and the expected inefficiency due to statistical fluctuations at low light-levels. The calibration reduces the uncorrected 30\% strip-to-strip response variations to approximately 8\%.
\item[Attenuation correction $A\left(i,x\right)$:] A radioactive source was used to map out the response of each scintillator module at many positions along each strip. This was done on a test-bench setup prior to installation of the modules in the Near and Far Detectors. The data were then fit to an empirical model of optical attenuation in the wavelength-shifting and clear optical fibers. The resulting parametrizations are used to correct the signals from cosmic-ray muons during the calculation of the uniformity calibration constants  $U\left(i\right)$ and also to correct the reconstructed shower energy for attenuation based on the reconstructed shower position. The relative size of the correction varies by about 30\% over the \unit[8]{m} length of a Far Detector scintillator strip when signals from both ends are added, and by about 50\% over the \unit[3]{m} length of a Near Detector strip.
\pend
\pstart
\item[Signal scale calibration $E$:] The overall scale of the signals is established by the detector's response to stopping muons. This provides the standard which fixes the absolute calibration of the Near and Far Detectors, allowing their signals to be compared to one another.  The calibration is done by tabulating the detector's response to muon crossings using only the portion of each track in which the muon energy is between \unit[0.5]{GeV} and \unit[1.1]{GeV}, deduced from the distance to the track endpoint. This energy window avoids the rapid variation in \dedx{} near the track's end. Prior to this calibration the signals have already been corrected for $D,A,U$ as described above. For each muon a correction is applied to account for the muon's path length in each scintillator plane. The mean response is then calculated for each individual strip and a single constant representing the entire detector is derived from the median over all strips.
\pend
\end{description}

\pstart
The effect of the uniformity and attenuation calibration stages is shown in Fig.~\ref{fig:muon_dedx}.  After calibration, the fully corrected signal is expressed in muon equivalent units (MEU) as defined by the signal scale ($E$) calibration procedure. One MEU corresponds to approximately 3.7 photoelectrons at the center of the Far Detector (7.5 photoelectrons when signals from both strip-ends are summed) and 5.4 photoelectrons at the beam center in the Near Detector. Based on muon stopping-power tables~\cite{Groom:2001kq} we find that one MEU corresponds to $\unit[2.00 \pm 0.02]{MeV}$ of muon energy loss in scintillator~\cite{Adamson:2006xv,jeff_thesis}. This calibration was independently derived at the Near and Far Detectors with a relative uncertainty of $\pm 2\%$. 
\pend

\pstart
\label{sec:shwr_calib}Data from a dedicated, test-beam calibration detector~\cite{Adamson:2006xv} are used to benchmark the hadronic and electromagnetic shower simulation done by \geant{}/\gcalor{}~\cite{mike_thesis,tricia_thesis}. The measurements demonstrate that the simulated response agrees with the data at the level of 1--5\%, depending on the energy and particle type. This level of agreement is inclusive of energy dependent offsets in the beam momentum of the test beam-lines and uncertainties in the simulation of energy loss upstream of the calibration detector. In light of the much larger uncertainty associated with final state interactions, no attempt was made to resolve these discrepancies when determining the conversion from MEU to GeV for neutrino induced showers. Instead a conservative uncertainty of 5.6\% is assigned to the absolute shower energy scale to account for the data/MC discrepancies from the calibration detector as well as the precision of the stopping muon calibration performed there. An energy dependent MEU to GeV conversion is then extracted from the simulation such that the reconstructed shower energy estimates the energy transferred to the hadronic system.
\pend











\section{CC Event Selection}

 
\label{sec:sandb}
\subsection{Event Classification}
\label{sec:evtsel}

\begin{figure}
\centering
  \subfloat[\label{fig:pdf1}]{\includegraphics[height=0.25\textheight,clip]{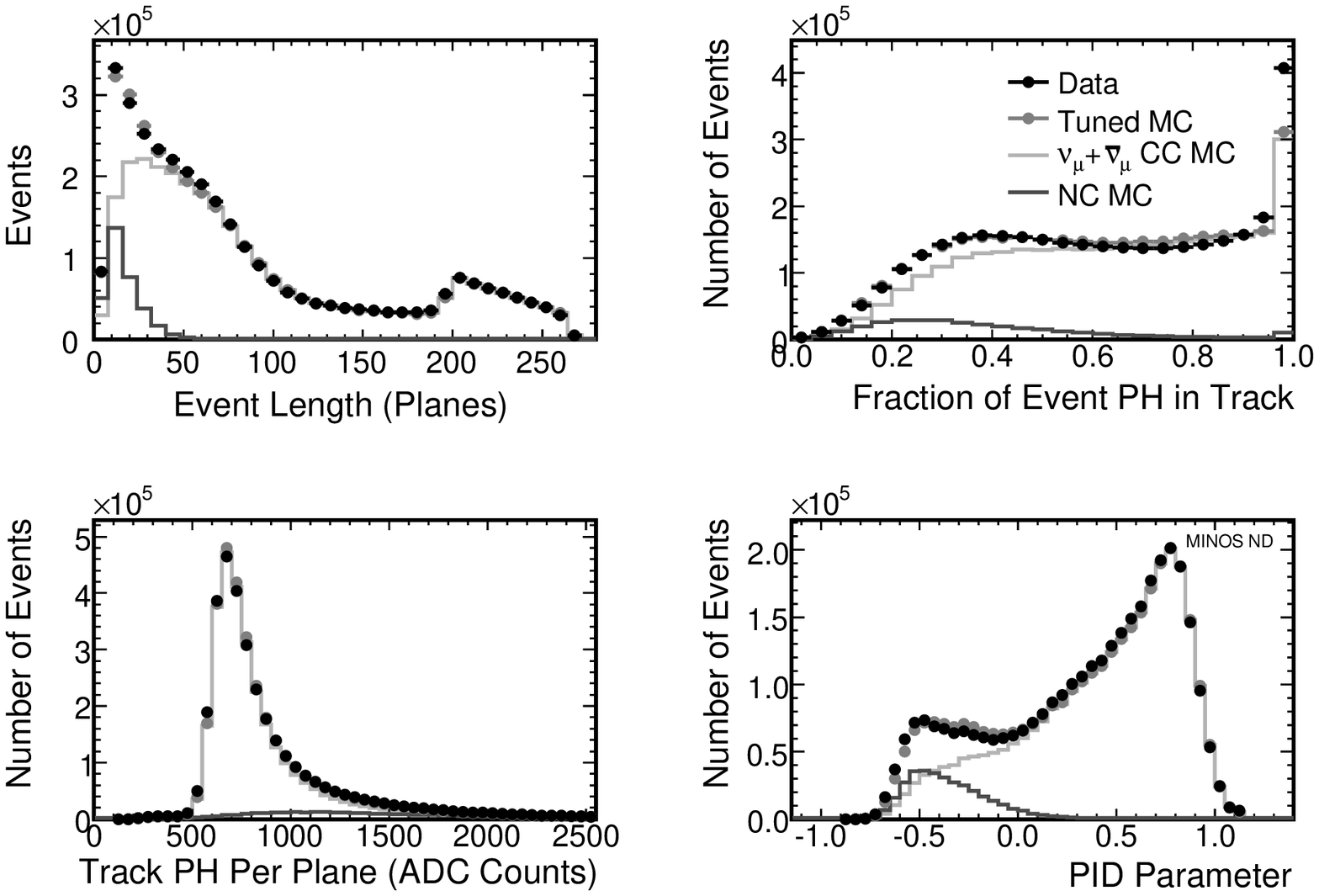}}\\
  \subfloat[\label{fig:pdf2}]{\includegraphics[height=0.25\textheight,clip]{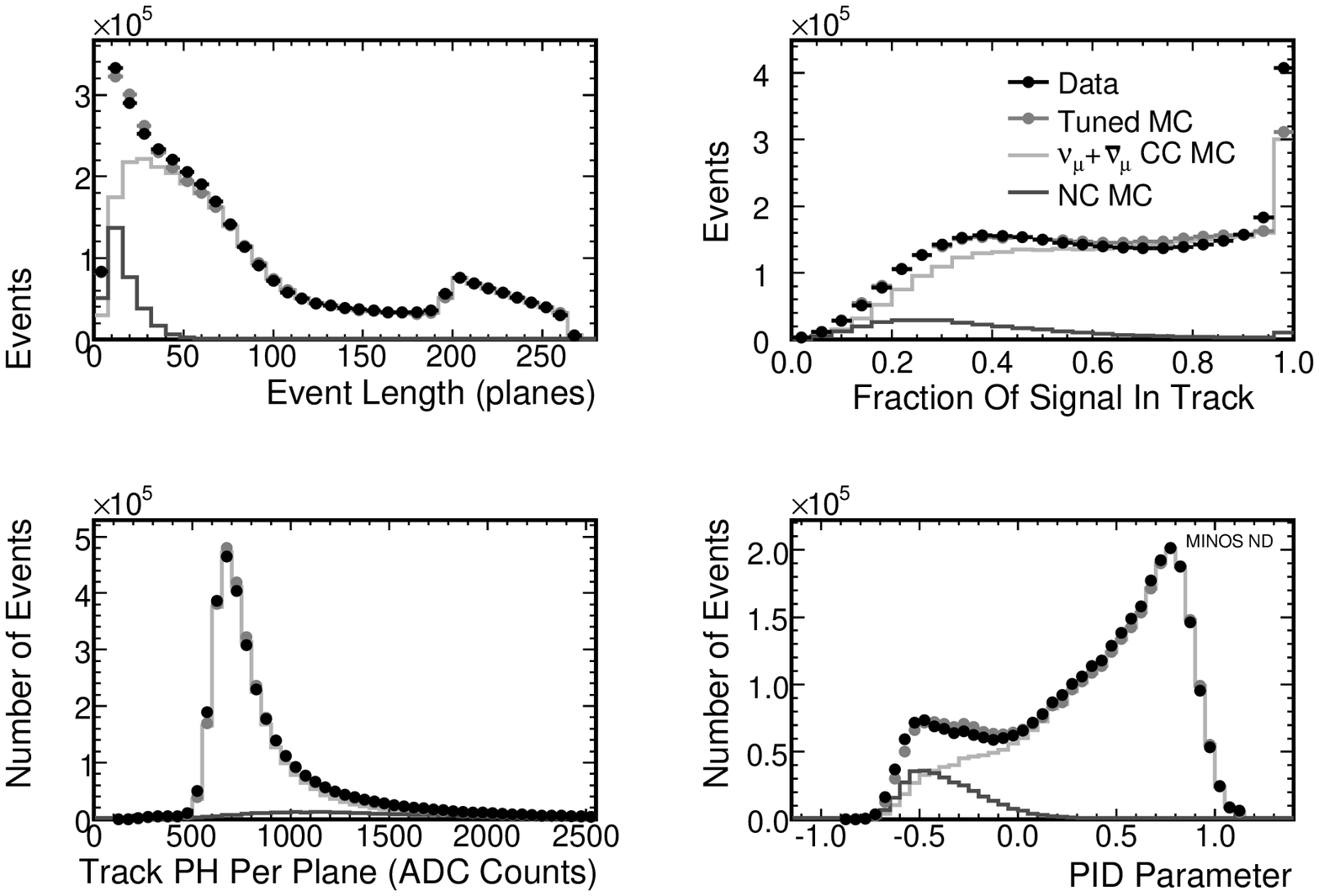}}\\
  \subfloat[\label{fig:pdf3}]{\includegraphics[height=0.25\textheight,clip]{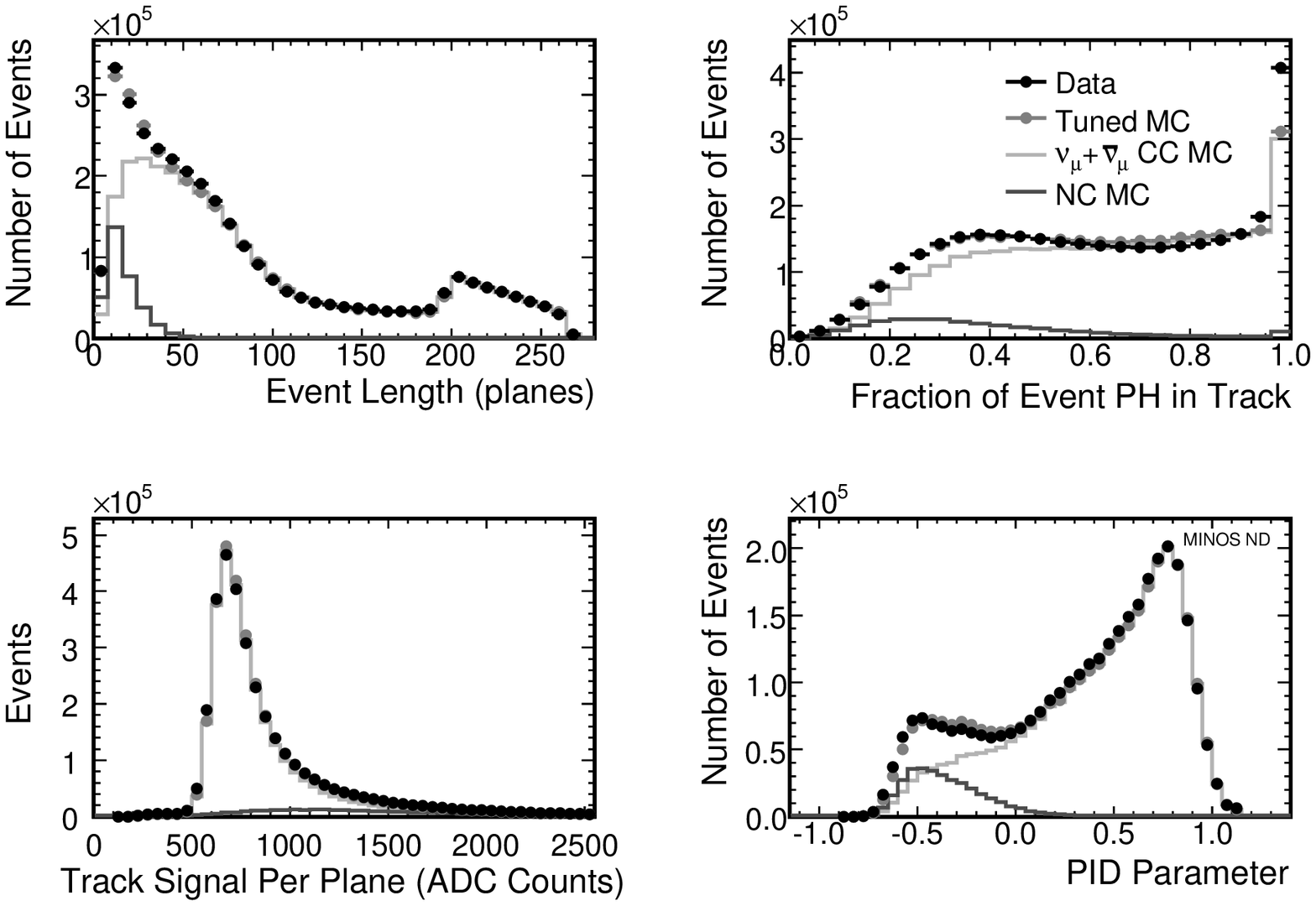}}
  \caption{Distributions of the variables used to the define the event separation parameter $S$ in Eq.~\ref{eq:P} and Eq.~\ref{eq:S} for events satisfying the criteria of Sec.~\ref{sec:reco}. Near Detector data collected in the LE10/185kA beam configuration is shown.  \label{fig:pdf} }
\end{figure}

\pstart
In the MINOS experiment, neutrino oscillations are expected to cause a deficit of \numucc{} events at energies $E_{\nu}\lesssim\unit[5]{GeV}$ for $\dm \approx \unit[2.5\times 10^{-3}]{\evmass{}}$ and our baseline of \unit[735]{km}. Neutral-current events which have been misidentified as \numucc{} tend to populate this energy range and could obscure the oscillation signal. We remove neutral-current events from the oscillation sample with a technique based on the event topology. The technique uses three probability density functions (PDFs), which are constructed for the following variables: (a) the event length, expressed in units of the number of detector planes, (b) the fraction of the total event signal in the reconstructed track, and (c) the average signal per plane induced by the reconstructed track. These quantities are related to the muon-range, the event inelasticity, and the average energy loss \dedx{} of the muon track and are distributed differently for \numucc{} and neutral-current events as shown in Fig.~\ref{fig:pdf}. The probability that a particular event is consistent with the \numucc{} or neutral-current PDFs is given by the product of the three individual probabilities
\pend
\begin{equation} \label{eq:P}
P_{CC,NC}=\prod_{i=1}^{3}f_{i}(x_{i})_{CC,NC} \quad,
\end{equation} 
\pstart
\noindent where the $f_{i}(x_{i})$ are the individual PDFs for \numucc{} and neutral-current events respectively. The PDF distributions are well modeled by the simulation in the regions in which the neutral-current and \numucc{} samples overlap and for neutrino energies above approximately \unit[10]{GeV} where the Monte Carlo simulation is better constrained by data from previous experiments and the neutrino flux has a weaker dependence on the energy. 
\pend

\begin{figure}
\centering
  \subfloat[\label{fig:pid_le}]{\includegraphics[height=0.25\textheight]{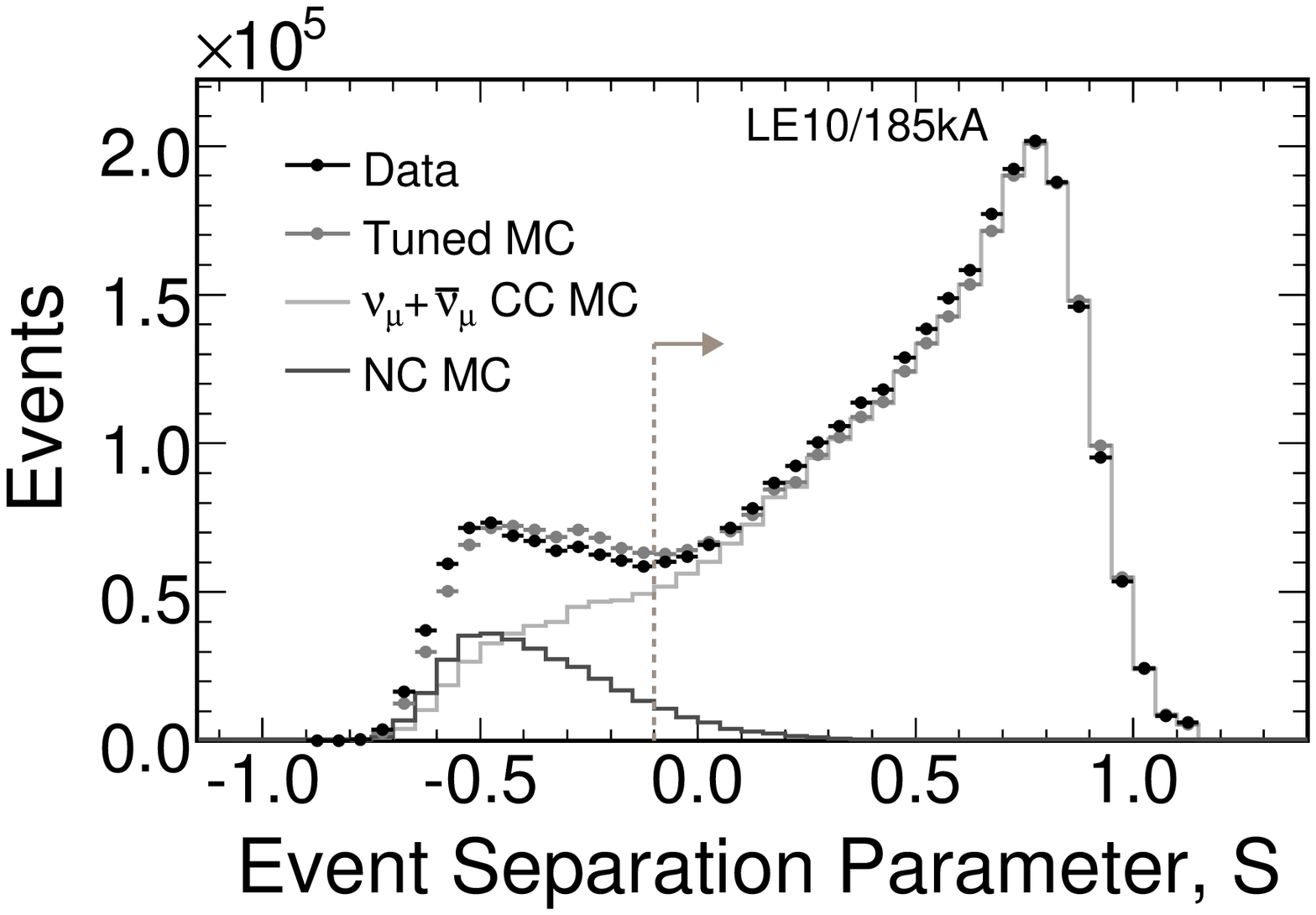}}\\
  \subfloat[\label{fig:pid_me}]{\includegraphics[height=0.25\textheight]{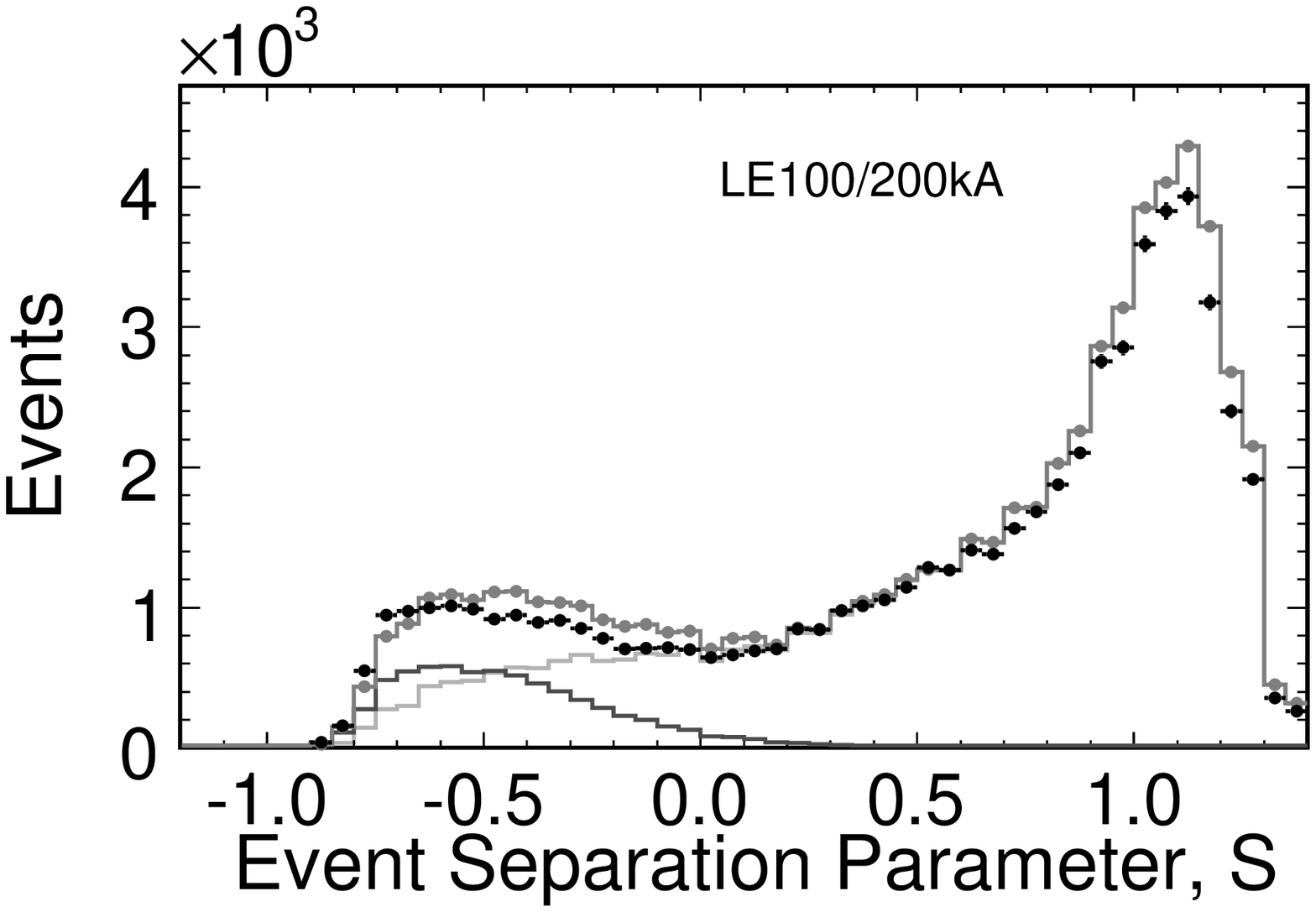}}\\
  \subfloat[\label{fig:pid_he}]{\includegraphics[height=0.25\textheight]{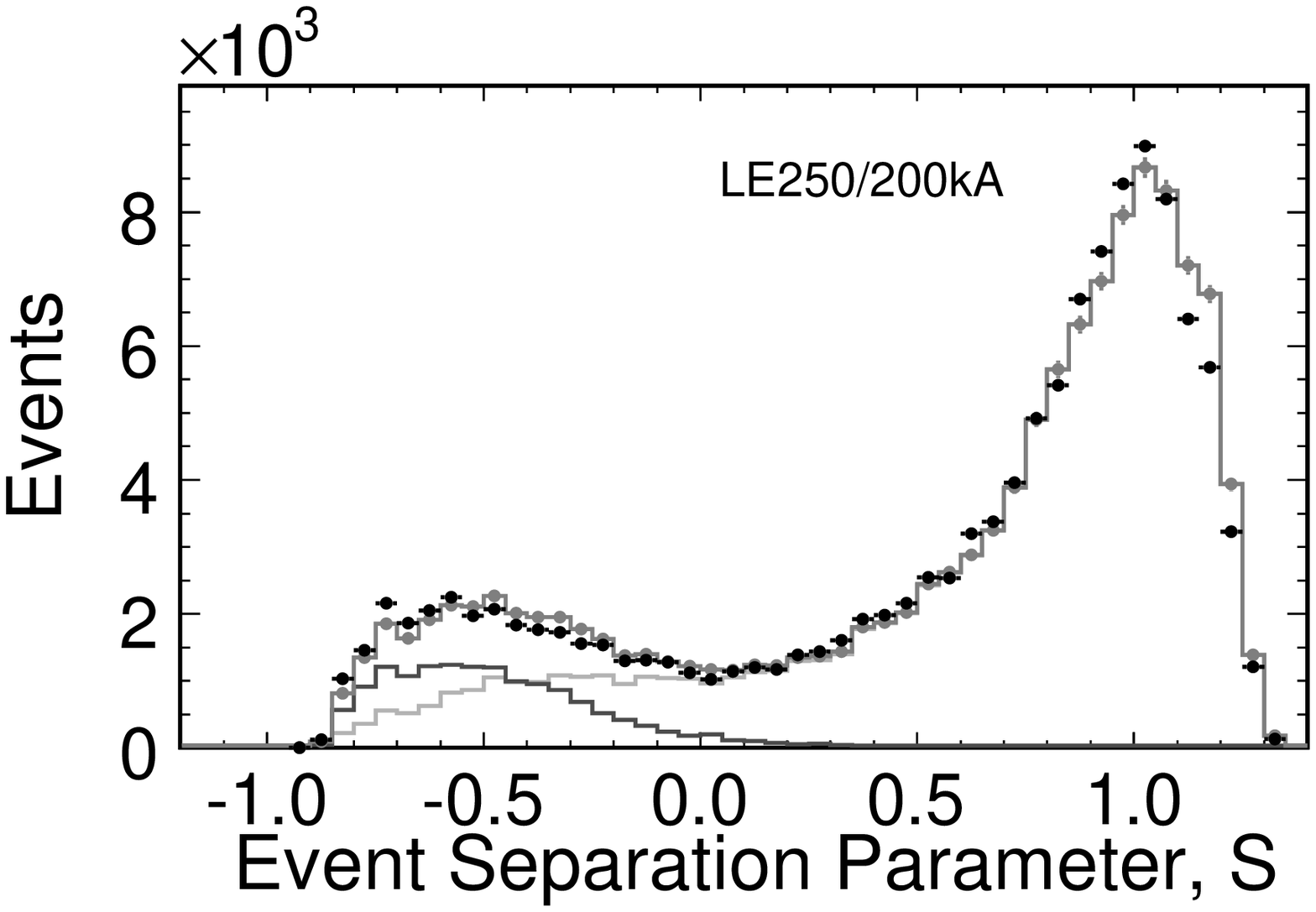}}
  \caption{ The event separation parameter $S$, as defined in Eq.~\ref{eq:S}, plotted for Near Detector data and simulation, and for three beam configurations in Tab.~\ref{tab:beam-config}.  \label{fig:pid} }
\end{figure}

\pstart
An event selection parameter $S$ is derived from $P_{CC}$ and $P_{NC}$ according to
\pend
\begin{equation} \label{eq:S}
S=-\left(\sqrt{-\ln(P_{CC})}-\sqrt{-\ln(P_{NC})}\right) \quad.
\end{equation}
\pstart
\noindent Events that are more likely to originate from the \numucc{} interactions give positive values, and those that are more likely to be neutral-current give negative values. Figure~\ref{fig:pid} shows the distribution of $S$ for data collected in the Near Detector in the LE10/185kA, LE100/200kA and LE250/200kA beam configurations. The behavior of the data is well-reproduced by the Monte Carlo simulation especially in the region $S>-0.1$ in which \numucc{} events dominate.

\pend

\begin{figure}
\centering
  \subfloat[\label{fig:eff_nd}]{\includegraphics[width=0.7\textwidth]{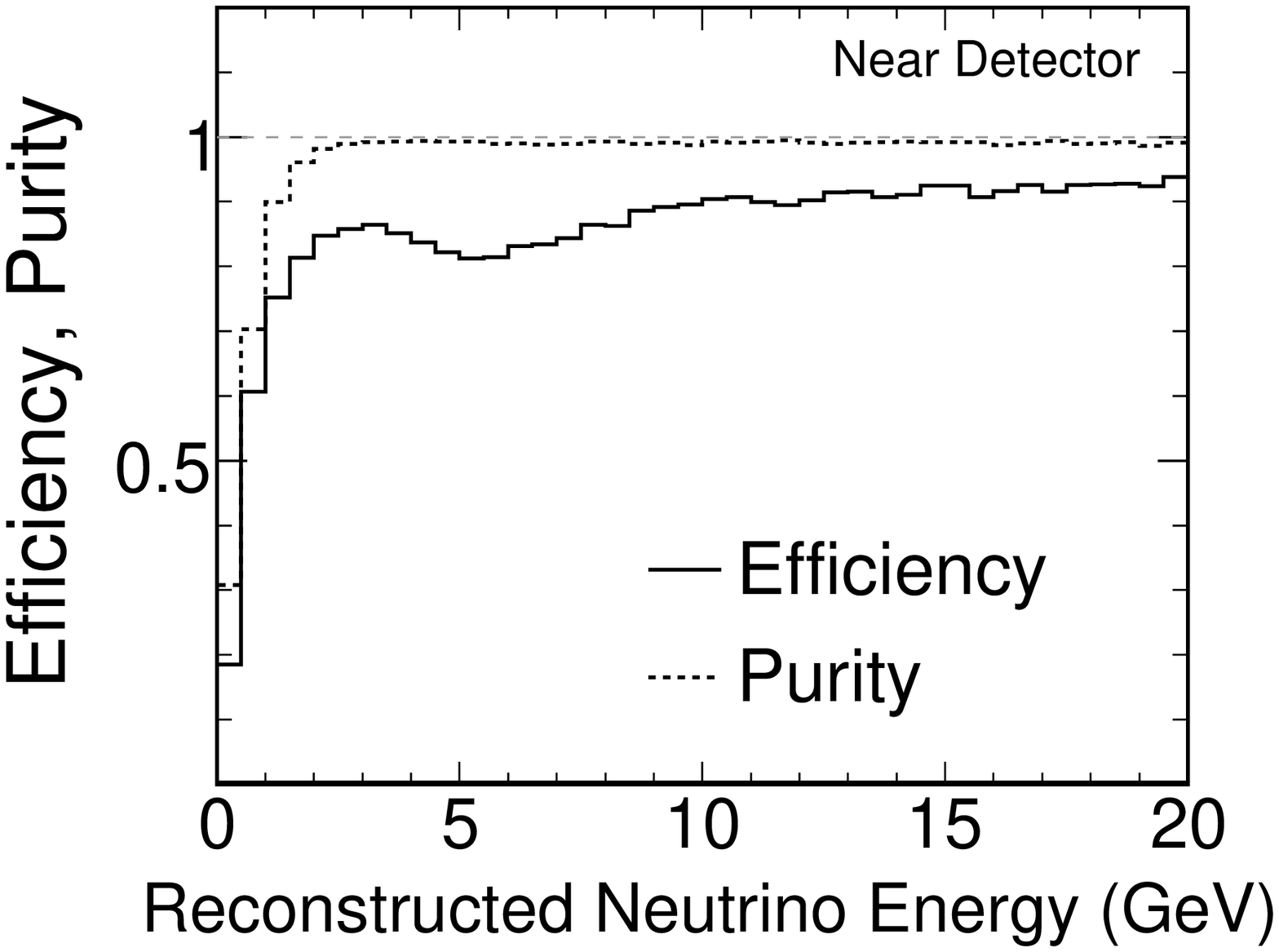}}
\\
  \subfloat[\label{fig:eff_fd}]{\includegraphics[width=0.7\textwidth]{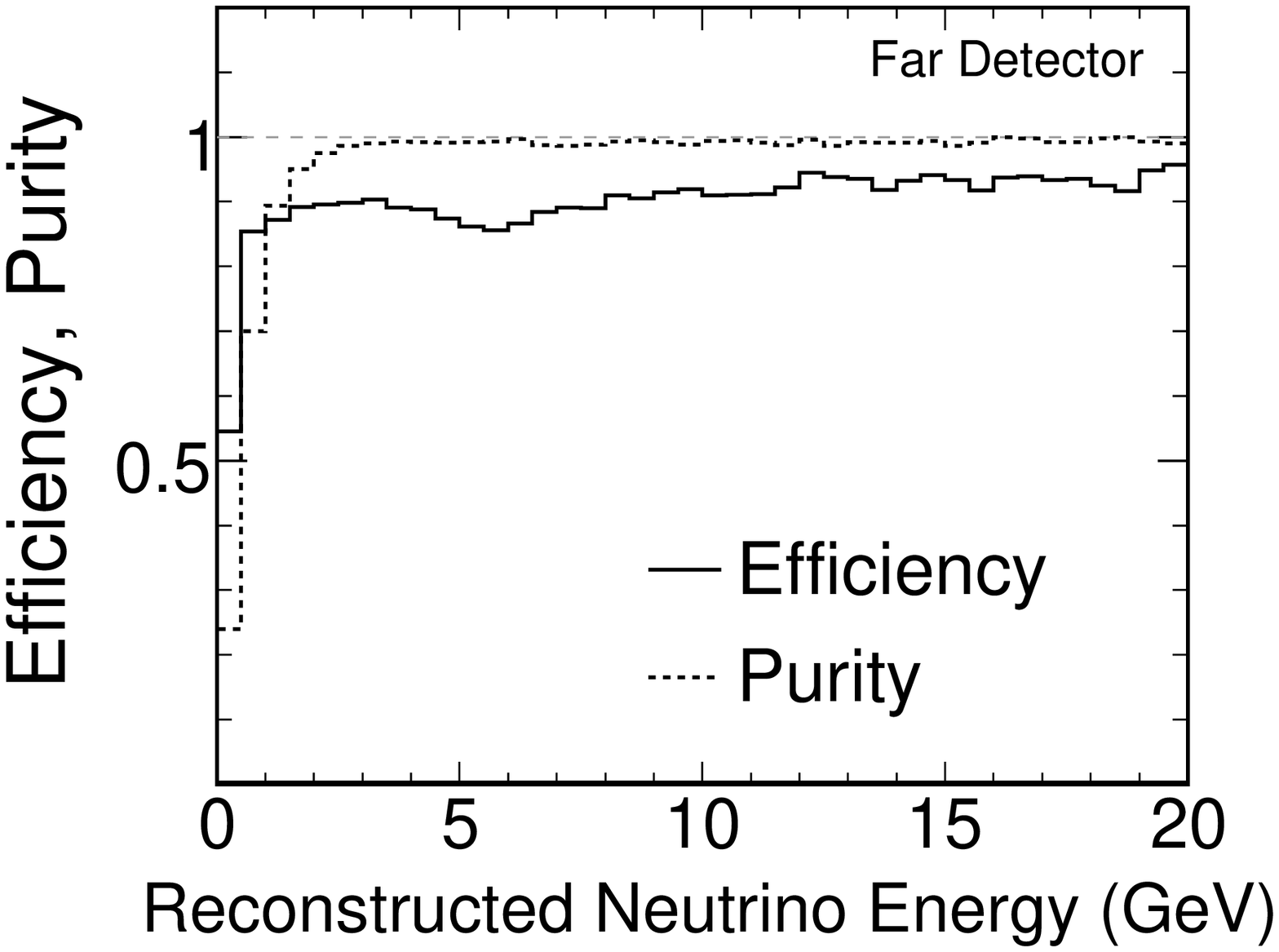}}
  \caption{Efficiency for identifying \numucc{} events using the event separation parameter defined in Eq.~\ref{eq:S}, along with the purity of the resulting sample. The efficiency is defined with respect to the events satisfying the criteria of Sec.~\ref{sec:reco}. \label{fig:pideffic} \label{fig:pid_eff} }
\end{figure}

\pstart
 We isolate an enriched sample of \numucc{} events for use in the oscillation analysis by requiring $S>-0.2$  in the Far Detector and $S>-0.1$ in the Near Detector. These values were chosen to optimize the statistical sensitivity to the oscillation parameters $\dm$ and $\st$ in the Far Detector, and to provide a selected event sample in the Near Detector that has the same purity, defined as the fraction of selected events that are true \numucc{} interactions, as the Far Detector.  Figure~\ref{fig:pid_eff} shows the selection efficiency, defined as the fraction of true \numucc{} events that have $S>-0.1$ in the Near Detector ($S>-0.2$ in the Far Detector), and the resulting purity of the sample in the two detectors. The efficiency is calculated with respect to the events passing the criteria of Sec.~\ref{sec:reco} and varies slowly as a function of reconstructed energy above \unit[0.5]{GeV}. The contamination from misidentified neutral-current interactions is greater at low reconstructed energy, hence the purity of the selected sample drops below \unit[1]{GeV}. The Monte Carlo simulation indicates that the average selection efficiency and purity (integrated over the unoscillated Far Detector LE010/185kA spectrum) are $90.0\%$ and $98.2\%$ respectively. 
\pend

\begin{table*}[]
\begin{center}

\begin{tabular}{lc}
\hline
\hline
Selection Criterion & Far Detector Events \\
\hline

Reco. track in fiducial vol. & 427\\

Data quality cuts & 408\\

Event Timing cut & 404\\

Beam Quality cuts & 390\\

Track direction cut & 384 \\

Track quality cuts & 365 \\

Negative track charge & 306 \\

Reco. energy $<$ 30 GeV & 275\\

$S>-0.2$ & 215\\  
\hline
\hline

\end{tabular}
\caption{Effect of selection criteria on the Far Detector neutrino event sample. The first eight criteria are described in Sec.~\ref{sec:reco}. In accordance with our analysis strategy, the total number of events passing all cuts was not revealed until all of the data quality checks were complete and our analysis method was fully defined.\label{tab:fdsel}}
\end{center}
\end{table*}


\begin{figure}[h]
\centering
\subfloat[\label{fig:fd_pid1}]{\includegraphics[width=0.45\textwidth]{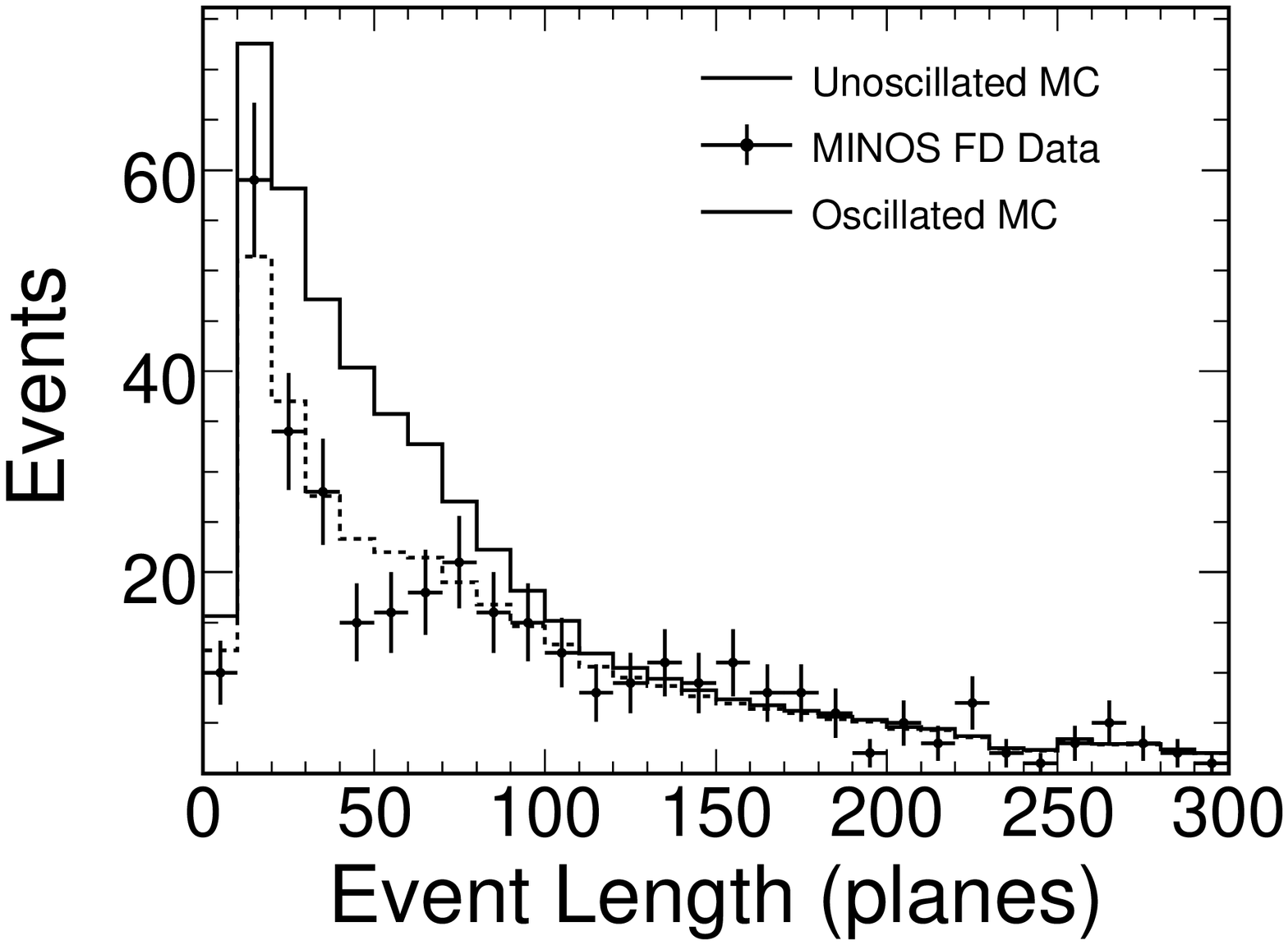}}
\hfill
\subfloat[\label{fig:fd_pid2}]{\includegraphics[width=0.45\textwidth]{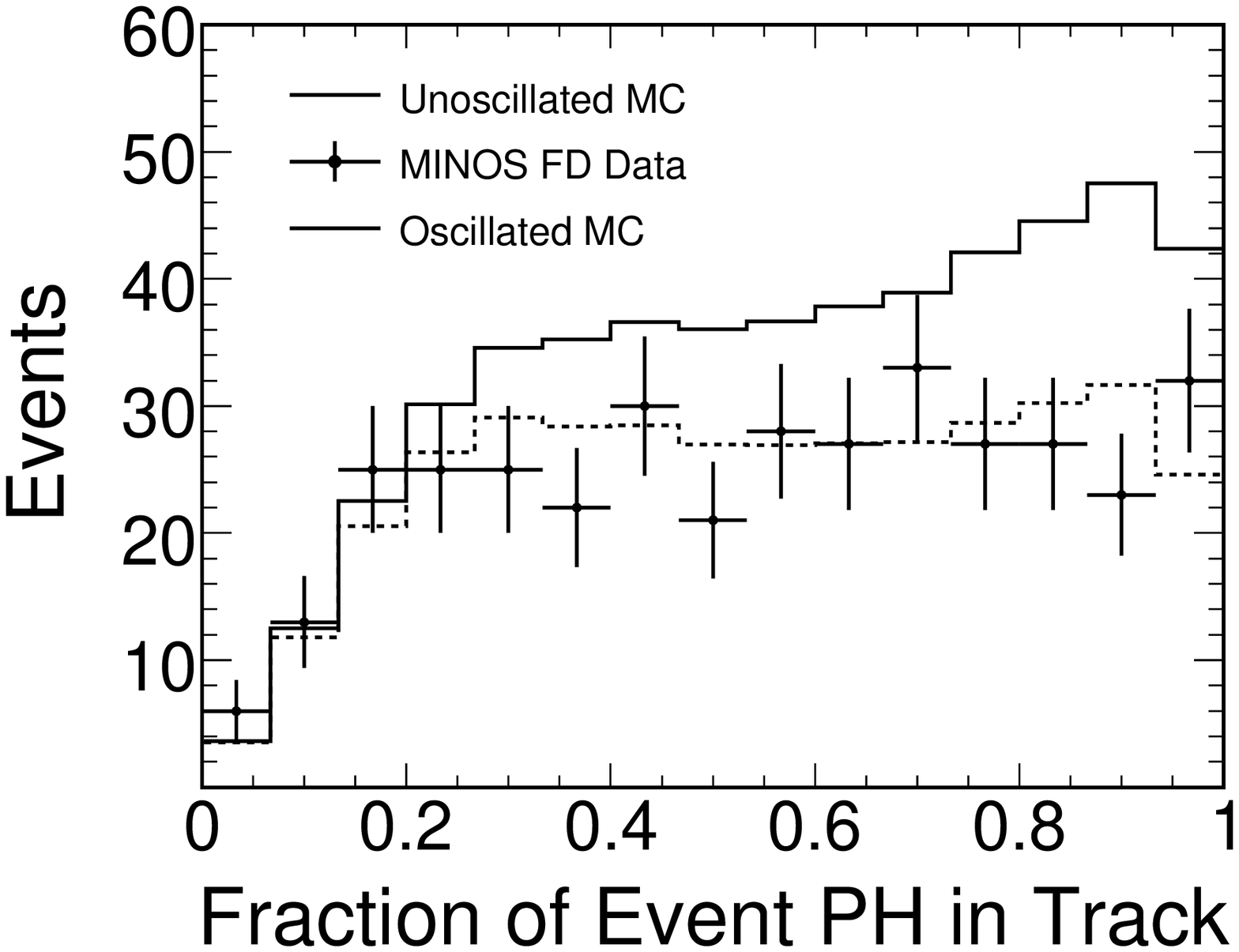}} \\

\subfloat[\label{fig:fd_pid3}]{\includegraphics[width=0.45\textwidth]{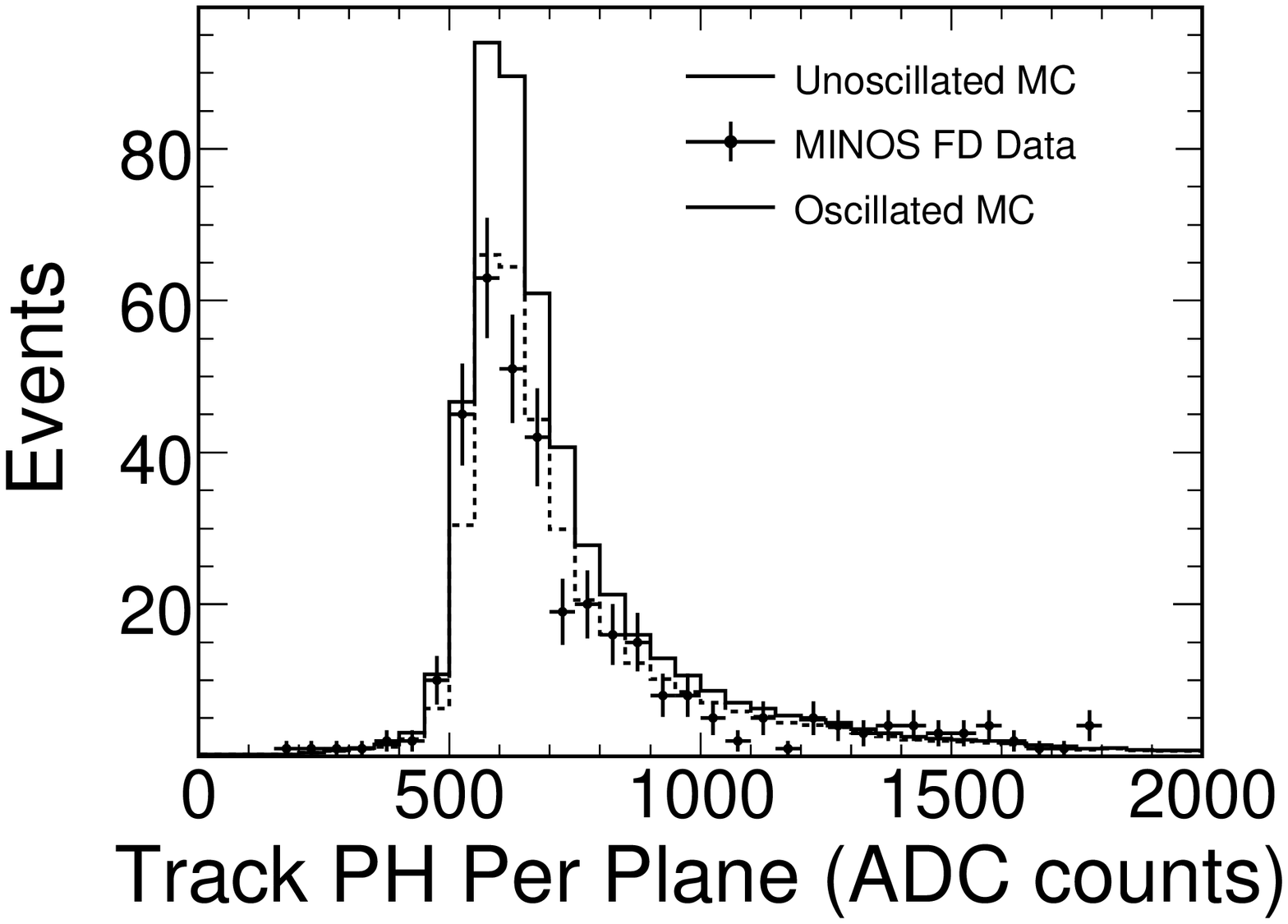}}
\hfill
\subfloat[\label{fig:fd_pid4}]{ \includegraphics[width=0.45\textwidth]{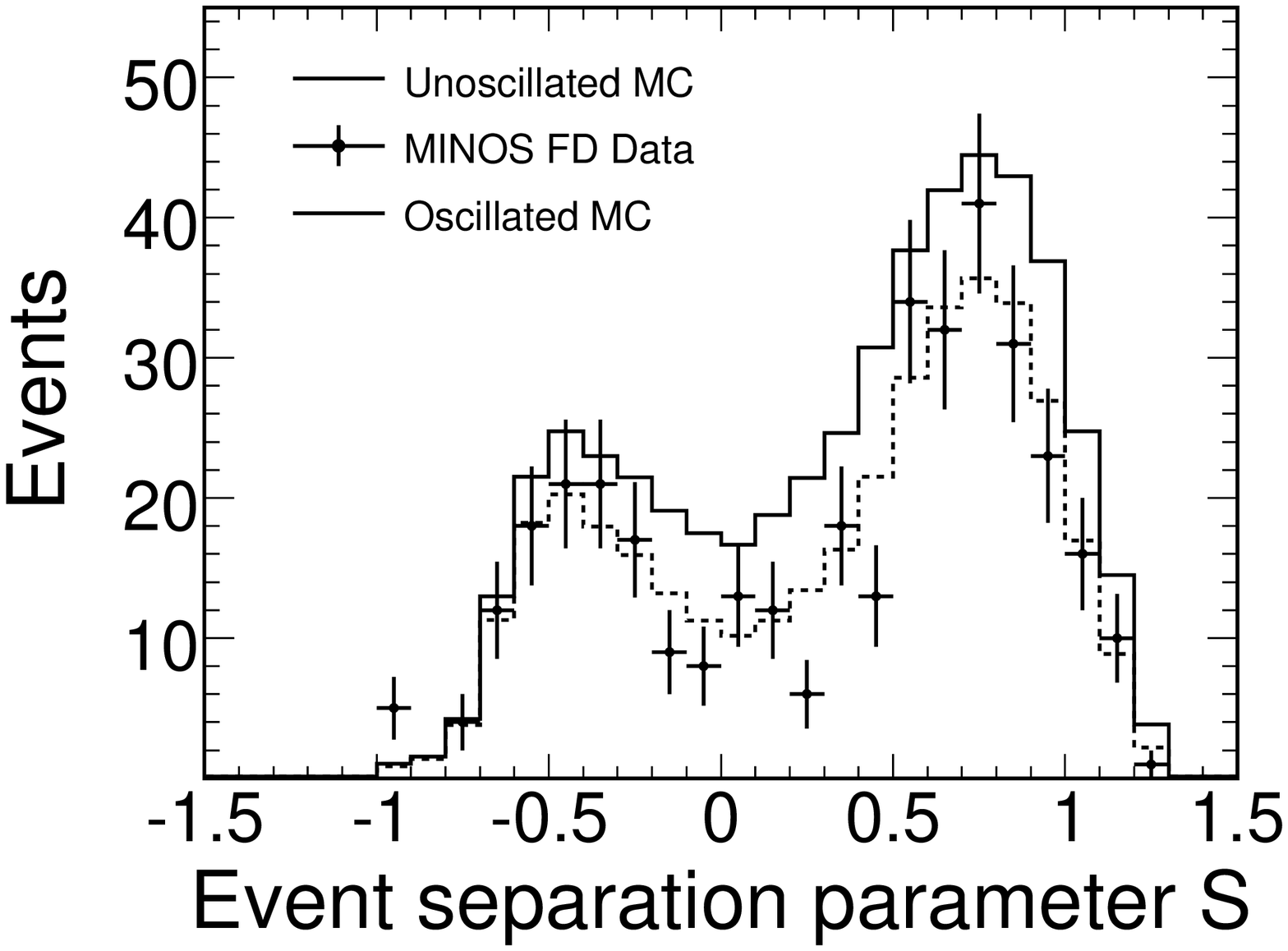}}

\caption{Distributions of the variables that are used in \numucc{} and neutral-current event classification, for Far Detector data and Monte Carlo simulation. The bottom-right plot shows the distribution of the event separation parameter $S$ that is derived from these three PDFs. Oscillations are accounted for according to the mixing parameters extracted from our data (see Sec.~\ref{sec:osc}). These distributions are shown here to illustrate the level of agreement between data and simulation for $S$ and its input variables. They were not examined until the analysis procedure had been fully defined.\label{fig:fd_pid} }
\end{figure}

\pstart
The PDFs used to discriminate between \numucc{} and neutral-current interactions depend on the \numu{} energy spectrum and are therefore sensitive to neutrino oscillations. As such, the discrimination technique was developed and validated by comparing the Monte Carlo simulation to data collected in the Near Detector. A fraction of the  Far Detector dataset was available for data quality checks. Examination of oscillation-sensitive distributions in the full Far Detector data set was only performed after these checks had been made and we had fully defined our analysis procedure. Table~\ref{tab:fdsel} shows the effect of our selection on those Far Detector events that are recorded in coincidence with \numi{} beam spills. After requiring $S>-0.2$ we retain 215 of the 275 events passing the criteria of Sec.~\ref{sec:reco}. The variables used in Eq.~\ref{eq:P} and Eq.~\ref{eq:S} are shown in Fig.~\ref{fig:fd_pid} for the 275 event sample. Two versions of the simulation are shown. The first does not account for neutrino oscillations while in the second we apply the effect of oscillations according to the mixing parameters extracted from our data using the procedure described in Sec.~\ref{sec:osc}. The measured deficit of events with respect to the unoscillated simulation is consistent with the partial disappearance of \numucc{} events at energies less than about \unit[10]{GeV} and is well described when we account for oscillations.
\pend


\subsection{Backgrounds}

\subsubsection{Neutral-Current Contamination}
\label{sec:nc}
The neutral-current background in the Far Detector is estimated from the Monte Carlo simulation as the number of neutral-current events that pass the event selection cut. These events comprise only $<2\%$ of the selected unoscillated \numu{} CC sample but tend to congregate in the lowest reconstructed energy bins, as shown in Fig.~\ref{fig:pideffic}, and so constitute a significant background to the \numucc{} oscillation signal. In this section, we describe how we have used Near Detector data to directly provide an estimate of the systematic error on the number of neutral-current events that are selected in the final Far Detector event sample.
\pend

\pstart
To handle uncertainties that arise from the modeling of hadronic showers in the Monte Carlo, showers from cleanly identified \numucc{}  events in the Near Detector with the muon tracks removed were used to model neutral current events.  The ratio between the track-removed shower $S$ distributions for data/MC provides a $S$-dependent Monte Carlo scaling factor for the shape of the neutral-current event selection parameter distribution. 
\pend

\pstart
 To determine the uncertainty on the overall neutral-current signal normalization, the event selection parameter distribution $S$ in the data was fitted to determine the amplitude of the \numucc{} and neutral-current signals.  In this fit, the shape of the charged-current component was obtained from the Monte Carlo simulation, while the neutral-current shape was scaled according to the data/MC ratio observed in the track-removed \numucc{} event sample.  The fits to the data for the signal amplitude and the data/MC comparison of the track-removed \numucc{} events were performed in six reconstructed energy slices.
\pend



\pstart

\begin{figure}
\centering
\includegraphics[width=\columnwidth]{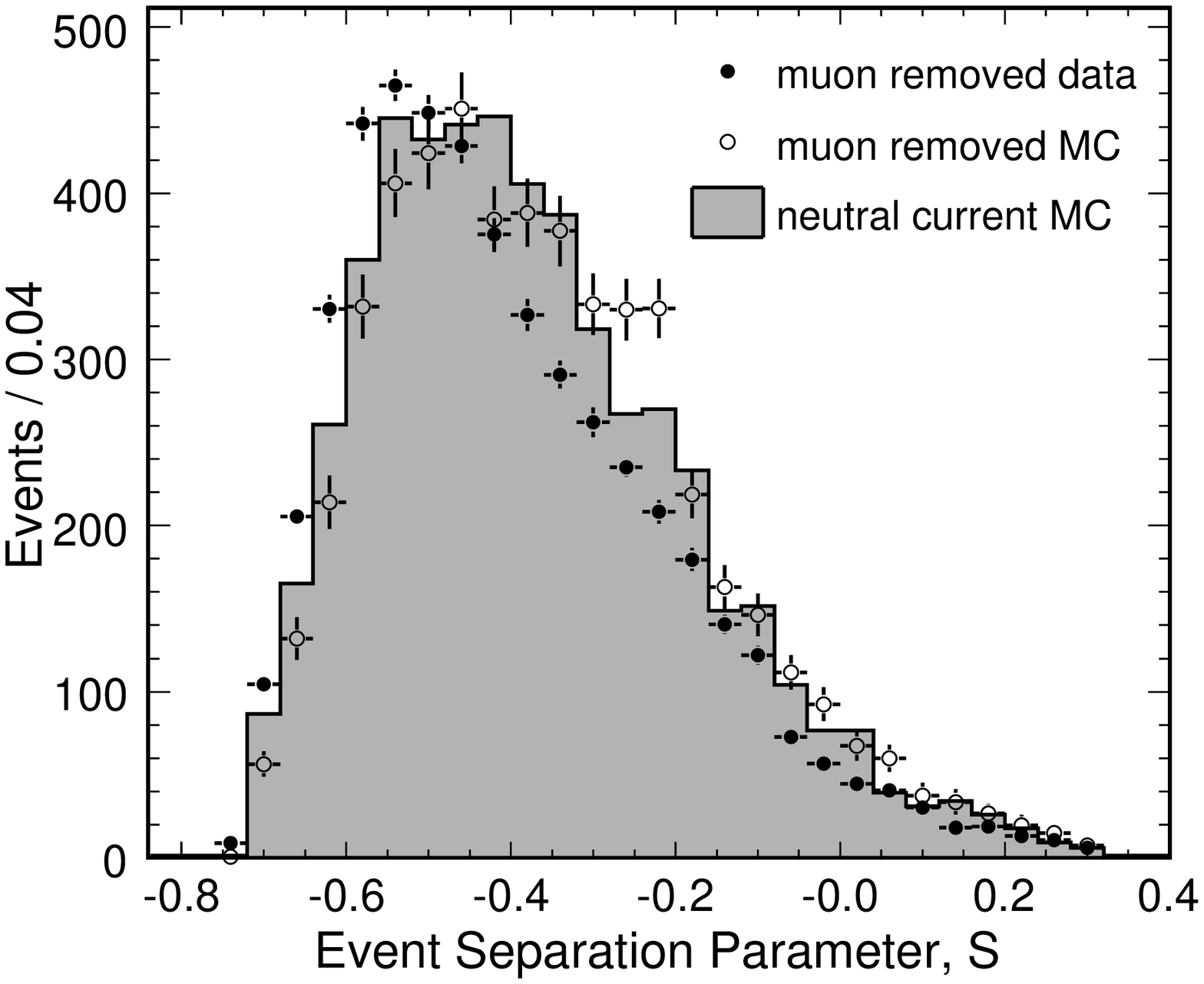}
  \caption{Distribution of the event separation parameter $S$ for true neutral-current Monte Carlo events (shaded histogram), for selected $\nu_\mu$  Monte Carlo events with the reconstructed muon track removed (open circles) and for selected Near Detector data events with the muon track removed (filled circles). \label{fig:mrpid} }
\end{figure}

A comparison of the event separation distribution $S$ for the simulated neutral-current and selected muon removed \numucc{} events yield reasonable agreement, as shown in Fig.~\ref{fig:mrpid}. However, the muon removed \numucc{} events from Near Detector data, which are represented by the black points in Fig.~\ref{fig:mrpid}, peak at lower values of $S$ than the corresponding Monte Carlo distribution. This discrepancy implies that the data contain shorter events and have higher average pulse height per plane than the Monte Carlo events.
\pend

\pstart
The ratio of the data to simulated $S$ distributions obtained from the muon-removed events is used to provide a reweighting of the shape of the $S$ distribution for neutral-current Monte Carlo events. This procedure is performed in six bins of reconstructed energy and in all cases, the reweighted distributions predict fewer neutral-current events for $S>-0.1$ than the {\it a priori} simulation, by an amount that ranges from 20-50\%, as shown in Table~\ref{table:nc_summary}. 
\pend

\begin{table}[]

\centering
\begin{tabular}{lc}
\hline
\hline
Energy Range\hspace{0.3in} &  Data/MC ratio \\ 
\hline 

0-1 GeV & 0.642 \\ 
1-2 GeV & 0.503 \\ 
2-4 GeV & 0.804 \\ 
4-6 GeV & 0.583 \\ 
6-12 GeV & 0.434 \\ 
12-30 GeV & 0.579 \\ 
\hline
\hline
\end{tabular}

\caption{The ratio of the number of neutral-current events with $S>-0.1$ in the simulation and the best fit to the data using the neutral-current shape derived from the track-removed \numucc{} events. The data indicates that the neutral-current background is over-predicted by the simulation. \label{table:nc_summary} }
\end{table}


\pstart

The origin of this difference is most likely due to an unknown combination of shower modeling and/or neutral-current cross-section uncertainties. We have thus assigned 50\%, the largest difference observed in the fit above $S=-0.1$, as the error on the neutral-current background.
\pend


\subsubsection{Cosmic Ray Background}

\pstart
We have estimated the background rate from cosmic-ray muons in the Far Detector selected sample using two independent methods. First, we examined the rate of selected events in the spill trigger window that are not within the expected $\unit[10]{\mu s}$ wide window around the time of true beam spill (see Fig~\ref{fig:fdtiming}). Secondly, we applied the \numucc{} selection criteria described above to a sample of $2.3\times 10^{6}$ spill triggers taken in anti-coincidence with the beam spill. Both of these methods yielded an upper limit (90\% C.L) of 0.5 background events in a $\unit[10]{\mu s}$ wide window around the time of true beam spill for an exposure of $\unit[1.27\times10^{20}]{POT}$. The background due to neutrino-induced \numucc{} interactions in the rock upstream of the Far Detector is estimated from the Monte Carlo simulation to be \unit[0.38]{events}. The rate of cosmic-ray muons is about 30 times larger in the Near Detector than it is in the Far Detector. The cosmic-ray background is negligible, however, due to the much greater ($\times 10^{6}$) neutrino flux at the Near Detector.
\pend


\section{Constraints on the Neutrino Flux From Near Detector Data}
\label{sec:skzp}

\begin{figure}
  \begin{centering}
    \includegraphics[width=\columnwidth]{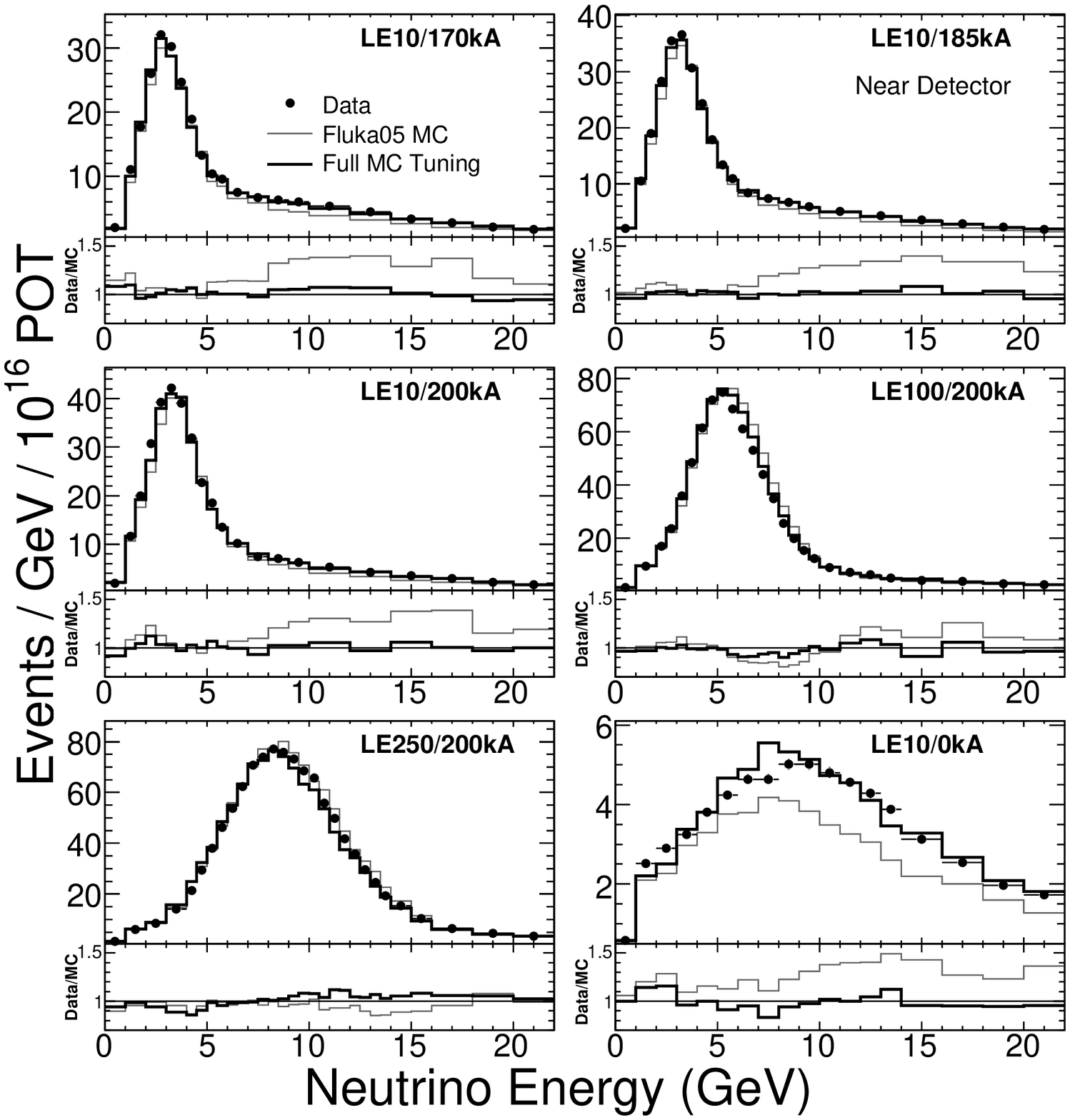}
   \caption{\numu{} charged-current energy spectra measured in the six beam configurations of Tab.~\ref{tab:beam-config} and compared with the Monte Carlo prediction. Two Monte Carlo predictions are shown: one (thin line) with the {\it ab initio} calculation based on \flukafive{}, the other (thick line) after constraining hadron production, focusing and detector parameters with the neutrino data. Panels along the bottom of each figure show the ratio of the measured and simulated spectra. \label{fig:6beamfit_all_enu}\label{fig:nd_specrat} }
  \end{centering}
\end{figure}

\pstart
Figure~\ref{fig:nd_specrat} compares the measured \numu{} charged-current energy spectrum with the Monte Carlo prediction for six different beam configurations. There are noticeable differences between data and the Monte Carlo calculation in all configurations but the magnitude of the discrepancies and the energy range over which they occur depends on the beam configuration. These observations suggest that a significant source of the disagreement between data and the Monte Carlo simulation may be due to inaccuracies in the calculation of the neutrino flux rather than mis-modeling of neutrino interactions or detector acceptance, since the latter depend most strongly on the energy of the incident neutrino while the former depends on the beam configuration.
\pend

\label{beammodeling}

\begin{figure*}
  \begin{centering}
    \includegraphics[width=0.35\textwidth]{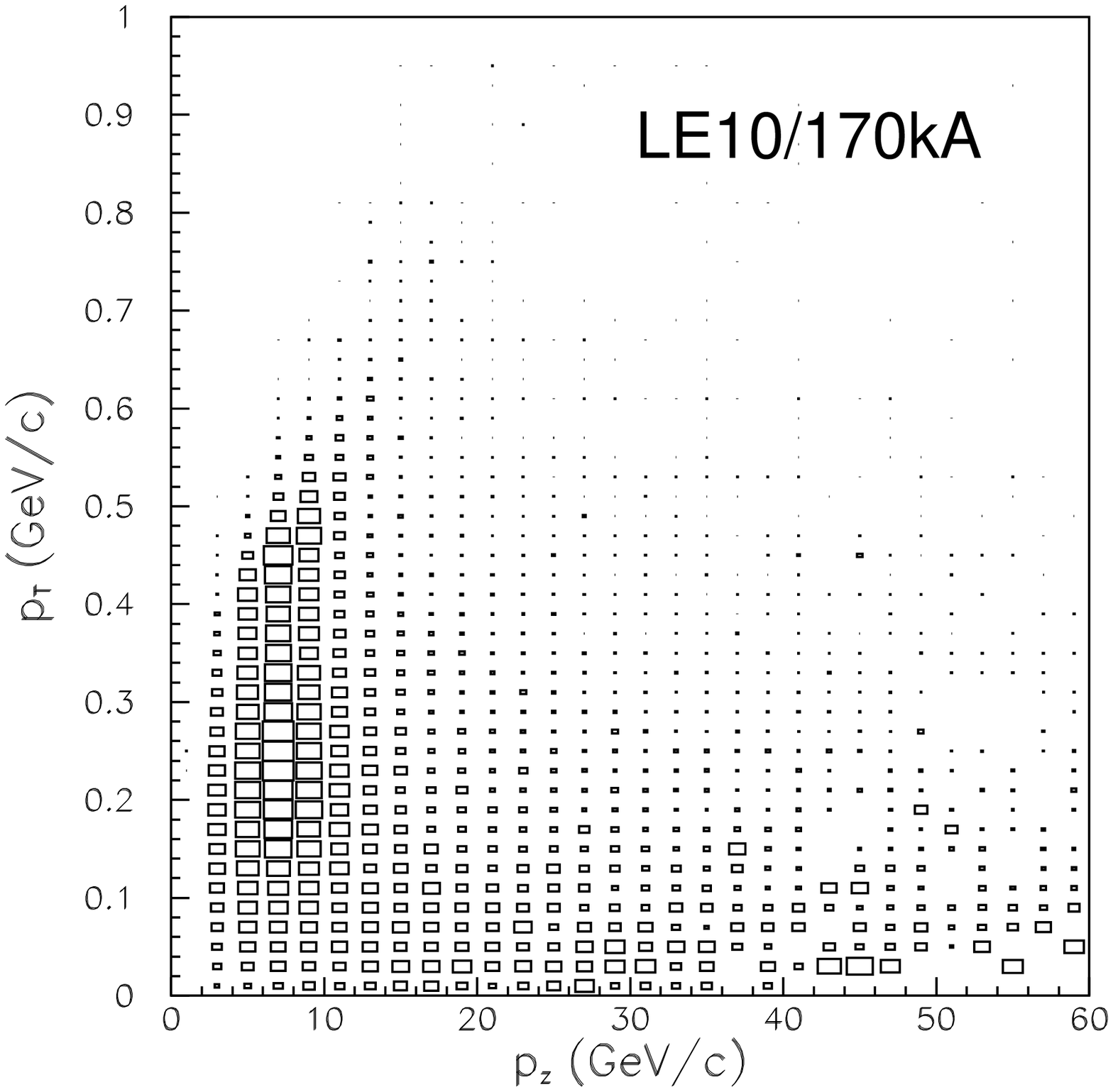}
    \includegraphics[width=0.35\textwidth]{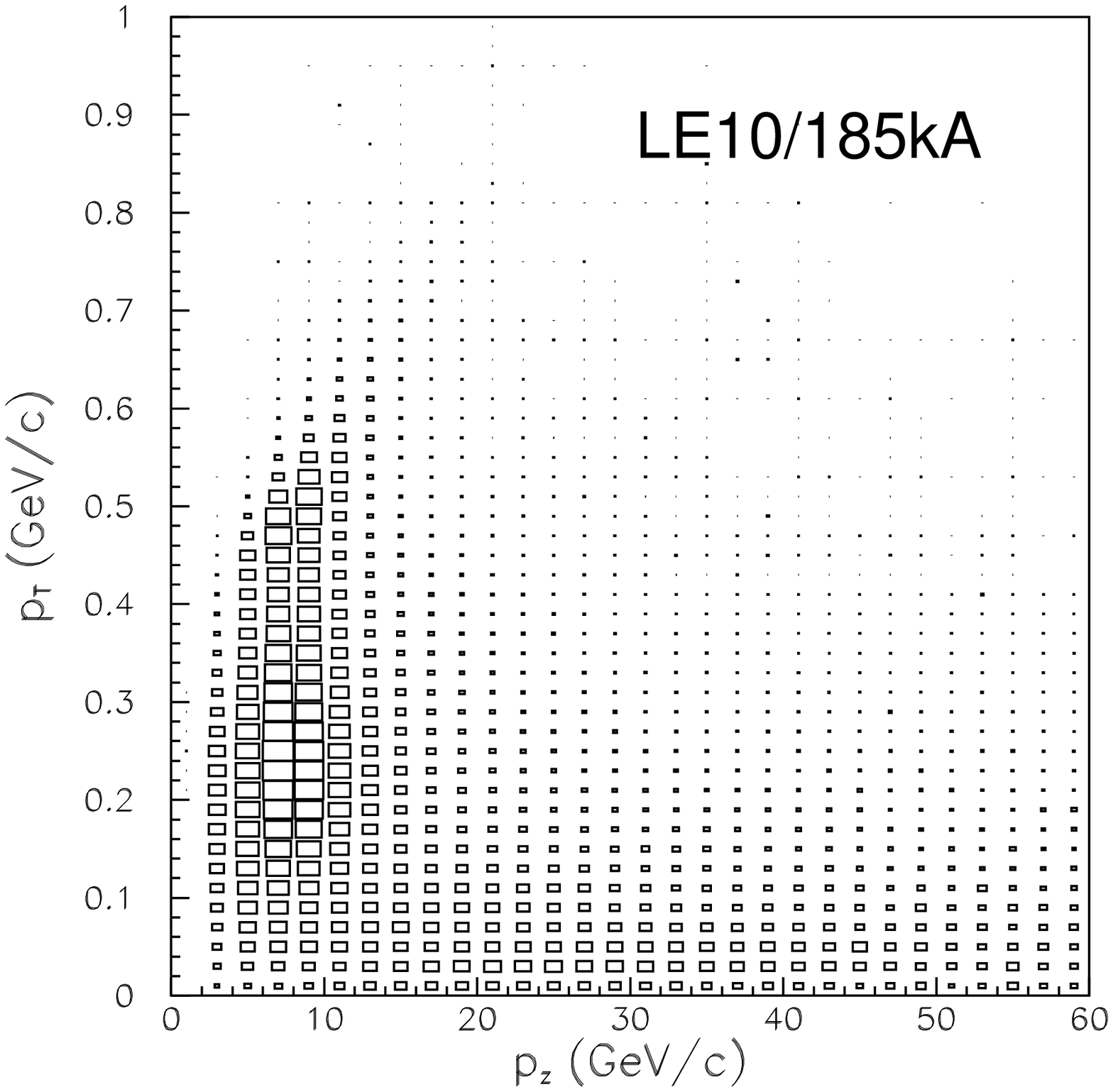}
\\
    \includegraphics[width=0.35\textwidth]{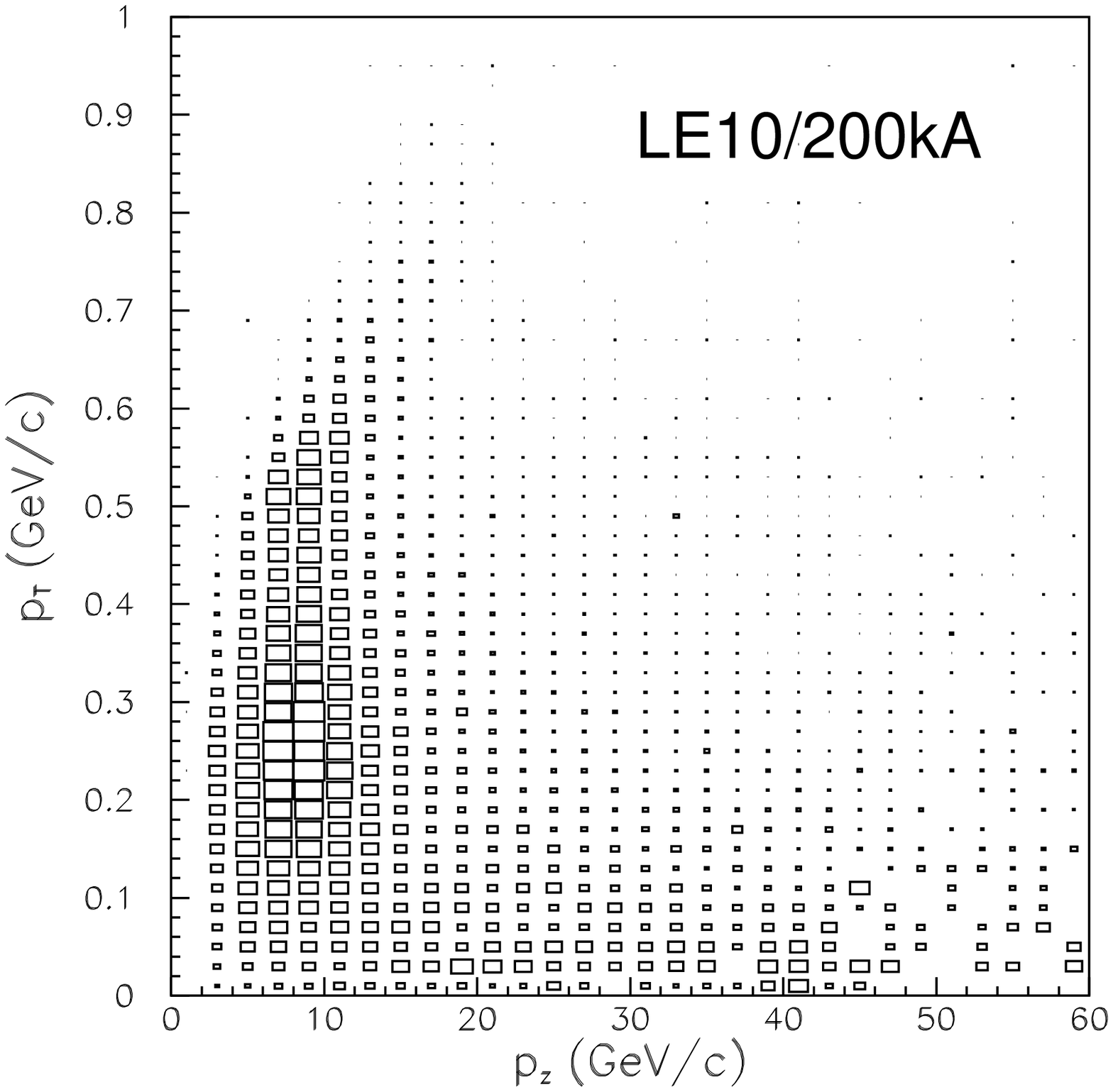}
    \includegraphics[width=0.35\textwidth]{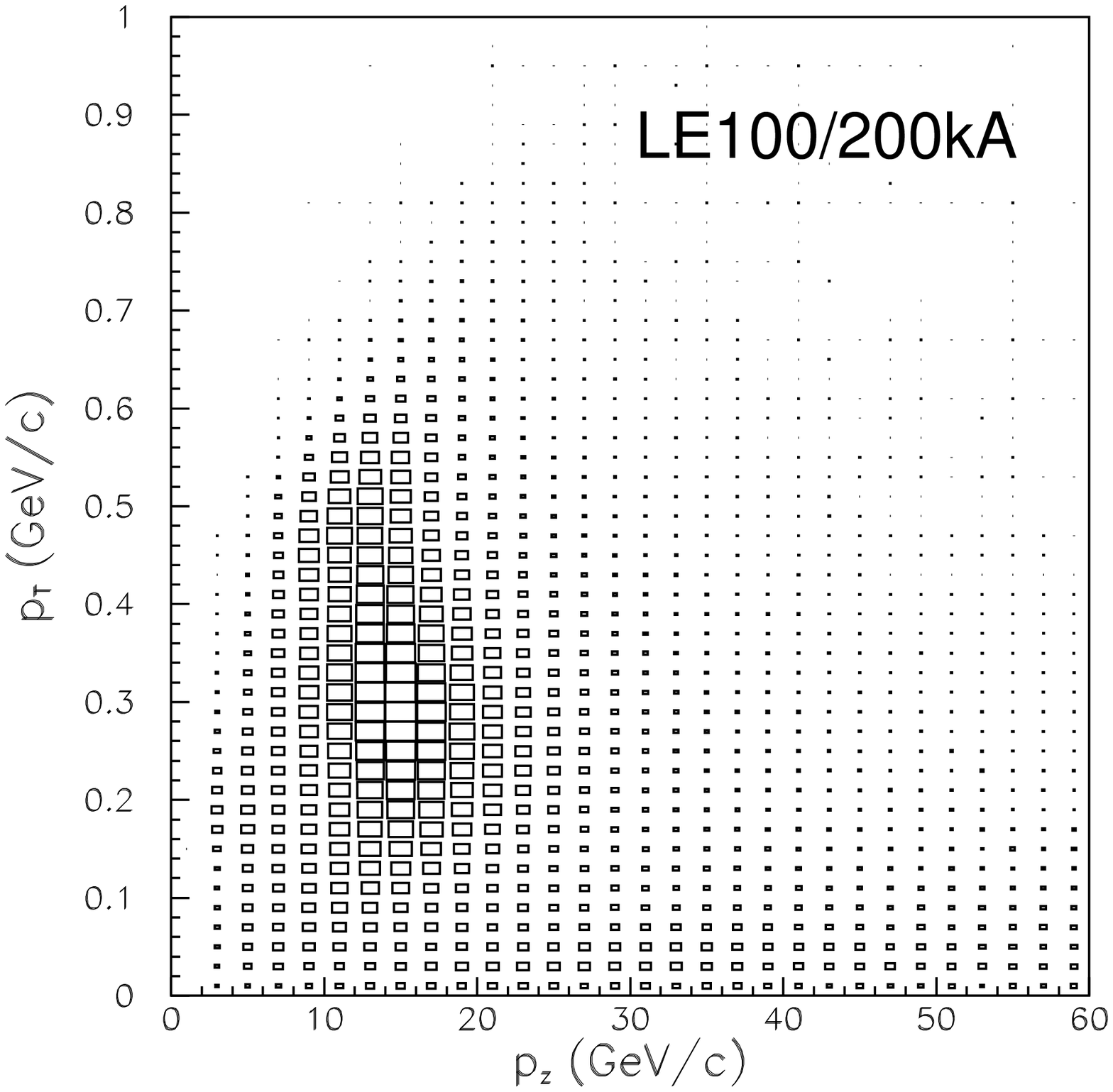}
\\
    \includegraphics[width=0.35\textwidth]{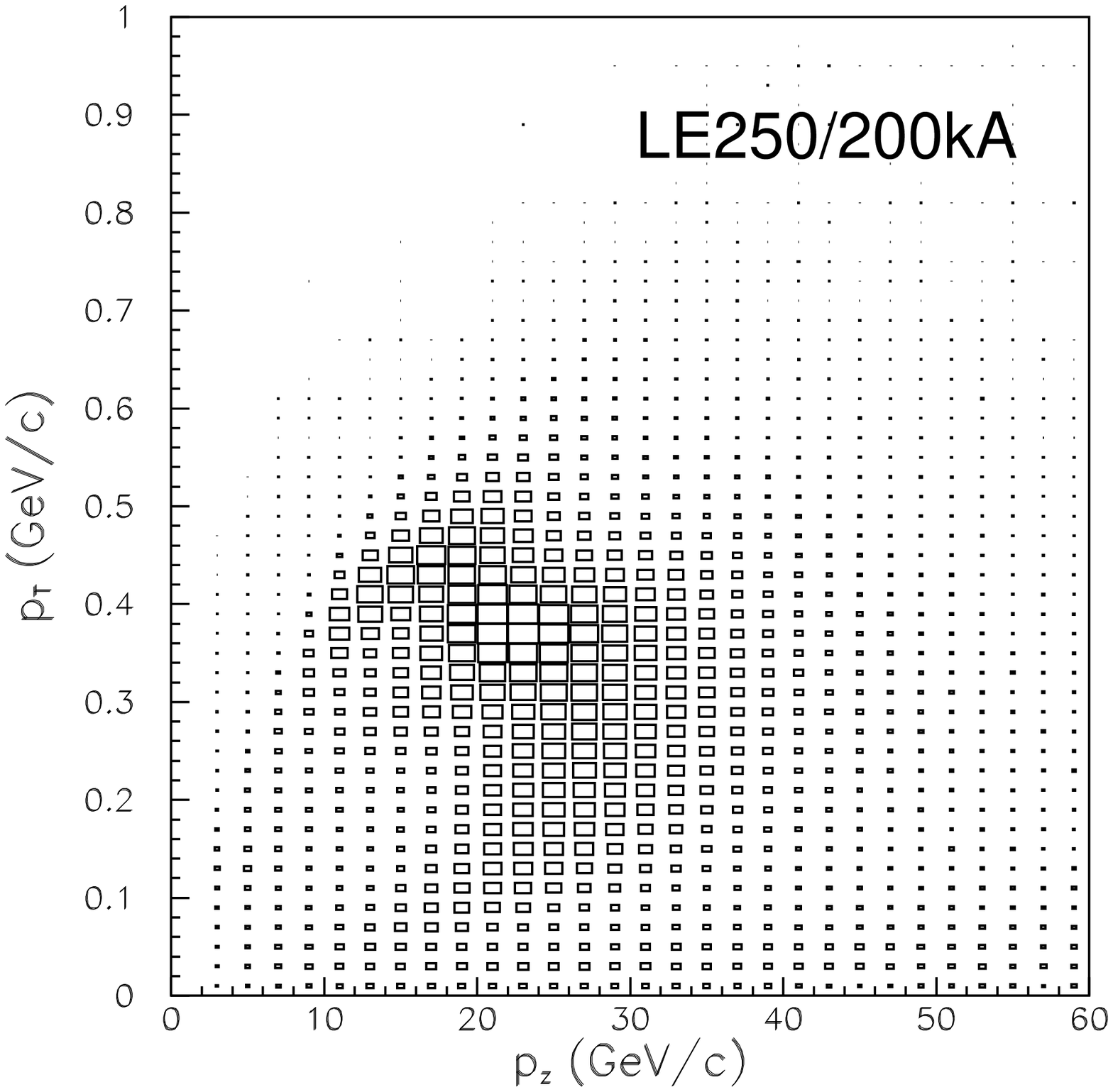}
    \includegraphics[width=0.35\textwidth]{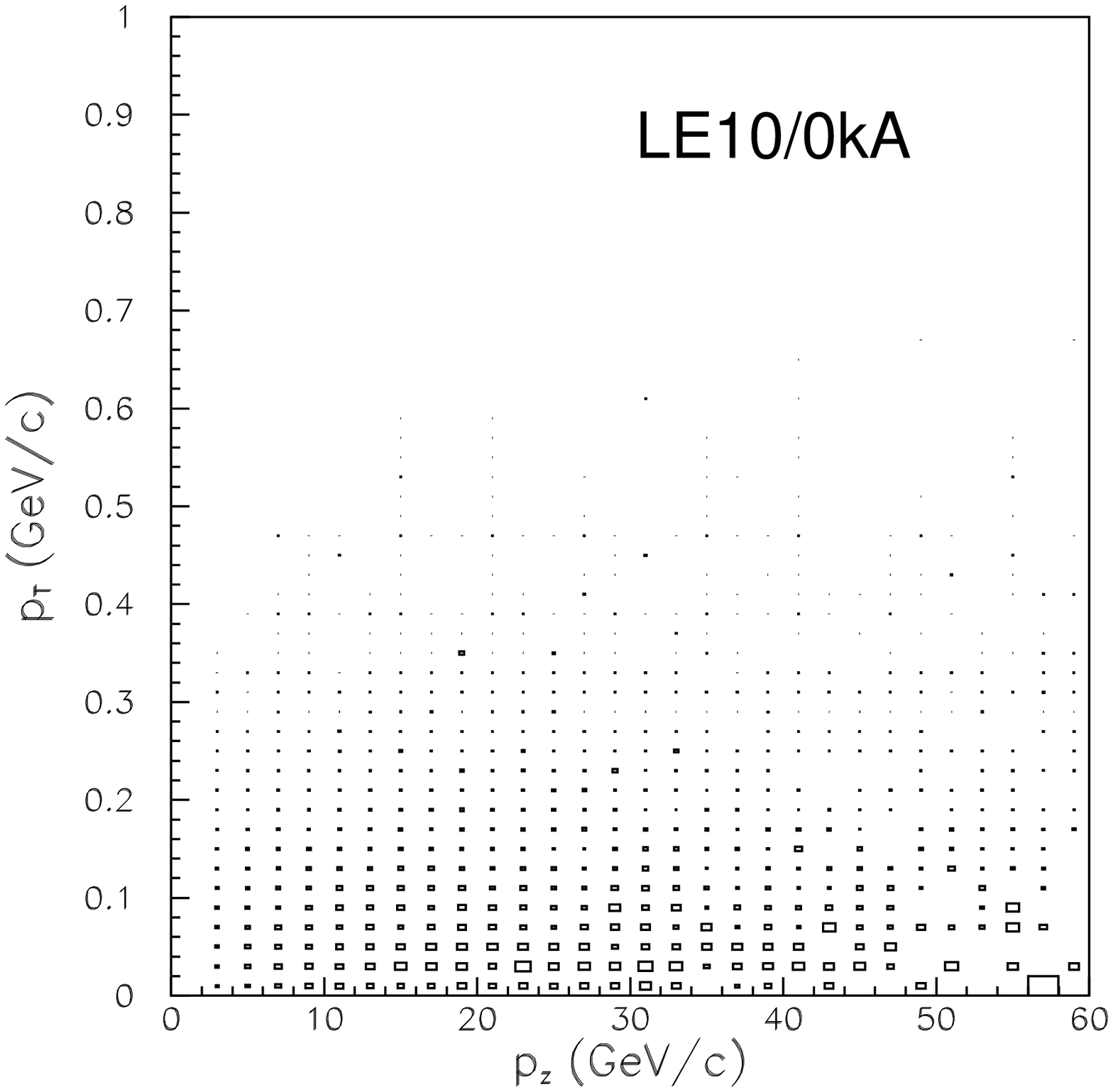}
      \caption{The yield, \hpxsec{}, of $\pi^+$ off the target that contribute to the \numu{} charged-current event rate in each of the beam configurations of Tab.~\ref{tab:beam-config}. Here \pt{} and \pz{} refer to the transverse and longitudinal momentum of pions leaving the \numi{} target before entering the focusing horns.  The box areas are proportional to the probability that a pion has the given (\pzpt{}) and results in a \numu{} charged-current interaction in the Far Detector. As is evident, each beam configuration samples a different region of (\pzpt{}).
      \label{fig:ptxf_allbeams}}
  \end{centering}
\end{figure*}


\begin{figure}
  \centering
  \includegraphics[width=\columnwidth]{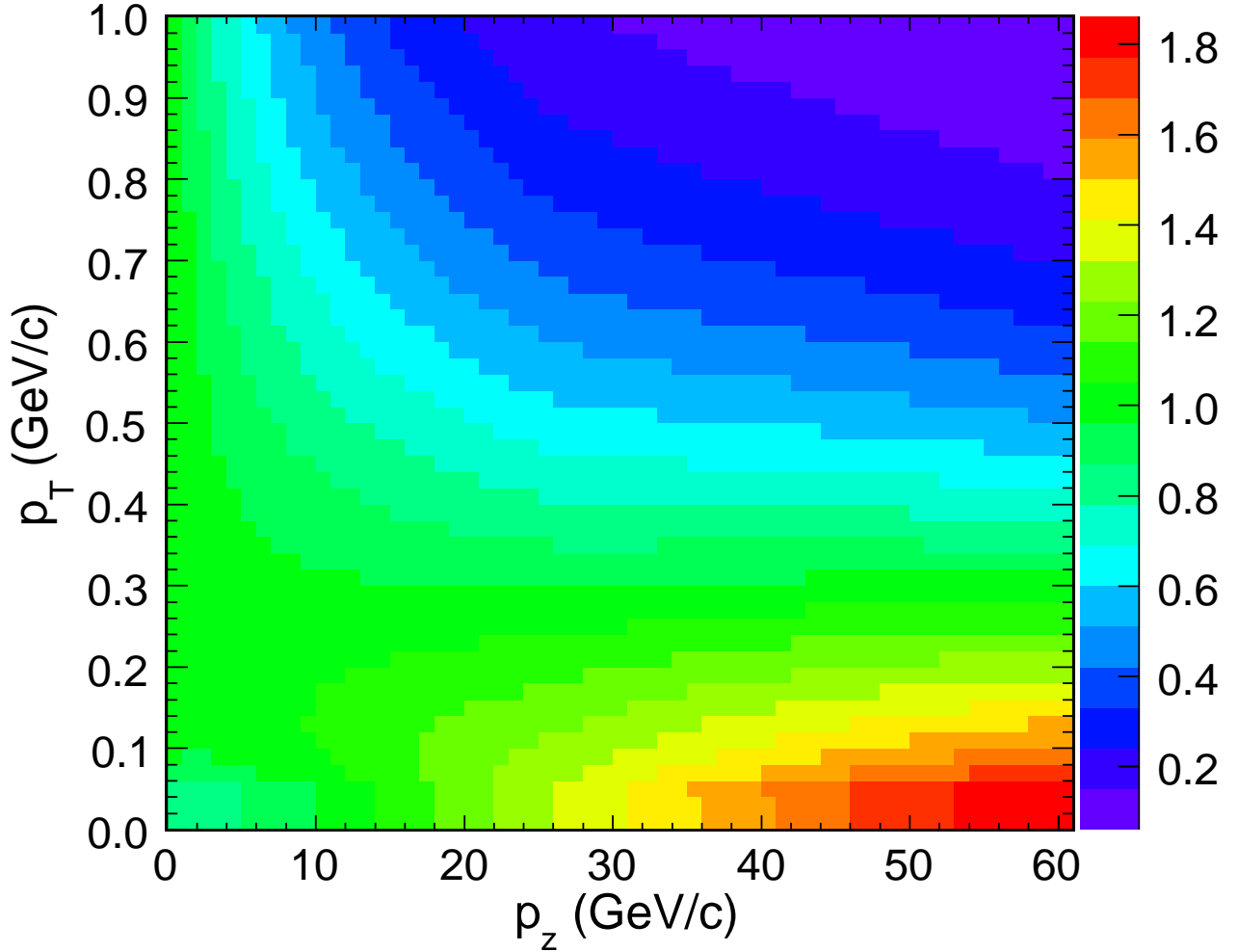}
  \caption{Pion production scale factors, as a function of \pzpt{}, derived by fitting Eq.~\ref{eq:skzp} to the Near Detector data according to the procedure described in the text. The scale factors are relative to the \flukafive{} prediction.}
  \label{fig:pzpt_weights}
\end{figure}

\begin{figure} 
  \begin{centering}
    \includegraphics[width=\columnwidth]{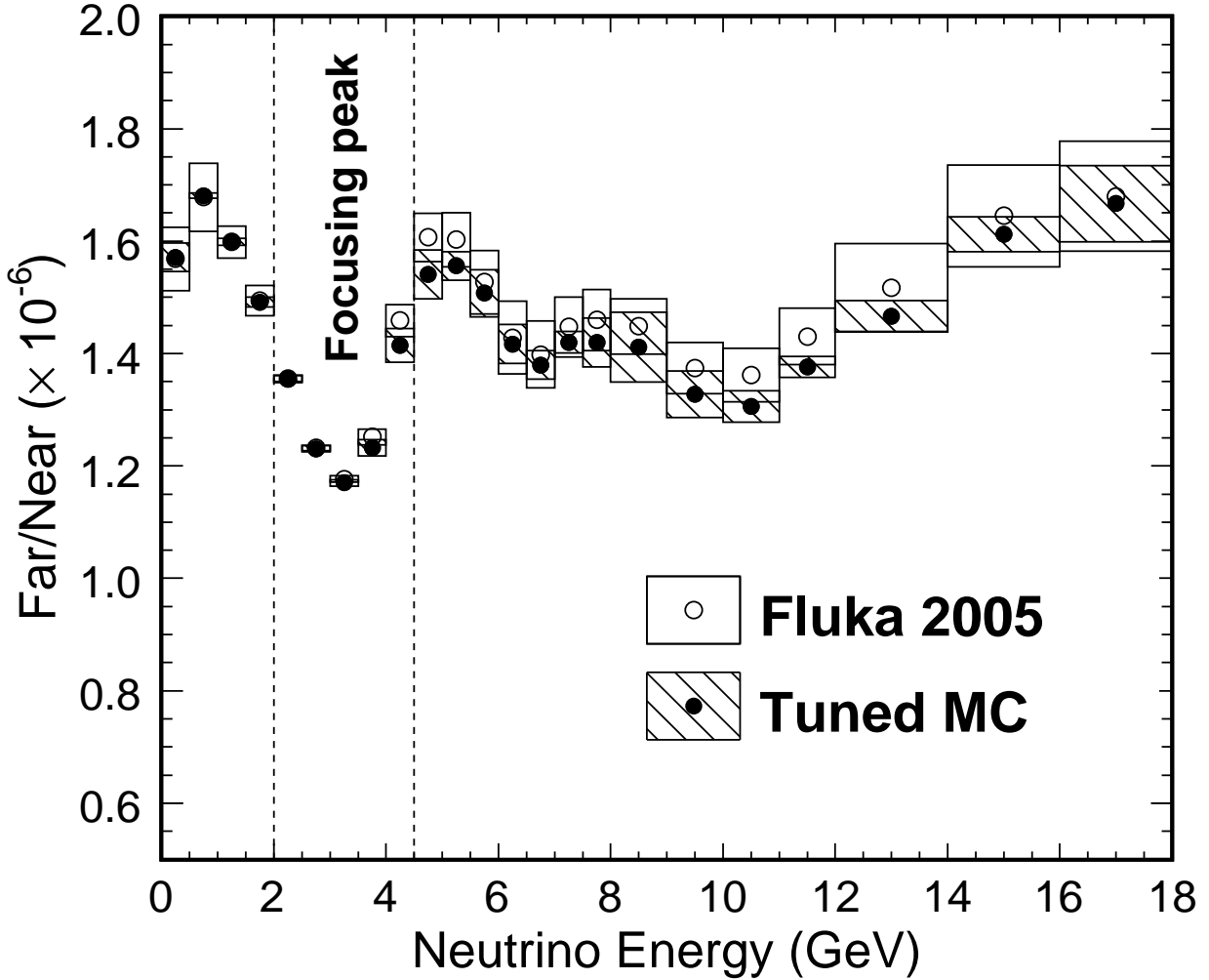}
    \caption{The predicted ratio (Far/Near) of \numu{} flux at the two detectors before and after the tuning procedure described in the text. The height of the box around each point denotes the uncertainty on the ratio.  For the points labeled ``Fluka 2005'' the uncertainties were calculated from the {\it a priori} uncertainties shown in Fig.~\ref{fig:miserr_fon} as well as the range of predictions from several hadron production codes~\cite{fluka4,mars1,malensek,bmpt}. For the points labeled ``Tuned MC'' the uncertainties are computed from the post-fit uncertainties on the functions $A',B'$ and $C'$ of Eq.~\ref{eq:skzp} as well as those on the additional parameters in Tab.~\ref{tab:fit_pars}. Unless specified otherwise, the tuned Monte Carlo is used in all further calculations. \label{fig:fn_before_after}}
  \end{centering}
\end{figure}

\begin{table}[t]
\begin{center}
\begin{ruledtabular}
\begin{tabular}{lcc}
parameter & initial uncertainty & best fit \\
\hline\hline
Horn 1 misalignment & \unit[1]{mm} & $\unit[0.1\pm 0.4]{mm}$  \\ 
POT Normalization & 2\% & $0.0\pm 0.8\% $\\
Horn Current Distribution $k$ & $\unit[0.17]{mm^{-1}}$ & $\unit[0.07\pm 0.05]{mm^{-1}}$\\
Shower Energy Offset & \unit[100]{MeV} & $\unit[24.7\pm 100.0]{MeV}$\\
Baffle scraping & 0.25\% & $ 0.1 \pm 0.1 \% $\\
Horn current calibration & 1\% & $0.9 \pm 0.3 \% $\\
Neutrino Energy Mis-calibration & 10\% & $-4.9 \pm 10.0\%$\\
Neutral-Current Contamination & 30\% & $-14\pm 7\%$\\
\end{tabular}
\end{ruledtabular}
\caption{Detector and beam modeling effects included when fitting the Monte Carlo to the Near Detector data. The effect of a $\unit[1]{\sigma}$ variation in each of the beam focusing parameters is shown in Fig.~\ref{fig:miserr_n}. The horn current distribution as a function of depth $z$ in the conductor is modeled as $I(z)\propto \exp(-k z)$ and we tabulate $k$ here.
\label{tab:fit_pars}}
\end{center}
\end{table}

%

\pstart
 As noted in Sec~\ref{sec:flux_uncertainties} uncertainties in the neutrino flux calculation arise from insufficient knowledge of hadron production off the \numi{} target as well as from several beam focusing effects described in Sec.~\ref{sec:flux_uncertainties}.  We constrain our flux calculation using \numu{} charged-current energy spectra measured in the Near Detector. The uncertainty in hadron production can be constrained because the position of the target and the horn current determine the region of pion (\pzpt{}) which contributes to \numu{} at the Near Detector. This is shown in Fig.~\ref{fig:ptxf_allbeams} for each of the beam configurations in which we collected data (see Tab.~\ref{tab:beam-config}).  Our method works by representing the underlying production yield, \hpxsec{}, as a parametric function which we use to tune the Monte Carlo in a \chisq{} fit to the Near Detector data.  We add terms to the \chisq{} which describe the influence that beam focusing and detector modeling uncertainties have on the \numu{} spectrum (see Fig.~\ref{fig:miserr_n}).  This technique is similar to those used in previous experiments~\cite{ahrens1986,allen1993,astier2003,mcfarland1998,zeller2001} but the multiple beam configurations allow us to selectively enhance different regions of \pz{} and \pt{}.
\pend

\pstart
The \flukafive{} prediction of $\pi^{+}$ yields off the \numi{} target is well described by the function
\pend
\begin{equation}
f(\pzpt{}) = \frac{\D^2N}{\D \pz{} \D p_T} = \left[A(\pz{})+ B(\pz{}) p_T \right]\exp\left(-C(\pz{})p_T^{3/2}\right)
\label{eq:skzp}
\end{equation}
\pstart
\noindent where  $A$, $B$ and $C$ are functions of \pz{}. Equation~\ref{eq:skzp} and $A$, $B$ and $C$ are similar to the functions advocated by~\cite{bmpt} but we have modified them to better describe the thick target yields. The \numu{} spectra are then fit by warping $A,B,C$ as linear functions of \pz{}. The fit outputs the warped functions $A',B',C'$ and we calculate hadron yield scale factors relative to \flukafive{} as
\begin{equation}
 w(\pzpt{})=\frac{f(\pzpt{};A',B',C')}{f(\pzpt{};A,B,C)}\quad .
\end{equation}
These scale factors are shown in Fig.~\ref{fig:pzpt_weights}.
\pend

\pstart
In our fit we include a term to constrain the mean transverse momentum of pions, $\langle p_T\rangle$, to the \flukafive{} value of \unit[364]{\mevc{}}, with an uncertainty of \unit[15]{\mevc{}}, obtained from the variation of independent hadron production calculations~\cite{fluka4,mars1,malensek,wang,ckp}. The $K^{+}$ yield does not contribute significantly to the \numu{} flux below \unit[30]{GeV} and was permitted to vary by a scale factor. We augment the fit to include terms which represent distortions of the \numu{} charged-current spectra caused by uncertainties in the horn current, horn position, distribution of current in the horn conductors, and proton beam scraping on the baffle. The effect of these uncertainties are shown in Fig.~\ref{fig:miserr_n}. We also include a 2\% uncertainty in the normalization due to proton on target counting.  To avoid forcing the fit to over-correct for detector acceptance uncertainties we include a 10\% error on the neutrino energy, a \unit[100]{MeV} uncertainty in the shower energy and a 30\% uncertainty in the normalization of the neutral-current contamination.
\pend


\pstart
The results of the fit are shown in Fig.~\ref{fig:6beamfit_all_enu}, Fig.~\ref{fig:pzpt_weights} and Tab.~\ref{tab:fit_pars}.  The agreement between the measured and calculated neutrino energy spectra in Fig.~\ref{fig:6beamfit_all_enu} is significantly improved with residual discrepancies reduced to 5\% or less in all beam configurations. The \chisq{} per degree of freedom improves from $3087/360$ to $569/345$.  Figure~\ref{fig:pzpt_weights} shows the scale factors applied to the hadron production yield \hpxsec{}. The most significant adjustment is made to the region $\unit[\pt{}\lesssim 150]{\mevc{}}$ and $\unit[\pz{}\gtrsim 15]{\gevc{}}$ to increase the flux of unfocused hadrons which dominate the flux for neutrino energies larger than $\unit[8]{GeV}$ in the LE10 beam configurations. Only modest adjustments are made in the region focused in the LE10 configurations ($\unit[100\lesssim \pt{}\lesssim 500]{\mevc{}}$, $\unit[\pz{}\lesssim 10]{\gevc{}}$ ). Hadrons produced with high \pt{} and \pz{} do not contribute to neutrinos in the \minos{} detectors and the scale factors are not constrained.  The best fit parameters, shown in Tab.~\ref{tab:fit_pars}, are all within $1\sigma$ of their initial values. Figure~\ref{fig:fn_before_after} shows the predicted ratio Far/Near of \numu{} charged-current energy spectra before and after the fitting procedure.  The fit improves our understanding of the relationship between the Far and Near spectra and we use the results in all further Monte Carlo calculations. Uncertainties in the ratio are shown as boxes which are reduced after the fit as a result of the constraints the neutrino data places on the functions $A,B,C$ of Eq.~\ref{eq:skzp} and the parameters of Tab.~\ref{tab:fit_pars}.  We have checked that alternate parametrizations of \hpxsec{} yield similar Far/Near ratios. This is important since the ratio characterizes the way in which the neutrino flux changes between the two detectors and indeed provides a way in which the Near Detector data can be used to predict the Far.
\pend



\section{Prediction of the Far Detector Spectrum}


\label{sec:extrap}
\pstart
The Near Detector measures the neutrino energy spectrum close to the source and before oscillations have occurred so as to mitigate the significant uncertainties in the neutrino flux, cross-sections and detector acceptance. As discussed in Sec.~\ref{sec:skzp} these data have been used to improve the flux calculation. The primary objective, however, is not to constrain the flux, cross-section or acceptance independently but instead to predict the energy spectrum at the Far Detector in the absence of neutrino oscillations.  Any differences between the prediction and the Far Detector data may then be interpreted as neutrino oscillations or some other hypothesis.
\pend

\subsection{Extrapolation Techniques}
\label{sec:extrap_methods}
\pstart
We refer to the process of predicting the Far Detector spectrum as ``extrapolation'' because of the crucial role played by the Near Detector data, and the expected similarity~\footnote{Modulo the ratio of solid angles $\Omega_{FD}/\Omega_{ND} = d^{2}_{ND}/d^{2}_{FD}\approx(\unit[1]{km}/\unit[735]{km})^{2}$ where $d_{ND},d_{FD}$ refer to the distance from the target to the detectors.} between the neutrino fluxes at the two sites (see Fig.~\ref{fig:near-far}, Fig.~\ref{fig:fn_before_after} and the discussion in Sec.~\ref{sec:expected_spectra}). The extrapolation process may then, in one approach, be viewed as making $\sim 30\%$\footnote{These adjustments are relative to the solid angle correction $\Omega_{FD}/\Omega_{ND}$.} adjustments based on the energy spectrum measured in the Near Detector in order to predict the Far Detector spectrum. These adjustments are made, in part, by using the Monte Carlo simulation but are relative between the two detectors and are thus less sensitive to uncertainties in the absolute flux, cross-sections and detector acceptance. We refer to this approach as ``direct'' extrapolation because the Near Detector data are used as measured without further constraining the simulation. In a second approach Near to Far Detector extrapolation may be viewed as using the Near Detector data to constrain the Monte Carlo calculation of the neutrino flux, cross-sections and detector acceptance. The improved simulation is then used to calculate the energy spectrum expected at the Far Detector. This approach is referred to as ``indirect'' extrapolation in contrast to the ``direct'' approach. In practice the two approaches are complementary and result in very similar predictions of the Far Detector spectrum. The present section describes the specific techniques that were developed to implement each approach. We have pursued a total of four techniques in order to better understand the extrapolation process and study the robustness of our oscillation measurement.
\pend

\subsubsection{Direct Extrapolation Methods}




\pstart
In the absence of oscillations the neutrino fluxes at the Far and Near Detectors are similar, but not identical, and the differences can be most transparently characterized in terms of the Far to Near flux ratio shown in Fig.~\ref{fig:fn_before_after}. The energy dependence in the ratio comes from two main sources~\cite{Szleper:2001nj,Kopp:2006ky}. First, the solid angle acceptance of the Near Detector varies as a function of the distance from the neutrino production point while the acceptance of the Far Detector is essentially constant. The two detectors subtend different solid angles and by Eq.~\ref{eq:angle} observe different neutrino angle distributions which, according to Eq.~\ref{eq:enu-eqn}, result in different neutrino energy distributions. Apertures in the beam-line (for example, the decay pipe) and under or over focusing in the horns enhance this effect by attenuating pions in certain energy ranges and locations along the beam-line. According to Eq.~\ref{eq:enu-eqn}, this introduces an energy dependent suppression of the neutrino flux which, because of the disparate angular acceptance, is not the same in the two detectors. Second, neutrinos produced by decays at finite radii from the beam axis will cover different angular ranges when intersecting the fiducial volumes of the Near and Far Detectors. This effect also introduces different energy distributions in the two detectors. These effects cause the peaks and dips in Fig~\ref{fig:fn_before_after}. Thus, the major sources of {\it relative} Near to Far differences in the neutrino flux (and hence, energy spectrum) are largely due to beam-line geometry, focusing, angular acceptance and decay kinematics. This suggests that the Monte Carlo simulation may be used, with relatively small uncertainties, to derive a transfer function that extrapolates the neutrino energy spectrum measured in the Near Detector to the Far Detector.
\pend

\pstart
The Far to Near flux ratio of Fig.~\ref{fig:fn_before_after} is itself nearly a transfer function except that it is expressed in true neutrino energy and does not account for detector energy resolution, small disparities in efficiency between the two detectors, and their different fiducial masses.  A suitable replacement, however, can be evaluated by applying the \numucc{} selection criteria of Sec.~\ref{sec:evtsel} to fully simulated events in both detectors to derive neutrino event rates, $\{n_{i}\}$ and $\{f_{i}\}$ in bins $i$ of reconstructed neutrino energy. The Near Detector data $\{N_{i}\}$ are then used to predict the Far spectrum:
\begin{equation}
F^{predicted}_{i}=N_{i}\times\frac{f_{i}}{n_{i}} \qquad \quad \mbox{``The F/N Method''}\quad .
\end{equation}
\noindent This technique is referred to as the ``F/N method'' but is equivalent to scaling each bin in the simulated Far Detector reconstructed neutrino energy spectrum by the ratio of the number of observed to expected events in the corresponding Near Detector reconstructed neutrino energy bin. The effect of neutrino oscillations may be accounted for by modifying $f_{i}$ according to Eq.~\ref{eq:oscprob}.
\pend


\begin{figure}
\centering
\subfloat[\label{fig:matrix_near}]{\includegraphics[width=0.45\textwidth]{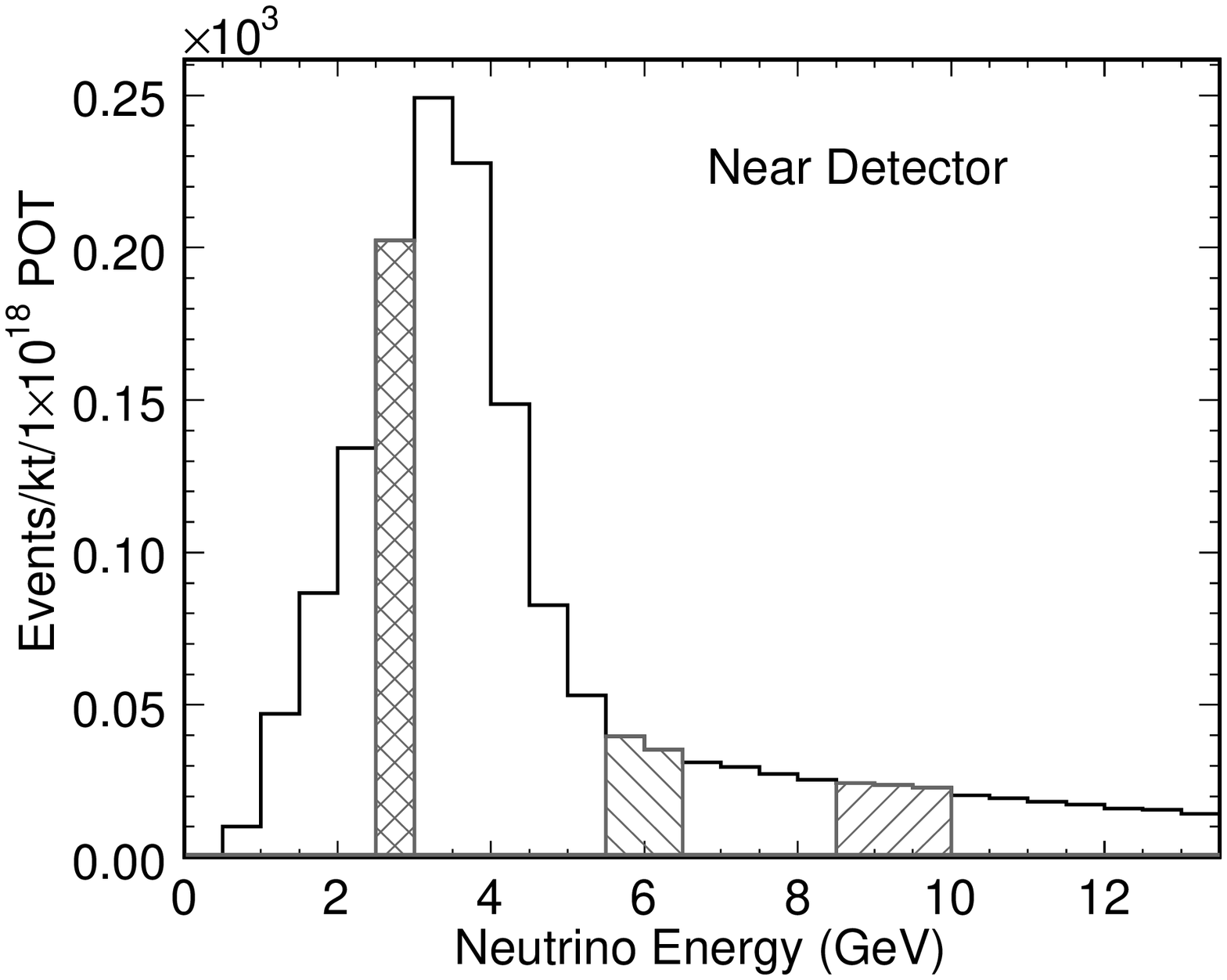}}
\subfloat[\label{fig:matrix_far}]{\includegraphics[width=0.45\textwidth]{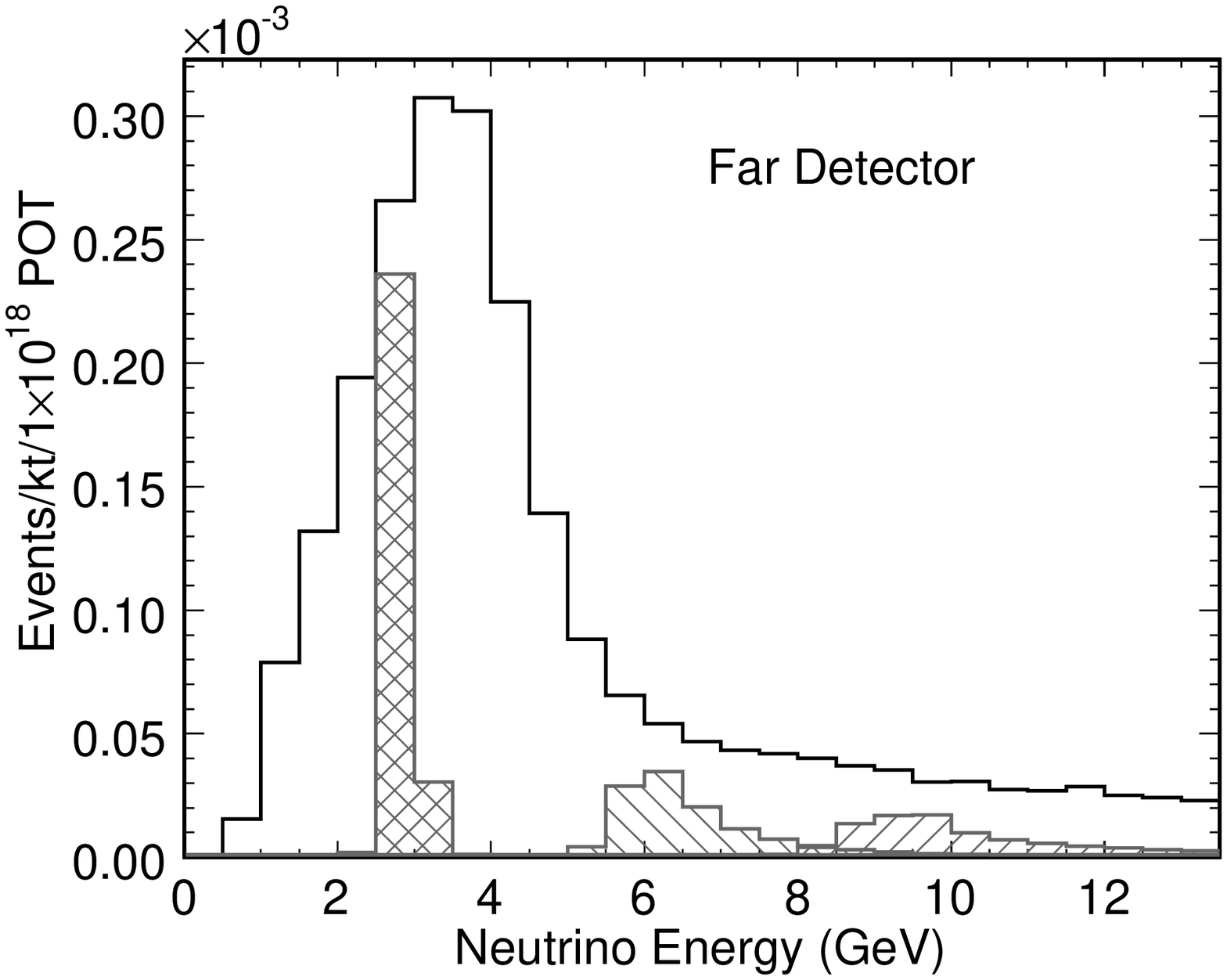}}
\caption{\label{fig:energy_spread} The relationship between the energy of neutrinos observed in the Near Detector with those observed in the Far Detector. Decays producing neutrinos with a given energy in the Near Detector would produce a range of energies in the Far Detector, yielding the energy smearing seen here. This relationship may be represented algebraically by treating the joint distribution in Fig.~\ref{fig:the_matrix} as a matrix, and the distributions \protect\subref{fig:matrix_near} and \protect\subref{fig:matrix_far} as column vectors as in Eq.~\ref{eq:matrix}. }
\end{figure}

\begin{figure}
\centering
\includegraphics[width=\columnwidth]{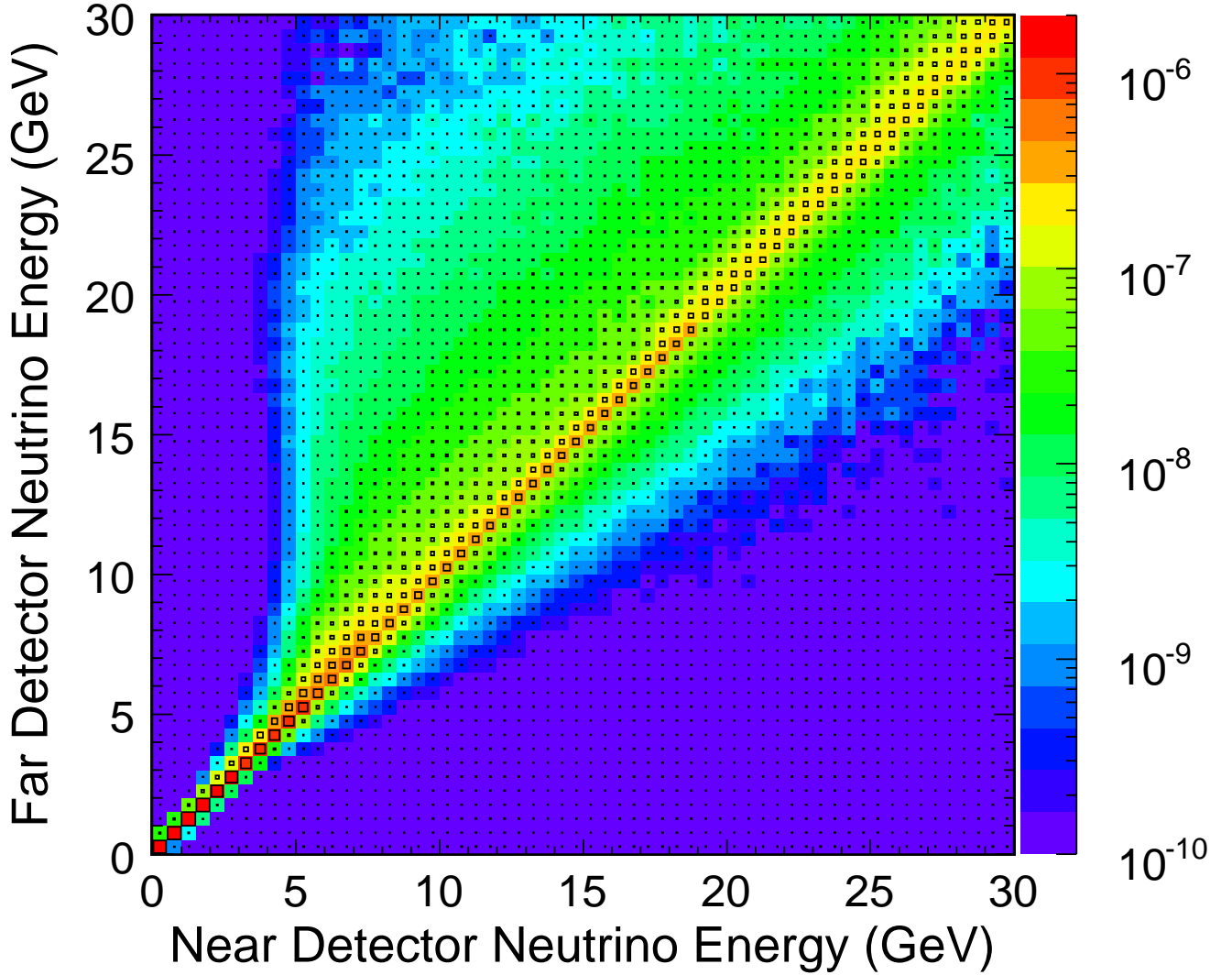}
\caption{\label{fig:the_matrix} The joint distribution of neutrino energies observed in the Near and Far Detectors. The contents of each cell represent the mean number of \numu{} events expected in the Far Detector for one event in the Near Detector.  This distribution may be treated as a matrix, as in Eq.~\ref{eq:matrix}, to relate the energy spectra measured in the Near Detector to those in the Far Detector. }
\end{figure}
\pstart
Neutrinos having a given energy in the Near Detector come from decays which would, collectively, yield neutrinos covering a range of energies in the Far Detector as indicated in Fig.~\ref{fig:near-far}, Eq.~\ref{eq:enu-eqn} and Eq.~\ref{eq:angle}. This effect is shown at several energies in Fig.~\ref{fig:energy_spread} and is again a consequence of the differences in angular acceptance of the two detectors, beam-line apertures, decay kinematics and off-axis decays. The fact that a single energy in the Near Detector corresponds to a range in the Far Detector suggests that the neutrino energy spectra may be related by a two-dimensional matrix rather than a one dimensional ratio.  This ``beam matrix'' $\{B_{ij}\}$ is shown in Fig.~\ref{fig:the_matrix}. Each cell represents the number of neutrinos expected in energy bin $i$ at the Far Detector for one neutrino in bin $j$ in the Near Detector. Neutrinos with energies in the range \unit[1-6]{GeV} are well focused and the contents of the cell along the diagonal is approximately proportional to the ratio of solid angles $\Omega_{FD}/\Omega_{ND} = d^{2}_{ND}/d^{2}_{FD}\approx(\unit[1]{km}/\unit[735]{km})^{2}\approx 1.8\times10^{-6}$. The matrix is constructed from the beam simulation using the known geometric acceptance of the two detectors. The energy dependence of the \numucc{} \xsec{} is included in the calculation but is most relevant for the small off-diagonal elements. As with the F/N ratio, the Far Detector spectrum predicted by the matrix is relatively insensitive to uncertainties in the hadron production calculation, neutrino cross-sections and detector effects.
\pend

\pstart
The matrix can only be employed after accounting for detector acceptance and inefficiencies. The Near Detector measurement is first corrected to remove non-  \numucc{} contributions. A second correction, derived from a sample of fully simulated events in the Near Detector,  is then applied to deconvolve the detector efficiency and energy resolution. Both corrections are derived from the simulation described in Sec.~\ref{sec:mc}. The resulting distribution estimates the true \numucc{} energy spectrum at the Near Detector. This distribution, organized in energy bins, is treated as an $m$-dimensional column vector $\{N_{i}\}$ and multiplied by the $m\times m$ dimensional matrix $\{B_{ij}\}$ to estimate the true \numucc{} energy at the Far Detector
\begin{equation}
\label{eq:matrix}
F_{i}=\sum_{j=0}^{m} B_{ij} N_{j} \qquad \qquad \mbox{``The Beam Matrix Method''} \quad .
\end{equation}
The effects of neutrino oscillations may be applied to the derived Far Detector spectrum after which detector energy resolution, efficiency, and non-\numucc{} backgrounds are added back in. The resulting spectrum may be directly compared to the Far Detector data.
\pend

\pstart
The direct extrapolation methods described here reduce, but do not entirely eliminate, the effect that uncertainties in hadron production, neutrino cross-sections and detector acceptance have on the prediction of the neutrino energy spectrum at the Far Detector. We will discuss these uncertainties in greater detail in Sec~\ref{sec:extrap_syst}.
\pend




\subsubsection{Indirect Extrapolation Methods}
\pstart
In these methods, a fit is performed to observed Near Detector distributions using a Monte Carlo dataset  with parametrized uncertainties due to beam modeling, neutrino cross-sections and neutrino energy scales. The result of this fit is a tuned Monte Carlo calculation which can then be applied to the Far Detector in order to obtain the predicted neutrino spectrum. These fits have the advantage that they can, in principle, separate systematic effects according to their effect on the fitted distributions with the disadvantage that the fitting functions may not adequately describe the data or (due to degeneracies in the fitted parameters) correctly describe the underlying physics.
\pend

\pstart
Two indirect extrapolation methods have been developed: ``NDFit'' and ``2DFit''. In both techniques a fit is performed to the measured Near Detector energy spectra in order to improve agreement between the observed distributions and those predicted from Monte Carlo simulation. Both methods use data collected in the six beam configurations of Tab.~\ref{tab:beam-config} with approximately equal numbers of events in each configuration. In total, approximately $2.5\times 10^{5}$ Near Detector data events are used. The fitting techniques are permitted to re-extract the neutrino energy scale and normalization parameters originally determined when tuning the flux calculation (see Tab.~\ref{tab:fit_pars}). The two techniques employ very similar fitting procedures. In the case of the NDFit method, the fit is performed to one-dimensional reconstructed neutrino energy histograms, whereas in the case of the 2DFit, the reconstructed neutrino energy distributions are sub-divided in bins of inelasticity, $y = 1-E_{\mu}/E_{\nu}$.  
\pend


\pstart
The NDFit method varies the parameters $\boldsymbol{\alpha}=\{\alpha_{j}\}$ to minimize
\pend
\begin{equation}\label{eq:ndfit}
\chi^{2}=\sum_{k}^{E_{\nu}\mathrm{-bins}} \frac{\left(N_{k}-n_{k}\left(\boldsymbol{\alpha}\right)\right)^{2}}{\sigma_{k}^{2}+{S_{k}}^{2}} +\sum_{j=1}^{\mathrm{syst}}\left(\frac{\Delta\alpha_{j}}{\sigma^{\alpha}_{j} }\right)^{2} \qquad \qquad \mbox{``The NDFit Method''} \quad ,
\end{equation}
\pstart
where $N_{k}$ and $n_{k}$ are the numbers of data and simulated events in bin $k$ of the reconstructed neutrino energy distribution. There are 38 energy bins for each of the six beam configurations (see Tab.~\ref{tab:beam-config} and Fig.~\ref{fig:6beamfit_all_enu}). The $\{\Delta\alpha_{j}\}$ are the deviations from nominal of the systematic parameters $\boldsymbol{\alpha}$, with associated uncertainties $\{\sigma^{\alpha}_{j}\}$. These are described below. The second term in Eq.~\ref{eq:ndfit} constrains the fit by increasing $\chi^{2}$ as the parameters are varied away from their nominal values. The error on the number of predicted events is the sum in quadrature of the statistical $\{\sigma_{k}\}$ and systematic $\{S_{k}\}$ uncertainties in the neutrino flux, where the latter are derived from the calculation described in Sec.~\ref{sec:skzp}. 
\pend

\pstart
 The five systematic parameters $\boldsymbol{\alpha}=\{\alpha_{1}\ldots\alpha_{5}\}$ describe uncertainties in the shape and normalization of the $\nu_\mu$ charged-current cross-section and the muon and hadron energy scales of the detector. These parameters are: $\alpha_1$, charged-current axial vector masses for quasi-elastic and resonance production processes, varied coherently; $\alpha_{2}$, cross-section scale factor for the non-resonant $1\pi$ and $2\pi$ production cross-section at invariant masses $W<\unit[1.7]{\gevcsq{}}$; $\alpha_3$, the absolute hadronic energy scale; $\alpha_4$, the absolute muon energy scale; $\alpha_5$, the overall normalization. The magnitude of these uncertainties is shown in Tab.~\ref{tab:ndfitres}.  The hadronic energy scale uncertainty is dominated by final state interactions (see Sec.~\ref{sec:mc} and \ref{sec:shwr_calib}). The muon energy scale was estimated from a comparison of range and curvature measurements for stopping muon tracks observed in the Near Detector. The normalization uncertainty accounts for the remaining uncertainties left after the improved flux calculation of Sec.~\ref{sec:skzp}.
\pend

\begin{figure*}
\centering
\subfloat{\includegraphics[width=0.33\textwidth,clip]{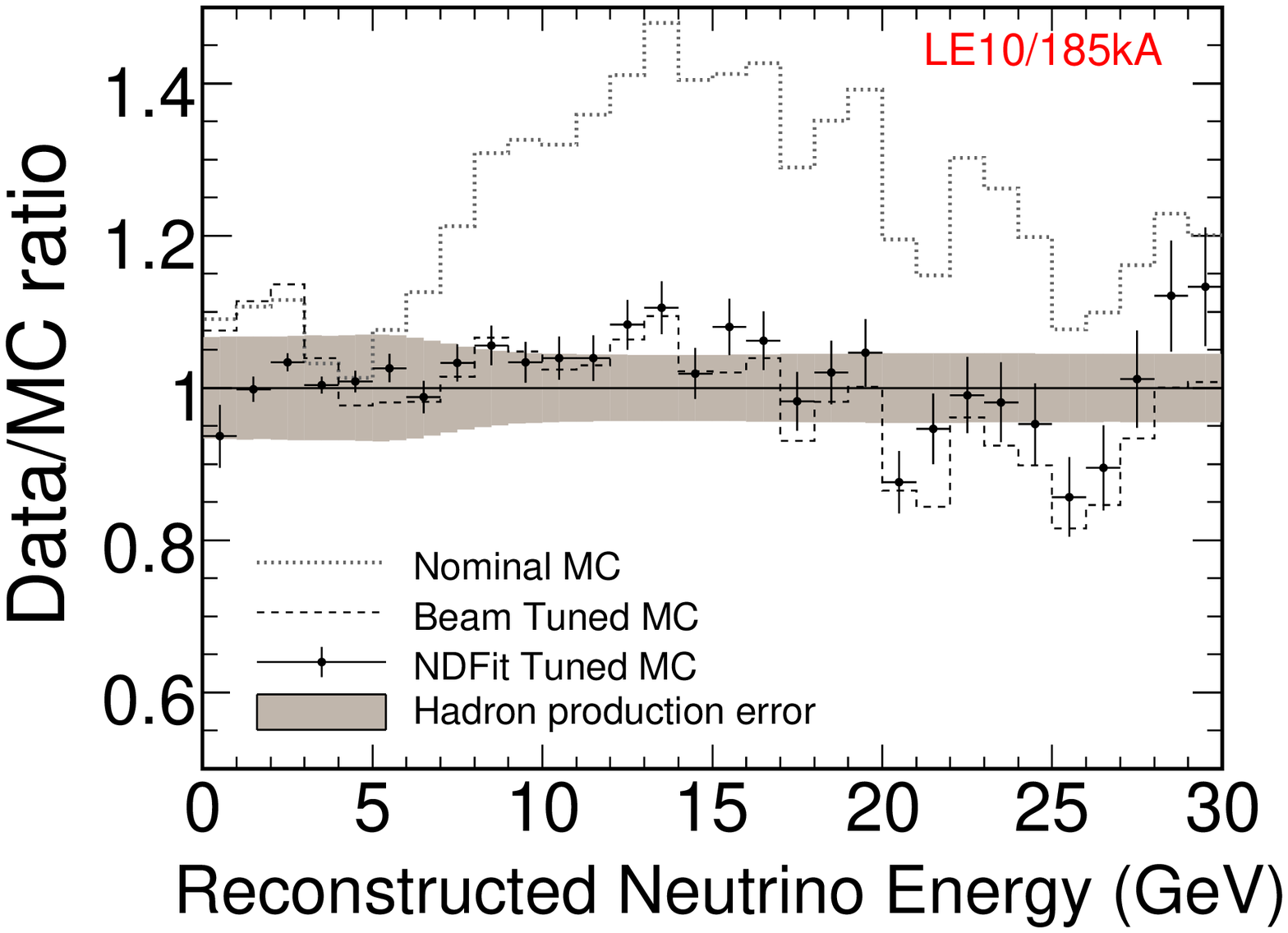}}
\subfloat{\includegraphics[width=0.33\textwidth,clip]{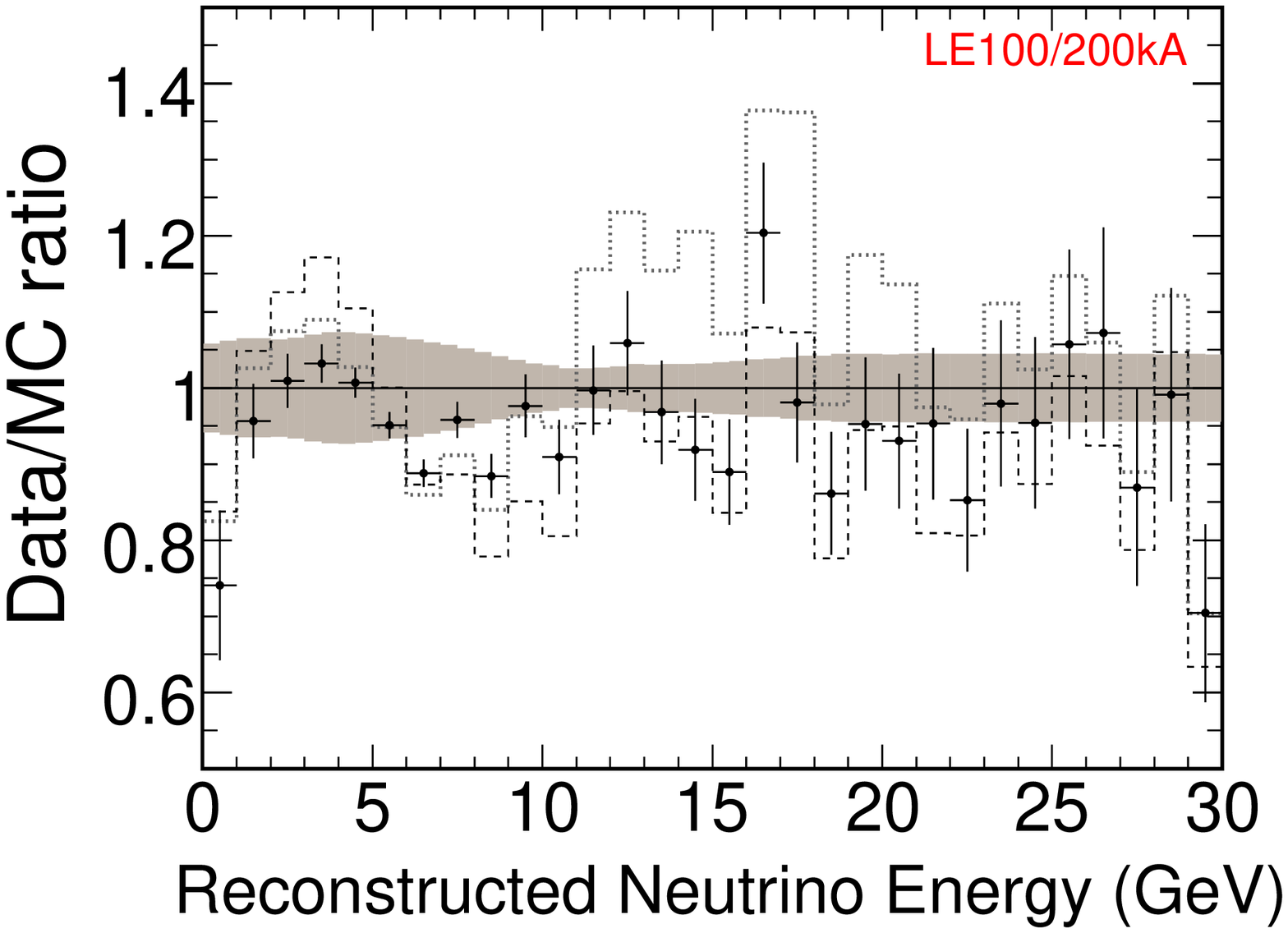}}
\subfloat{\includegraphics[width=0.33\textwidth,clip]{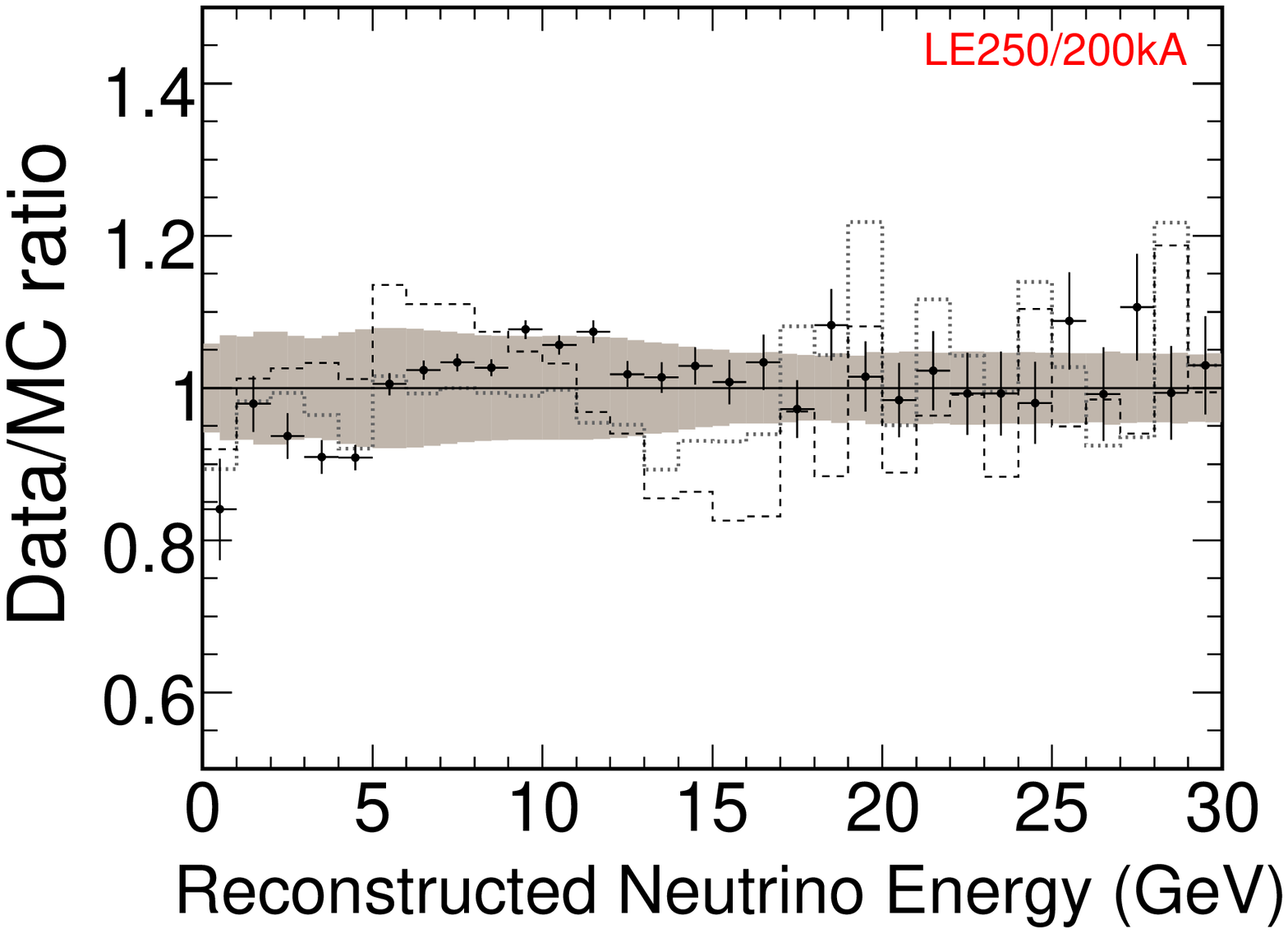}}

\caption{\label{fig:ndfit} Ratio of Near Detector data to Monte Carlo simulation as a function of reconstructed neutrino energy for three beam spectra before and after the NDfit. The dotted gray histograms show the ratio relative to the nominal simulation, the dashed black lines show the ratio after applying the improved hadron production calculation (Sec.~\ref{sec:skzp}), and the black points with error bars show the ratio after the best-fit NDfit parameters are also applied to the simulation. The errors on the data points represent statistical uncertainties and the shaded regions represent the residual systematic uncertainty from the beam-re-weighting procedure.}
\end{figure*}

\pstart
The result of the NDfit is shown in Fig.~\ref{fig:ndfit}. Here the ratio of observed over simulated reconstructed neutrino energy spectra are shown before and after the fit for three of the six beam configurations. The large excursions in the ratio formed with the nominal simulation are substantially reduced by the improved hadron production calculation of Sec.~\ref{sec:skzp}. Further improvements are observed after minimizing Eq.~\ref{eq:ndfit}, especially in the 0-\unit[5]{GeV} region of the LE10/185kA configuration, and around the focusing peak of the LE100/200kA and LE250/200kA spectra. Initial excursions in the data/MC ratio of $\sim30\%$ have been reduced to $\sim 10\%$ or less in all six spectra. Remaining deviations are generally within the uncertainties of the hadron production calculation.
\pend

\pstart
Table~\ref{tab:ndfitres} shows the best-fit values of the five systematic parameters used in the NDfit. All five parameters lie within one standard deviation of their nominal values. It is important to note that the results do not constitute a direct measurement of these quantities, as strong correlations exist between the fit variables. The value of $\chi^2$ at the best-fit point is 186.0 for 228 degrees of freedom, of which the penalty terms contribute four units of $\chi^{2}$ for five parameters.
\pend
 
\begin{table}[h]
\begin{center}
\begin{ruledtabular}
\begin{tabular}{clcc}
\multicolumn{2}{c}{Systematic parameter}  & Uncertainty &  Best-fit \\
\hline
$\quad \alpha_1$ & CC+RES axial-vector mass & $5\%$ & $+5\%$ \\

$\quad \alpha_2$ & non-resonant 1 and 2$\pi$ production & $20\%$ & $+12\%$\\

$\quad \alpha_3$ & Hadronic energy scale & $10\%$ & $-4\%$ \\

$\quad \alpha_4$ & Muon energy scale & $2\%$ & $-3\%$ \\

$\quad \alpha_5$ & Normalization  & $5\%$ & $-2.5\%$ 

\end{tabular}
\end{ruledtabular}

\caption{Systematic parameter values returned from the NDFit (Eq.~\ref{eq:ndfit}).\label{tab:ndfitres}}
\end{center}
\end{table}







\pstart
The 2DFit method treats the Near Detector data as a function of both the reconstructed neutrino energy $\enu{}$ and inelasticity $y$ in order to allow degeneracies between muon and hadron energy scale parameters to be broken. In the fit, the joint distribution of reconstructed neutrino energy and $y$ is divided into a two dimensional grid with 40 energy bins spaced between \unit[0]{GeV} and \unit[30]{GeV} and eight reconstructed $y$ bins between zero and one. Data from each of the six beam configurations is tabulated separately, yieldings a total of 1920 analysis bins which are employed simultaneously in the fit. The fit attempts to minimize
\pend
\pstart
\begin{equation} \label{eq:2dfit}
\chi^{2}=\sum_{k}^{E_{\nu}\mathrm{-bins}}\sum_{l}^{y\mathrm{-bins}} \frac{\left(N_{k,l}-n_{k,l}\left(\boldsymbol{\beta}\right)\right)^{2}}{\sigma_{k,l}^{2}} +\sum_{j=1}^{\mathrm{syst}}\left(\frac{\Delta\beta_{j}}{\sigma^{\beta}_{j} }\right)^{2} \qquad \qquad \mbox{``The 2DFit Method''} \quad ,
\end{equation}
\pstart
\noindent where the variables $N_{k,l}$ and $n_{k,l}$ refer to the contents of the \enu{},$y$ bins for the selected \numucc{} data and Monte Carlo samples respectively. Like the $\alpha_{j}$ in Eq.~\ref{eq:ndfit}, the $\beta_{j}$ are parameters which account for the systematic uncertainties in the Monte Carlo simulation and are varied in the fitting procedure to improve agreement with the data.
\pend

\pstart
The $n_{k,l}$ are implicit functions of the parameters $\boldsymbol{\beta}$ which account for the following systematic uncertainties: $\beta_1$, the normalization of the neutral-current background; $\beta_2$, an energy dependent variation in the event rate dictated by the uncertainty in the tuned hadron production calculation; $\beta_3$, the absolute hadronic energy scale; $\beta_4$ the absolute muon energy scale; In addition, the true neutrino energy versus $y$ distribution for Monte Carlo events is divided into a two dimensional grid composed of eight neutrino energy bins (0-2, 2-4, 4-6, 6-8, 8-10, 10-15, 15-20, $>\unit[20]{GeV}$) and five $y$ bins (0.0-0.2, 0.2-0.4, 0.4-0.6, 0.6-1.0). This yields an additional 40 fit parameters, $\beta_5\ldots\beta_{44}$, which allow the normalization of the Monte Carlo to vary independently in each bin with an uncertainty of 20\%. This uncertainty is comparable to the {\em a priori} uncertainties in the neutrino flux and charged-current cross-sections. Table~\ref{tab:2dfitres} shows the magnitude of the systematic parameters. In total, 45 fit parameters are varied to minimize Eq.~\ref{eq:2dfit}.
\pend




\begin{table}[h]
\begin{center}
\begin{ruledtabular}
\begin{tabular}{clcc}
\multicolumn{2}{c}{Systematic parameter}  & Uncertainty &  Best-fit \\
\hline
$\quad \beta_1$ & Neutral-current background & $50\%$ & $-37\%$ \\

$\quad \beta_2$ & Flux uncertainty  & energy dependent  & $0.88\sigma$ \\

$\quad \beta_3$ & Hadronic energy scale & $10\%$ & $+2.5\%$\\

$\quad \beta_4$ & Muon energy scale & $2\%$ & $-4.8\%$ \\


$\quad \beta_5\ldots\beta_{44}$ & $(E_{\nu},y)$ normalization factors  & $20\%$ & 0.64--1.61
\end{tabular}
\end{ruledtabular}

\caption{Systematic parameter values returned from the 2DFit (Eq.~\ref{eq:2dfit}). The uncertainty on the flux parametrized by $\beta_2$ is energy dependent, approximately $\pm 9\%$ ($\pm 6\%$) for $E_{\nu} \lesssim \unit[6]{GeV}$ ($E_{\nu} \gtrsim \unit[6]{GeV}$). The fit parameter is expressed in units $\sigma$ of the uncertainty. The parameters $\beta_5\ldots\beta_{44}$ take values between 0.64 and 1.61 with a mean of 1.02 and an r.m.s. of 0.21, approximately equal to the {\it a priori} uncertainty. \label{tab:2dfitres}}
\end{center}
\end{table}

\begin{figure}[h]
\begin{center}
  \includegraphics[width=\columnwidth]{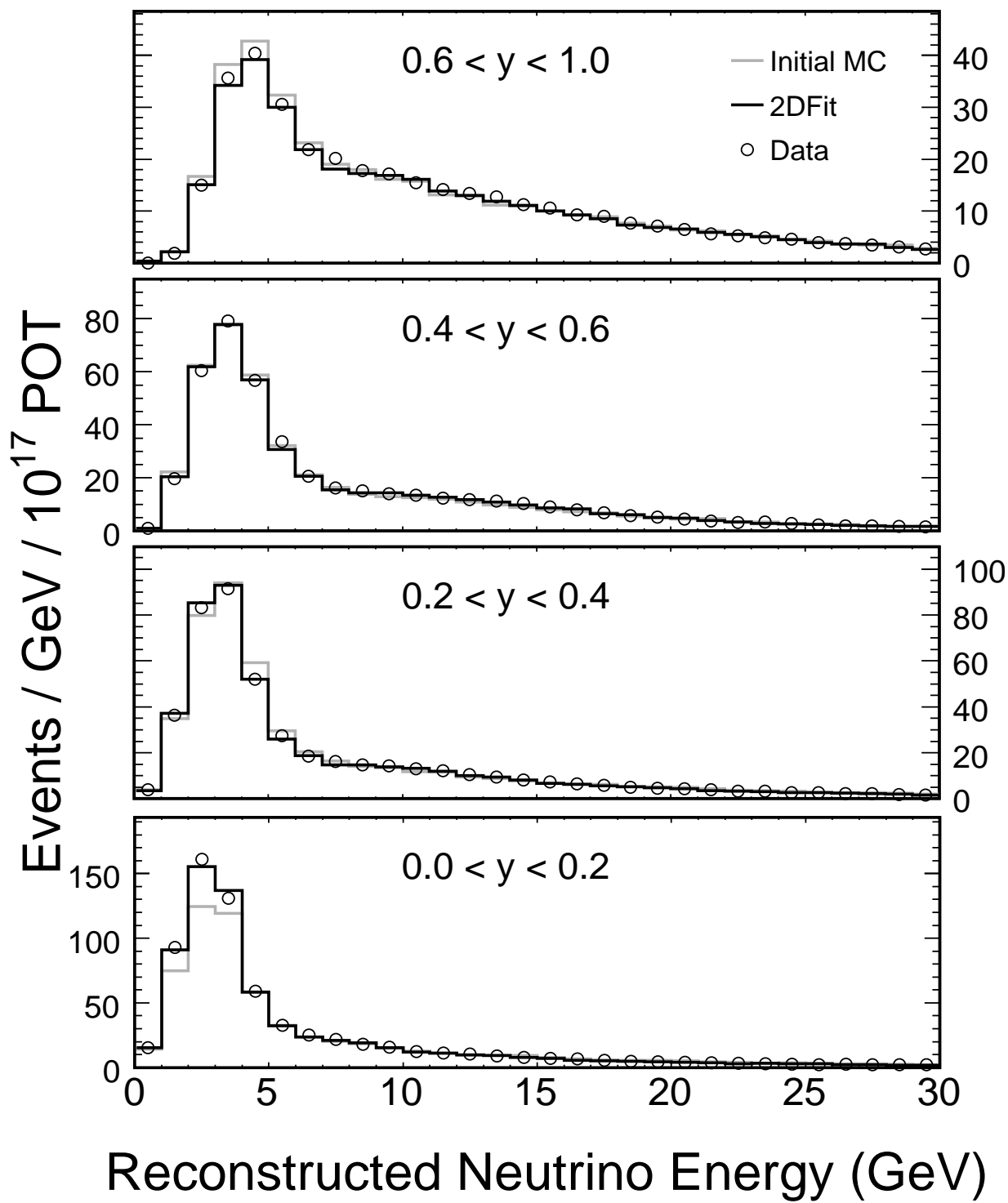}
  \caption{Near Detector reconstructed neutrino energy spectra collected in the LE10/185kA beam configuration and for different reconstructed $y$ regions. Open circles show the data. The Monte Carlo simulation is shown before and after the tuning procedure performed by the 2DFit technique (Eq.~\ref{eq:2dfit}). To simplify the presentation we have grouped the eight reconstructed $y$ bins used in the fit into four bins here.}
   \label{fig:2dfit_nd_le10}
\end{center}
\end{figure}

\pstart
Figure~\ref{fig:2dfit_nd_le10} shows the neutrino energy distribution from the LE10~\unit[185]{kA} configuration split into four ranges of $y$. The value of $\chi^{2}$ at the best-fit point is 2606.3 for 1919 degrees of freedom, compared to a $\chi^{2}$ of 5432.1 before the fit. The best-fit values of the 40 $(E_{\nu},y)$ normalization factors range from 0.64 to 1.61, with a mean of 1.02 and an r.m.s. of 0.21 (approximately equal to the {\it a priori} uncertainty). The best-fit values for the other parameters are -37\% for neutral-current normalization and -4.8\% and +2.5\% for muon and shower energy scales, respectively. The overall normalization was reduced according to a $0.88\sigma$ shift in the tuned hadron production calculation.
\pend

\subsection{Predicted Far Detector spectrum}

\begin{figure}[h]
\begin{center}
\subfloat[\label{fig:4spectra}]{\includegraphics[width=0.7\columnwidth,clip]{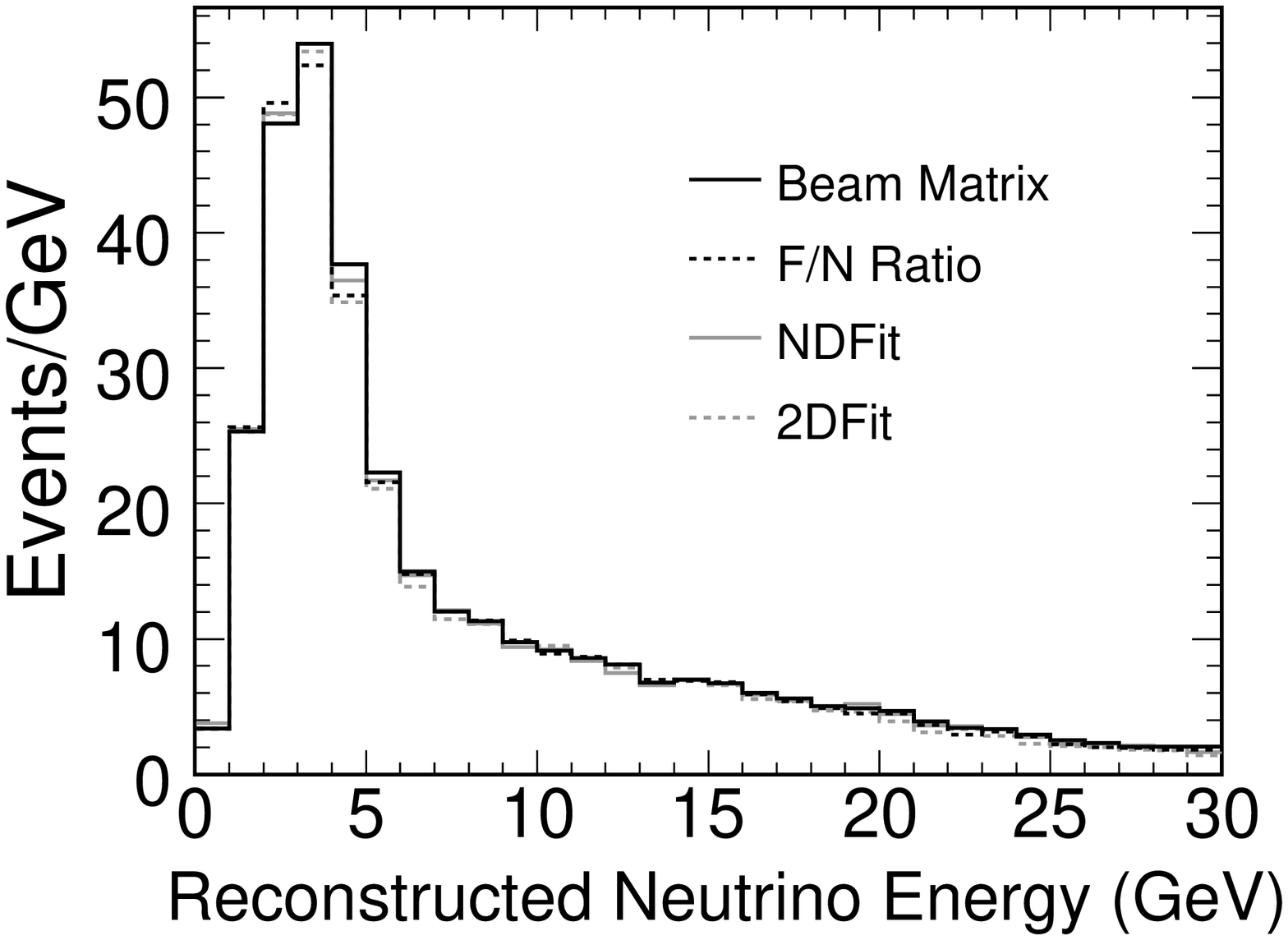}}\\
\subfloat[\label{fig:4ratios}]{\includegraphics[width=0.7\columnwidth,clip]{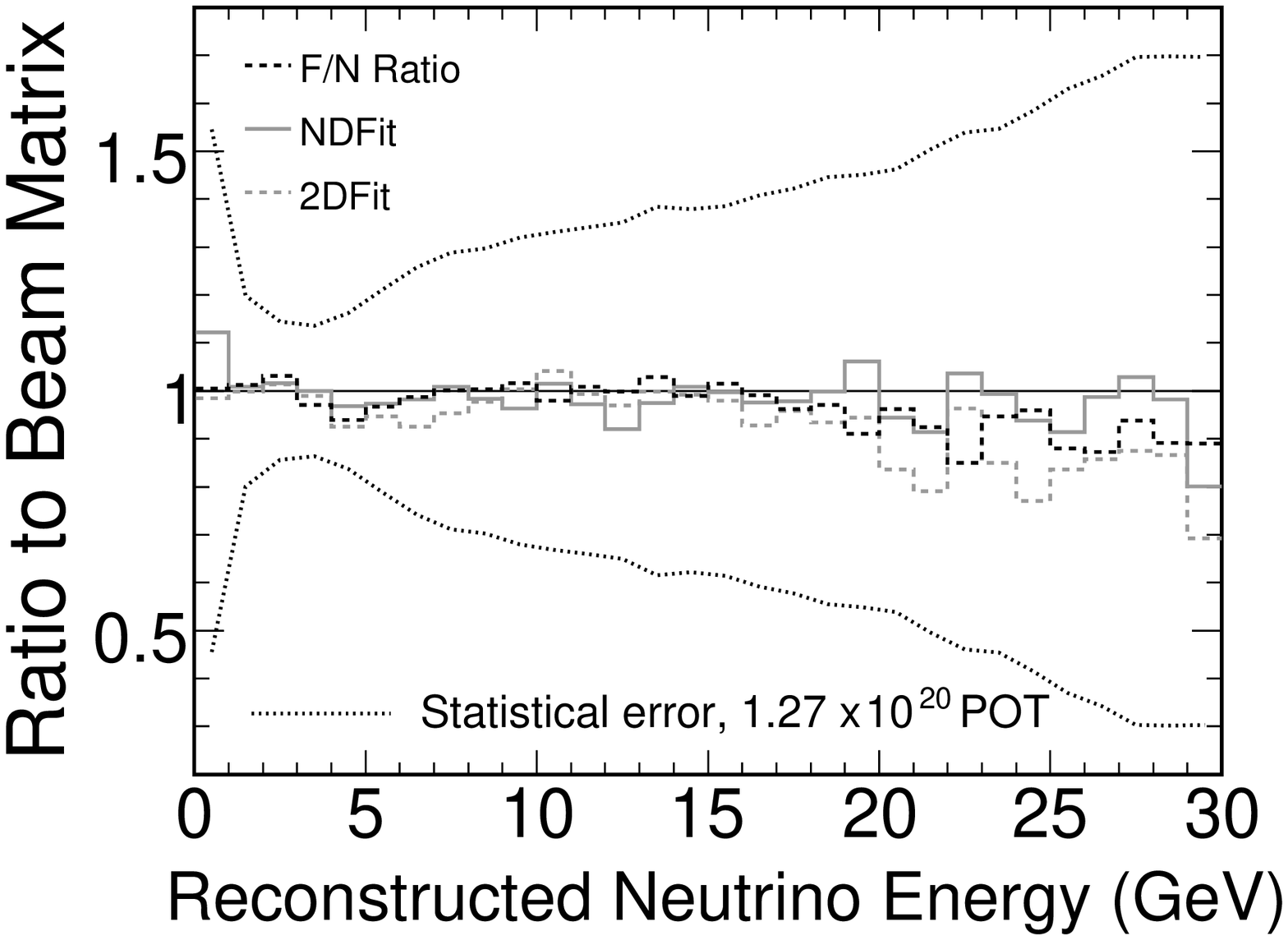}}
\caption{\label{fig:spectra_4methods} Neutrino energy spectra at the Far Detector in the absence of neutrino oscillations as predicted by the four extrapolation methods.}
\end{center}
\end{figure}

\pstart
Figure~\ref{fig:4spectra} shows a comparison of the predicted Far Detector unoscillated visible energy spectra obtained from the four extrapolation methods, for an exposure of $1.27\times10^{20}$ protons on target. The ratios of the three other spectra relative to the Beam Matrix prediction are shown in Fig.~\ref{fig:4ratios}. The agreement between the predicted spectra is better than $\pm 5\%$ in the region 1-\unit[15]{GeV}, where the oscillation signal is expected to lie. The spread on the predicted spectra is significantly smaller than the statistical error when the data are binned in \unit[1]{GeV} wide energy bins. Table~\ref{tab:fdprednum} shows the number of Far Detector events predicted between 0-30 GeV for each extrapolation method. The systematic errors on the predicted rate, which are described below, are dominated by the $4\%$ relative normalization error. 
\pend

\begin{table*}[]
\begin{center}
\begin{ruledtabular}
\begin{tabular}{cc}
Extrapolation Method & Predicted Number of Far Detector Events \\
\hline

Beam Matrix & $336.0\pm 14.4$ \\

Far/Near Ratio & $330.0\pm 14.2$ \\

NDFit & $332.7\pm 14.3$ \\

2DFit & $323.9\pm 13.9$ \\

\end{tabular}
\end{ruledtabular}

\caption{Predicted numbers of Far Detector events for the four extrapolation methods for an exposure $\unit[1.27\times10^{20}]{POT}$. The systematic errors on the predictions, which are dominated by the 4\% relative normalization uncertainty, are also shown.\label{tab:fdprednum}}
\end{center}
\end{table*}



\subsection{Sensitivity to systematic errors for the different extrapolation methods}

\label{sec:extrap_syst}
\pstart
In this section we describe the various sources of systematic uncertainty and their impact on the neutrino spectrum predicted at the Far Detector.  Each source was considered separately and Monte Carlo datasets were generated with $\pm1\sigma$ shifts of the systematic parameter applied along with oscillations according to $\dm=\unit[2.72\times10^{-3}]{\evmass{}}$ and $\st=1$, chosen to agree with our best fit to the data.  These ``mock datasets'' were produced for both Near and Far Detectors and were analyzed in the same way as the actual data. In particular the tuning procedure of Sec.~\ref{sec:skzp} was run on each dataset and the extrapolation methods were able to use the Near Detector mock data to predict the spectra at the Far Detector. These predictions were then used to fit the Far Detector mock data for neutrino oscillations, allowing us to examine the way in which systematic uncertainties affect our measurements of \dm{} and \st{}. The largest sources of uncertainty were identified with this procedure and then accounted for when fitting the Far Detector data for neutrino oscillations.
\pend

\subsubsection{Systematic error sources}
\begin{figure}[h]
\centering
\subfloat[\label{fig:syst_ccma}]{\includegraphics[width=0.45\textwidth]{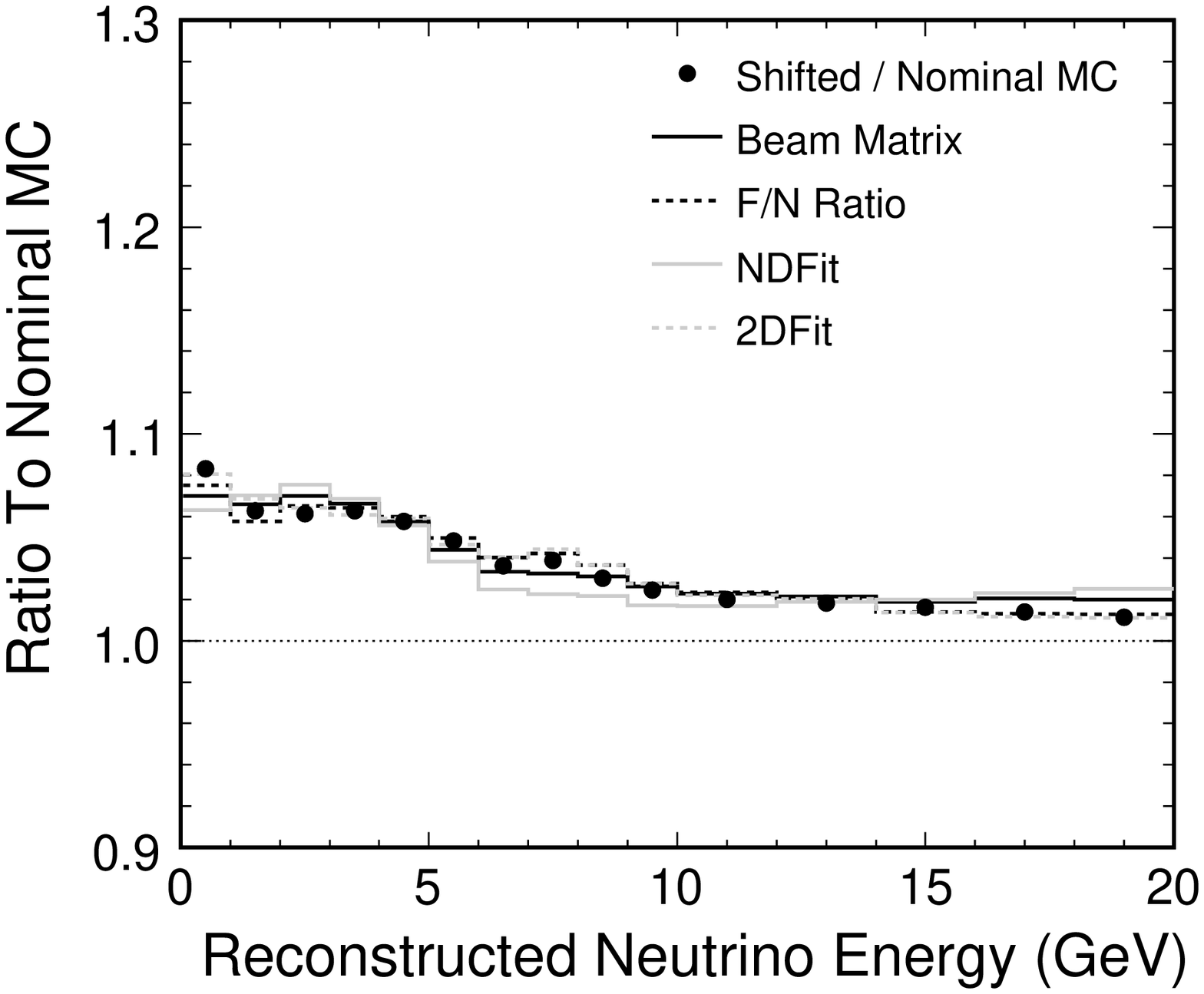}}
\hfill
\subfloat[\label{fig:syst_eshw}]{ \includegraphics[width=0.45\textwidth]{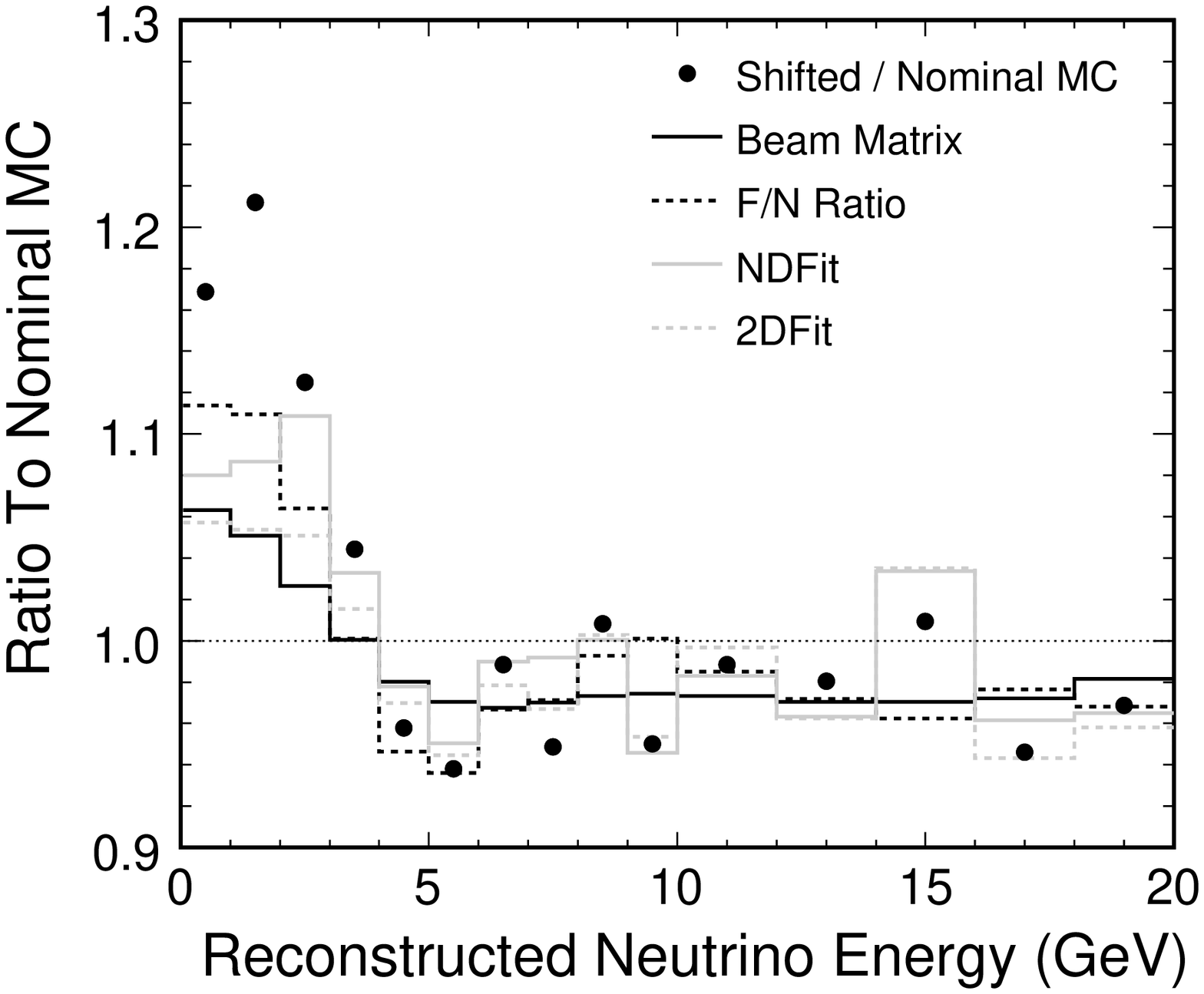}}
\\
\subfloat[\label{fig:syst_skzp}]{\includegraphics[width=0.45\textwidth]{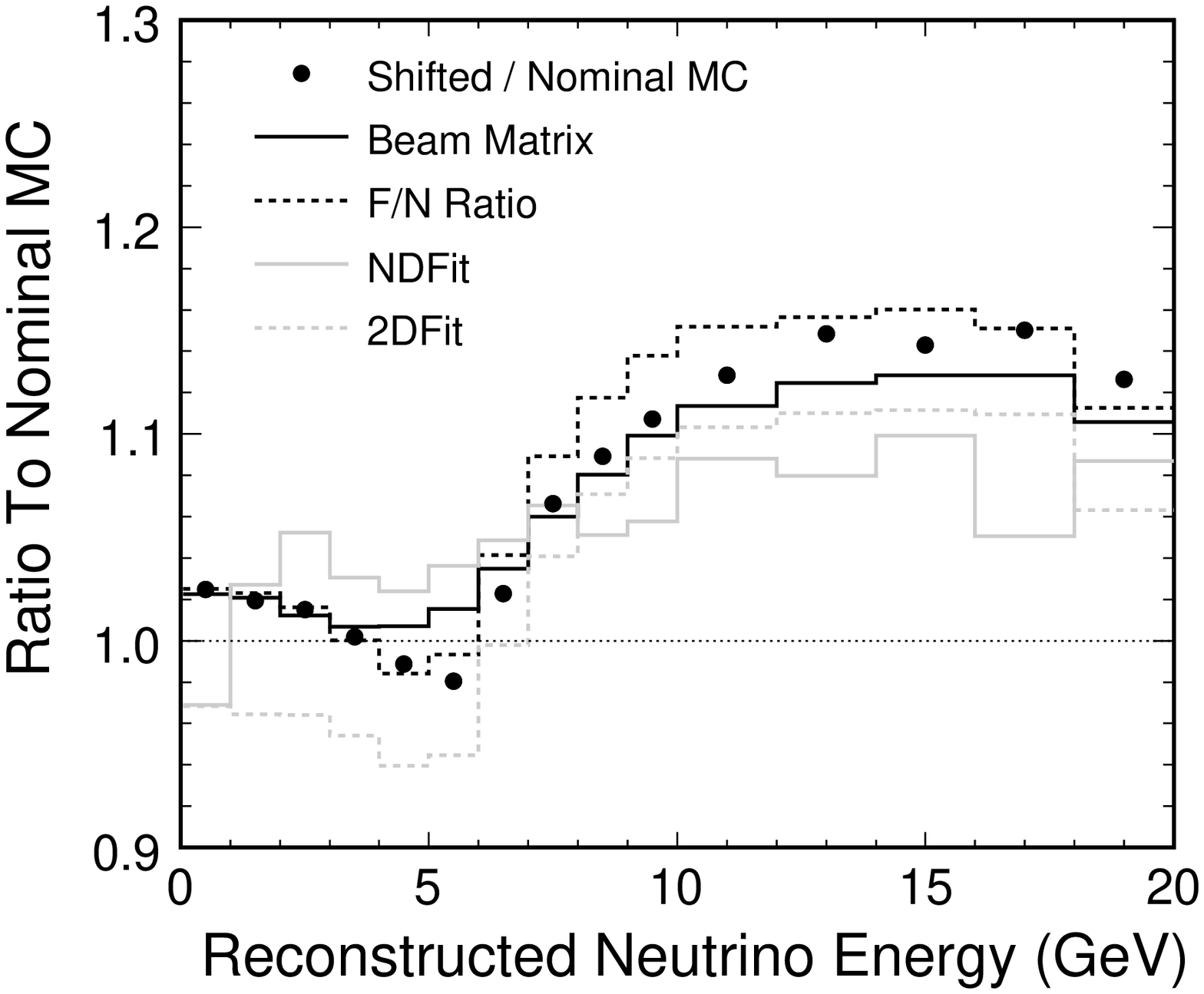}} 
\hfill
\subfloat[\label{fig:syst_nc}]{\includegraphics[width=0.45\textwidth]{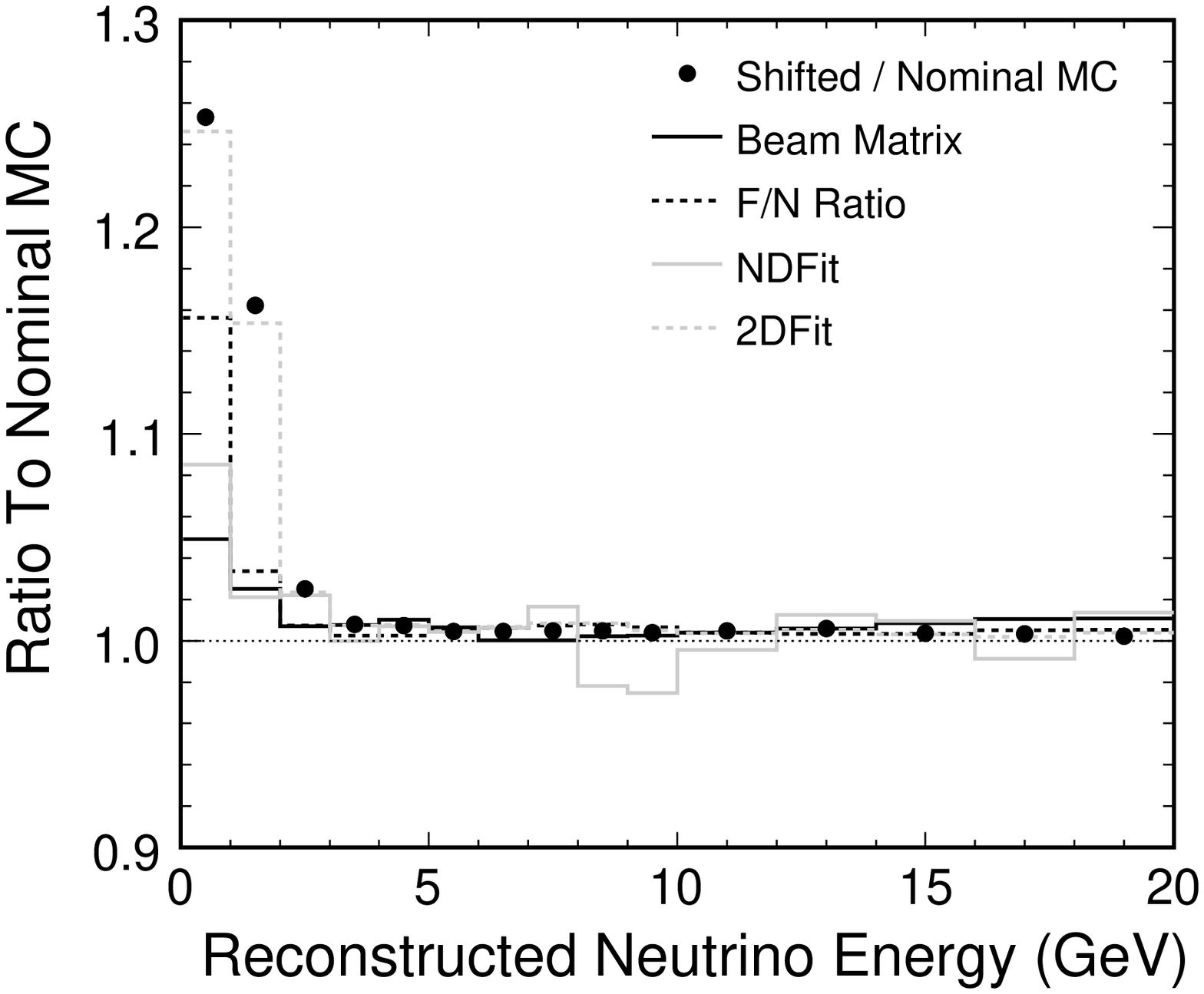}}

 \caption{Effect of systematic uncertainties on the predicted Far Detector energy spectrum. The following systematic shifts are shown: \protect\subref{fig:syst_ccma} $+10\%$ increase in \numu{} charged-current quasi-elastic and resonance production cross-sections;  \protect\subref{fig:syst_eshw} $-10\%$ shift in reconstructed shower energy scale; \protect\subref{fig:syst_skzp} $+50\%$ uncertainty on the beam tuning correction; \protect\subref{fig:syst_nc} $+50\%$ increase in neutral-current background rate. Further details are described in the text. \label{fig:systnceshw}}
\end{figure}

\pstart
Figure~\ref{fig:systnceshw} illustrates the effect that four representative sources of uncertainty have on the Far Detector neutrino energy spectrum, and the way in which the extrapolation methods are able to account for them with Near Detector measurements. Black points show the ratio of reconstructed energy spectra at the Far Detector, where the denominator corresponds to the nominal simulation and the numerator was generated with a particular systematic uncertainty source shifted with respect to the default value. Deviations from unity indicate the way in which the systematic uncertainties modify the neutrino energy spectrum. The size of the deviations does not necessarily indicate the potential effect on \dm{} and \st{} because the extrapolation methods use similarly shifted Near Detector Monte Carlo samples as ``mock data''. Their prediction of the ratio, shown as lines, would follow the black points if the systematic shift could be completely corrected with Near Detector data. In that case one would extract the same oscillation parameters as were put into the study and the particular source of uncertainty would not affect the results. We describe the sources of uncertainty below. 
\pend

\pstart
{\em Charged-current cross-sections:} We assign a $\pm10\%$ uncertainty in the axial vector masses for quasi-elastic and resonance production processes. Fig.~\ref{fig:syst_ccma} shows the effect on the FD spectrum when a $+10\%$ shift in those quantities is applied in the simulation. The predicted spectra from the four extrapolation methods successfully reproduce the shift.  This is expected, since cross-section changes are common to both Near and Far detector events, the Near to Far extrapolation methods should provide a significant cancellation of such uncertainties. We have also conducted a similar study to verify that our analyses are able to correct for the $\pm20\%$ uncertainty in the magnitude of the non-resonant $1\pi$ and $2\pi$ production cross-sections for invariant masses $W<\unit[1.7]{\gevcsq{}}$. See Sec.~\ref{sec:mc} for further details.
\pend

\pstart
{\em Final-state interactions:} As described in Sec.~\ref{sec:mc} we estimate that uncertainties in final-state interactions cause a  $\pm10\%$ uncertainty on the shower energy scale. Figure~\ref{fig:syst_eshw} shows the effect that a $-10\%$ shift in the energy scale has on the Far Detector neutrino spectrum. The distortion of the spectrum below \unit[10]{GeV} is significant. Because the same shift is present in both Near and Far Detectors, the predicted spectra from the four methods provide an improved description of the distorted spectrum and thus reduce the effect of this systematic error. However, the neutrino energy and \dm{} are coupled in Eq.~\ref{eq:oscprob} and a change in the energy scale may be incorrectly interpreted as a shift in \dm{}. It is therefore necessary to explicitly account for energy scale uncertainties when extracting the oscillation parameters from our data.
\pend


\pstart
{\em Beam-related uncertainties:} The beam tuning procedure described in Sec.~\ref{sec:skzp} applies corrections to the (\pzpt{}) distribution in the Monte Carlo that result in changes in the visible energy spectrum of approximately $10\%$ around the peak in the LE10/185kA beam configuration and about $30\%$ in the high energy tail. We have conservatively assigned a $50\%$ error on the magnitude of this correction as an estimate of the uncertainty on the beam tuning. The effect of this uncertainty on the FD spectrum is shown in Fig.~\ref{fig:syst_skzp}. There is a significant shift in the high-energy tail of the spectrum ($\enu{} > \unit[5]{GeV}$) where the beam tuning correction is largest. The plot shows that the direct extrapolation methods are in general more robust than the fitting methods to this type of uncertainty, as they are much less dependent on the requirement that the tuned Monte Carlo closely match the data in order to provide an accurate prediction of the FD spectrum.
\pend

\pstart
{\em Normalization:} We estimate an uncertainty of $\pm4\%$ in the relative normalization of the energy spectra measured in the Near and Far Detectors. The estimate is composed from a $2\%$ uncertainty in the fiducial mass of both detectors, a $3\%$ uncertainty in the relative Near and Far Detector reconstruction efficiencies, estimated from a visual scan of Near and Far Detector data and Monte Carlo events, and a $1\%$ uncertainty in the live time.
\pend


\pstart
{\em Neutral-current background:} As described in Sec.~\ref{sec:nc}, we estimate an uncertainty of $\pm 50\%$ on the rate of neutral-current events misidentified as \numu{} charged-current. Figure~\ref{fig:syst_nc} shows that the effect of this uncertainty is largest in the lowest reconstructed energy bins ($E_\nu <\unit[2]{GeV}$) and is only partially corrected by the F/N, Beam Matrix and NDFit extrapolation methods. The 2DFit method includes the neutral-current rate as a fit parameter and recovers the correct prediction in this special case. In the general case however, where several systematic effects will be present, no method will be able to completely correct for this source of uncertainty.
\pend



\pstart
{\em Shower energy scale calibration:} Based on a comparison of test beam measurements in the Calibration Detector and the Monte Carlo simulation (see Sec.~\ref{sec:shwr_calib}) we estimate a $6 \%$ uncertainty in the response to single pions and protons. The uncertainty in the response to electrons and muons is negligible. As described in Sec.~\ref{sec:calib}, stopping muons were used to cross-calibrate the two detectors with a relative uncertainty of $\pm 2\%$.
\pend

\pstart
{\em Muon energy scale:} A $2\%$ muon energy scale uncertainty is assumed based on studies of fully contained muon tracks reconstructed in the Near Detector. The difference between the momentum obtained from the track range and the momentum obtained from a fit to the curvature of the track due to the magnetic field of the detector was examined for both real and simulated events, as a function of the range momentum. The deviation of this quantity between data and simulation was approximately 2\%, which was taken as an estimate of the uncertainty on the magnetic field calibration in the Near Detector, and hence the error on the relative muon energy scale between Near and Far Detectors. Our estimate utilizes the muon range/energy relation which has an uncertainty of approximately 2\% based on a material assay of the detectors and a comparison of our muon stopping power tables with those in~\cite{Groom:2001kq}. 
\pend

\pstart
As discussed in Sec.~\ref{sec:mc}, our magnetic field was recalibrated after we performed this analysis. This predominantly affects the reconstruction of muons with a momentum larger than $\sim \unit[7]{\gevc{}}$ which arise from neutrinos outside the energy region in which we observe oscillations.
\pend


\pstart
{\em CC selection efficiency:} We varied the requirement on the event selection parameter $S$ (see Eq.~\ref{eq:S}) for mock data sets by $\pm 0.02$ while holding the cut applied in the nominal Monte Carlo dataset constant. This changes both the number of true charged-current and neutral current events that are classified as \numu{} charged-current. The magnitude of this shift was obtained from a comparison of the data and Monte Carlo $S$ distributions in the Near Detector. 
\pend
\pstart
We used mock datasets including the systematic uncertainties listed above to examine the accuracy with which the different methods are able to predict the Far Detector spectrum and study the \st{} and \dm{} extracted from our oscillation fits. As described below, this procedure was used to explore the capabilities of the different extrapolation techniques and identify the most important sources of systematic uncertainty.


 
\pend
\subsubsection{Choice of primary analysis method}

\pstart
 The indirect extrapolation methods (NDFit, 2DFit) attempt to adjust the Monte Carlo prediction of the neutrino energy spectrum in the Near Detector to improve agreement with the data and then use the improved Monte Carlo calculation to predict the Far Detector spectrum. These procedures are not able to arrive at a perfect description of the data. In particular, discrepancies of $\pm5\%$ are present when data collected in the LE10/185kA beam configuration are compared with the Monte Carlo prediction after the best-fit NDFit and 2DFit systematic parameters have been applied. We also notice that somewhat larger discrepancies are still present in the LE100/200kA beam configuration. If the major source of these distortions is caused by uncertainties in the neutrino flux or neutrino \xsecs{} then the direct (Beam Matrix and the F/N) extrapolation methods simply translate the measurements to the Far Detector and the discrepancies do not affect the oscillation measurement. As a consequence we have found that the predictions made by the direct methods are less sensitive to the absolute agreement between data and the Monte Carlo simulation in the Near Detector than are the those made by the indirect methods.
\pend




\pstart 
In light of these issues, we have decided to use one of the direct extrapolation methods to obtain our primary oscillation result. We chose the Beam Matrix method as our primary extrapolation technique because it had smaller systematic errors than the F/N ratio method when all sources of uncertainty were considered.  The results that we will present in Sec.~\ref{sec:osc} therefore use the Beam Matrix method to predict the Far Detector spectrum and perform the oscillation fit. As a cross-check of these results, we present the best fit oscillation parameters and allowed regions obtained from the other three extrapolation methods as well.
\pend



\subsubsection{Systematic Uncertainties in the Oscillation Fit}
\pstart
The three largest contributions to the systematic error for the Beam Matrix method are a) the uncertainty in the relative normalization of the energy spectra measured in the two detectors, b) uncertainties in the absolute hadronic energy scale, and c) uncertainties in the neutral current background rate. The systematic shifts calculated for the other sources of uncertainty are small in comparison. The magnitude of these shifts is summarized in Tab.~\ref{tab:oscsyst}. As expected from the above discussion, because uncertainties due to beam modeling and cross-sections are common to the two detectors they largely cancel out in the extrapolation. The three largest uncertainties are included as systematic nuisance parameters in the Far Detector oscillation fit. By fitting for these systematic parameters simultaneously using both Near and Far Detector data, the effect of these uncertainties is substantially reduced, due to significant cancellations of these errors between the two detectors.  
\pend



\section{Oscillation Analysis}
\label{sec:osc}
\pstart
With an exposure of $\unit[1.27\times 10^{20}]{POT}$, a total of 215 beam-coincident events with reconstructed energies below \unit[30]{GeV} are selected as \numucc{} in the Far Detector. Assuming no oscillations, the predicted number of Far Detector events in the same energy range for this exposure is $336\pm14$. The error quoted here is dominated by the 4\% systematic error on the overall normalization. The deficit corresponds to a significance of 5.2 standard deviations, where both statistical and systematic errors on the total rate are taken into account. In this section, we describe an oscillation analysis of the observed Far Detector reconstructed energy spectrum. We present results obtained using the extrapolated Far Detector spectrum from the Beam Matrix method, and compare these with results from the other three extrapolation methods.
\pend

\begin{figure}[h]
\begin{center}
\subfloat[\label{fig:bm_fit}] {\includegraphics[width=0.6\columnwidth,clip]{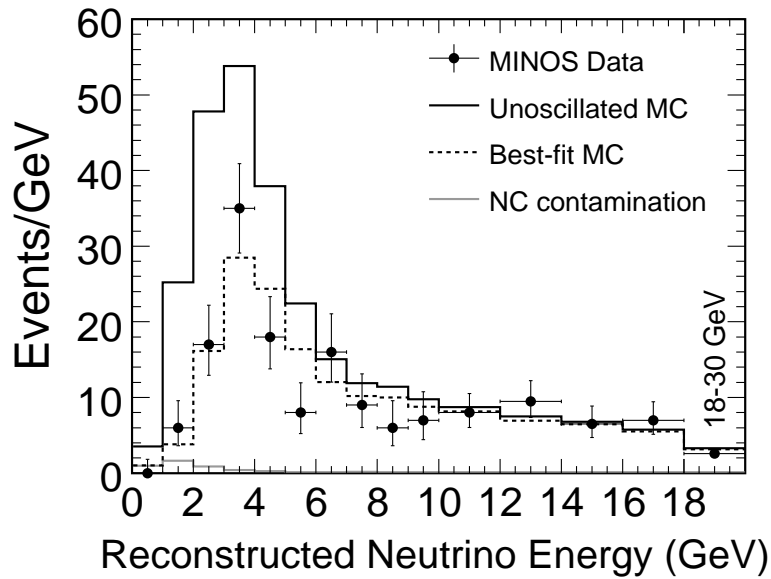}}\\
\subfloat[\label{fig:bm_ratio}] {\includegraphics[width=0.6\columnwidth,clip]{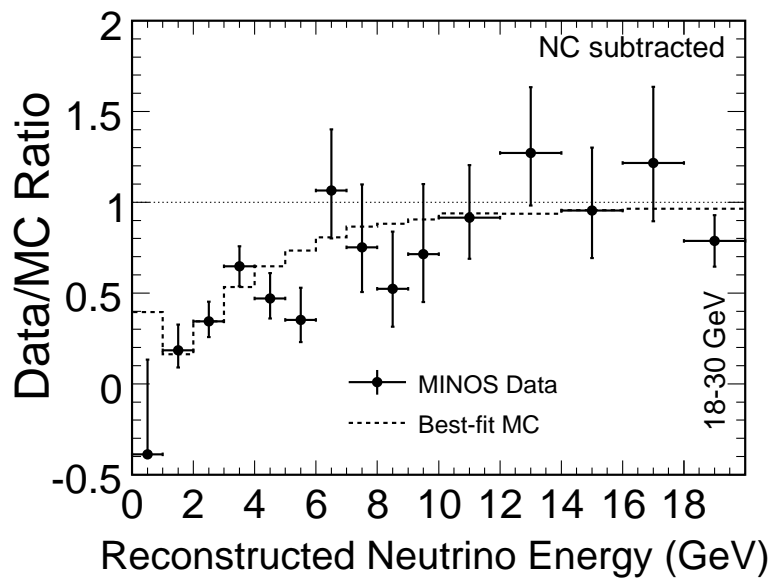}}
\caption{\label{fig:spectra_data} \protect\subref{fig:bm_fit} The reconstructed energy spectra of selected Far Detector events with the Far Detector unoscillated prediction (solid histogram) and best-fit oscillated spectrum (dashed histogram) overlaid. The predicted neutral-current background in the selected sample is shown in gray. The right-most bin in this distribution contains all events between 18 and \unit[30]{GeV}. The asymmetric error bars on the data points represent the 68\% C.L. Poisson errors on the numbers of observed events. \protect\subref{fig:bm_ratio} The ratio of the observed spectrum to the unoscillated Far Detector prediction, where the expected neutral-current background has been subtracted.}
\end{center}
\end{figure}

\pstart
Figure~\ref{fig:bm_fit} shows the reconstructed energy distribution of the selected Far Detector events. A fit to these data is performed to extract the mixing parameters $\dm$ and $\st$, within the context of two-flavor $\nu_\mu\leftrightarrow\nu_\tau$ oscillations (Eq.~\ref{eq:oscprob}).  We minimize the following statistic, $\chi^{2}=-2\ln\lambda$, where $\lambda$ is the likelihood ratio~\cite{PDBook}:
\pend
\begin{widetext}
\begin{equation}
\chi^{2}(\dm,\st,\alpha_{j})=\sum_{k=1}^{15}{2(N_{k}^{exp}-N_{k}^{obs})+2N_{k}^{obs}\ln(N_{k}^{obs}/N_{k}^{exp})}+\sum_{j=1}^{3}\left({\Delta\alpha_{j}\over\sigma_{\alpha_{j}}}\right)^{2},
\end{equation}
\end{widetext}
\pstart
\noindent where $N_{k}^{obs}$ and $N_{k}^{exp}$ are the numbers of observed and expected events in bin $k$ of the reconstructed energy distribution, the $\alpha_{j}$ are fitted systematic parameters, with associated errors $\sigma_{\alpha_{j}}$. 
\pend




\pstart
The three leading systematic uncertainties identified in Sec.~\ref{sec:extrap_syst} are included in the fit as nuisance parameters, with Gaussian distributed errors. These parameters are the relative normalization between the Far and Near Detectors, with a 4\% uncertainty; the absolute hadronic energy scale with a 11\% uncertainty, and a 50\% uncertainty in the neutral-current background rate for the selected sample. Since the energy scale and neutral-current background uncertainties are common to the two detectors, their effect significantly cancels in the extrapolation of the energy spectrum from the Near to the Far Detector. In order to account for this in the fit, these parameters are varied simultaneously for both Near and Far Detectors, and the Monte Carlo reconstructed energy distributions are modified  accordingly. A new Far Detector predicted spectrum is then obtained for every value of these systematic parameters as they are varied in the oscillation fit.
\pend




\pstart
Figure~\ref{fig:bm_fit} shows the reconstructed neutrino energy spectrum measured at the Far Detector, the unoscillated Far Detector predicted spectrum obtained from the Beam Matrix method, along with the predicted spectrum weighted by the best-fit oscillation parameters. Figure~\ref{fig:bm_ratio} shows the ratio of data to the unoscillated Monte Carlo prediction as a function of reconstructed energy, and the predicted ratio for the best-fit oscillation parameters. The contamination from misidentified neutral-current events, which is shown by the gray histogram in Fig.~\ref{fig:bm_fit}, is subtracted from both the predicted and observed distributions in Fig.~\ref{fig:bm_ratio}. The shape of the data distribution is well modeled by the oscillation hypothesis.
\pend


\begin{figure}[h]
\begin{center}
\includegraphics[width=\columnwidth]{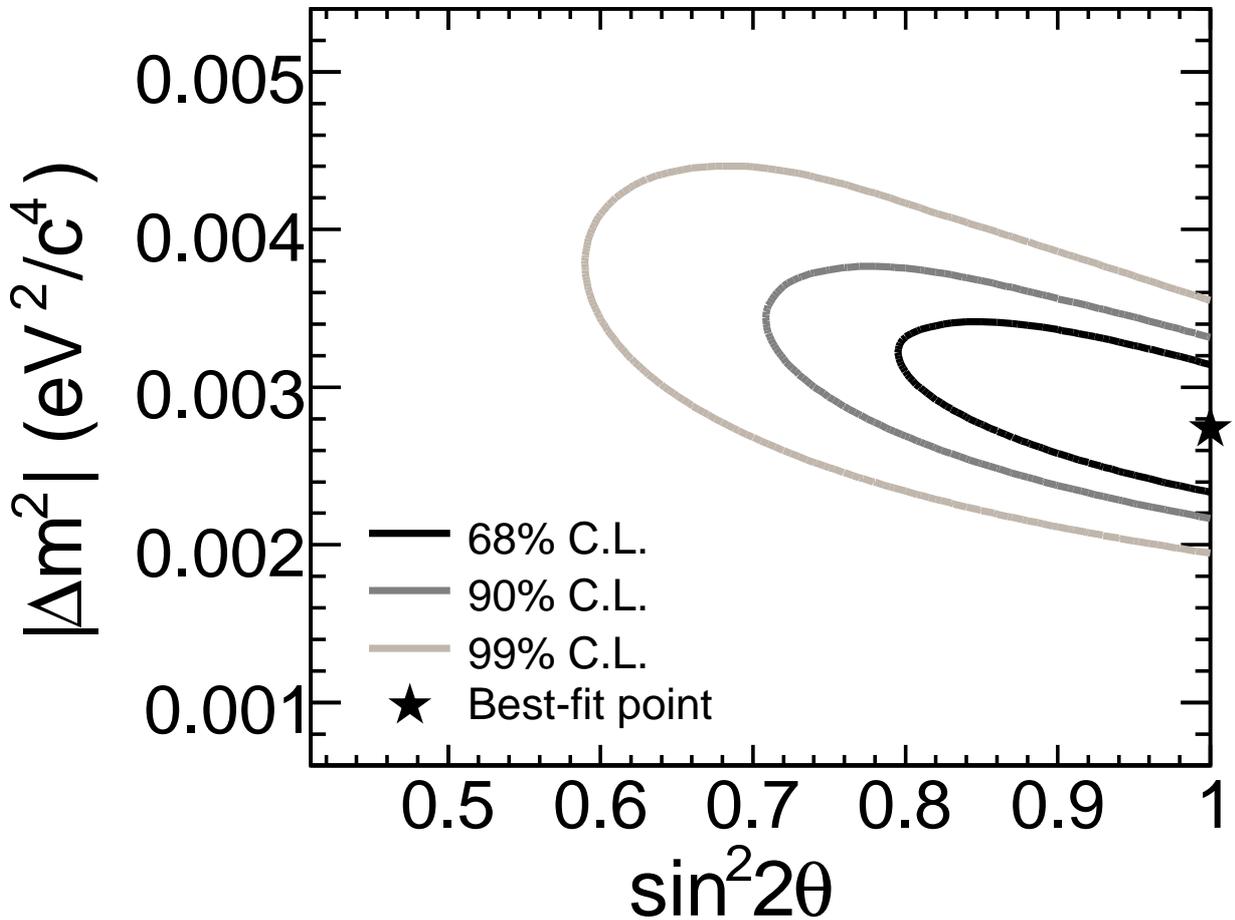}
\caption{\label{fig:region_cl} Allowed regions at 68,90,99\% C.L. in the $\dm,\st$ plane from a fit to the Far Detector reconstructed energy spectrum using the Beam Matrix extrapolation method. The best-fit point, which occurs at $\dm=\unit[2.74\times10^{-3}]{\evmass{}}$ and $\st=1$, is represented by the star.}
\end{center}
\end{figure}

\pstart
The allowed regions at 68, 90, 99\% C.L. in the $\dm,\st$ plane from a fit to the 215 Far Detector selected data events using the Far Detector predicted spectrum are shown in Fig.~\ref{fig:region_cl}. Here the confidence level intervals are obtained using the Gaussian approximation ($\Delta\chi^2=2.3,4.6,9.2)$. These confidence-level intervals were found to be in good agreement with those obtained from a study using the unified approach of Feldman and Cousins~\cite{Feldman:1997qc}. The best-fit parameters are $\dm=\unit[2.74\times10^{-3}]{\evmass{}}$ and $\st=1$, where the fit has been constrained to the region $\st\leq1$. The allowed ranges of  these parameters at 68\% C.L. and for 1 d.o.f. are $2.54\times10^{-3}<\dm<\unit[3.18\times 10^{-3}]{\evmass{}}$ and $\st>0.87$. The $\chi^2$ at the best-fit point is 20.3 for 13 degrees of freedom, which corresponds to a $\chi^2$ probability $P(\chi^2,\mathrm{d.o.f.})=8.9\%$. It has been verified using Monte Carlo experiments that the probability of 8.9\% is valid for the relatively low statistics of the data sample. The $\chi^2$ for the null oscillation hypothesis is 104 for 15 degrees of freedom. 
\pend

\pstart
The contribution of $\nu_\tau$ charged-current events from $\nu_\mu\leftrightarrow\nu_\tau$ oscillations is included in the fit. At the best-fit point, the expected number of $\nu_\tau$ charged-current events in our sample is 0.78 events. The expected background from oscillated $\nu_e$ charged-current events at our best-fit point is $<0.3$ events, for $\nu_\mu\leftrightarrow\nu_e$ mixing with $\sin^{2}2\theta_{13}<0.2$~\cite{Apollonio:2002gd}. This background, and the 0.38 events expected from rock muon interactions, are not included in the fit.
\pend


\begin{figure}[h]
\begin{center}
\includegraphics[width=0.45\columnwidth]{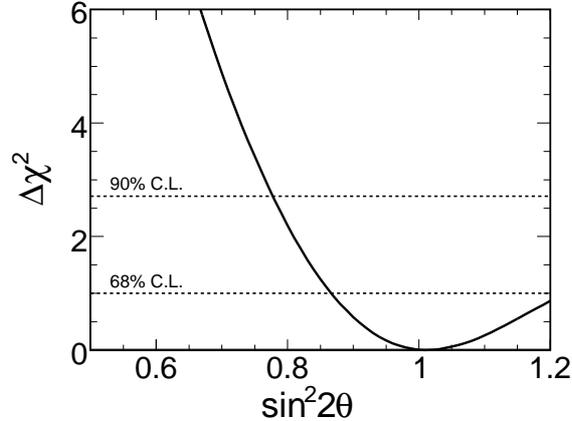}
\caption{\label{fig:1dproj_unphys} One-dimensional projection of the $\Delta\chi^2$ surface for $\st$ using the Beam Matrix extrapolation method, for a fit where the physical boundary at $\st=1$ has been removed. }
\end{center}
\end{figure}

\pstart
A fit to the Far Detector energy spectrum where the physical boundary constraint ($\st\leq1$) is removed yields best-fit parameters  that are very slightly in the unphysical region: $\dm=\unit[2.72\times10^{-3}]{\evmass{}}$ and $\st=1.01$, with a best-fit $\chi^2/d.o.f=20.3/13$. The one-dimensional projection of the $\chi^{2}$ surface for $\st$, where the value of $\chi^{2}$ has been minimized at each point with respect to $\dm$, is shown in Fig.~\ref{fig:1dproj_unphys}.
\pend


\begin{figure}[h]
\begin{center}
\includegraphics[width=\columnwidth]{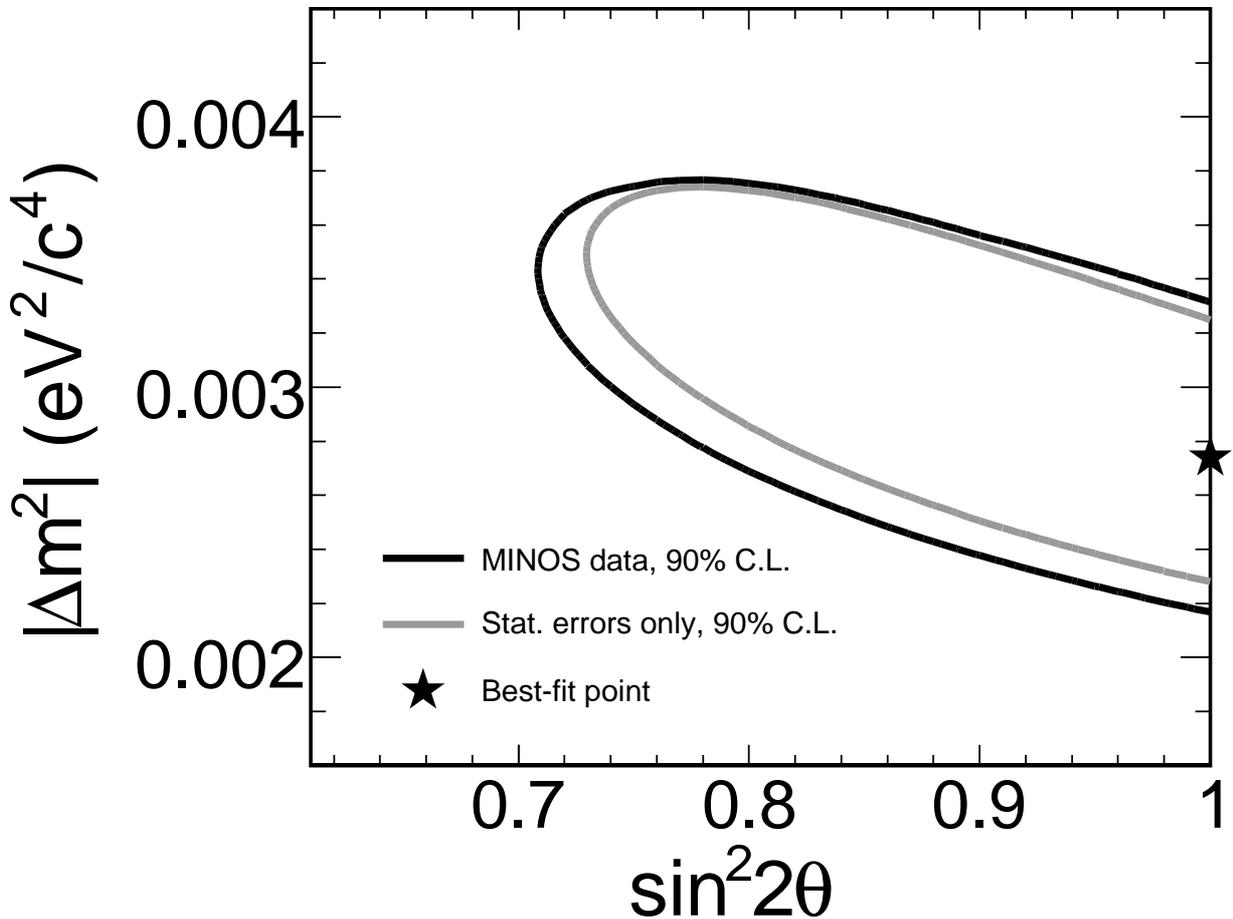}
\caption{\label{fig:region_statsyst} Comparison of the 90\% C.L. regions obtained by considering both statistical and systematic errors (black contour) and statistical errors only (gray contour). The best-fit point for the fit including statistical and systematic errors is shown by the star.}
\end{center}
\end{figure}

\pstart
Figure~\ref{fig:region_statsyst} shows the effect of systematic errors on the measurement of the oscillation parameters. The figure shows the 90\% C.L. allowed regions obtained from fits to the Far Detector data assuming statistical errors only, and statistical and systematic errors combined. The 90\% C.L. allowed region increases in size by approximately 10\% in both $\dm$ and $\st$ when systematic errors are also taken into account. This indicates that, for this exposure, the measurement errors on the oscillation parameters are limited by statistical uncertainties. Table~\ref{tab:oscsyst} shows the systematic shifts on the best-fit point for the various sources of systematic error considered in Section~\ref{sec:extrap_syst}.  The shifts due to beam and cross-section uncertainties are negligibly small. As an additional check we have repeated the analysis without the improved flux calculation of Sec.~\ref{sec:skzp} and find that \dm{} changes by $\unit[2\times10^{-5}]{\evmass{}}$, consistent with the uncertainty quoted in Tab.~\ref{tab:oscsyst}. The shifts due to the three largest systematic errors combined (a)$-$(c) in Tab.~\ref{tab:oscsyst}) are approximately three times smaller than the statistical errors on $\dm$ and $\st$.  The magnetic field recalibration discussed in Sec.~\ref{sec:mc} predominantly affects the portion of the neutrino spectrum where oscillations do not occur and may be closely approximated by scaling the strength of the field used in analysis of the data. This results in shifts in the best fit \dm{} and \sintwo{} that are small in comparison with the major systematic errors in Tab.~\ref{tab:oscsyst}.
\pend

\begin{table}[h] 
\begin{center} 
      \begin{tabular}{llcc}
	\hline
	\hline
	Uncertainty &&\dmsq & \sintwo \\
	  && $(10^{-3}\,\evmass{})$ & \\
	\hline
	(a) Normalization & $(\pm 4\%)$          & 0.050 & 0.005\\
	(b) Abs. hadronic $E$ scale & $(\pm 11\%)$ & 0.057 & 0.048\\
	(c) NC contamination &$(\pm 50\%)$       & 0.090 & 0.050\\
	\hline
	(d) Beam uncertainties &    & 0.015 & $<$0.005\\
	(e) Cross sections  &       & 0.011 & 0.005\\
	All other systematics &                  & 0.041 & 0.013\\
	\hline
Statistical Error &   & 0.35 & 0.13\\
        \hline
	\hline
      \end{tabular}
\caption{Systematic shifts on the measurement 
	of \dmsq{} and \sintwo{} for various sources of systematic error, calculated using Monte Carlo generated with the best-fit oscillation parameters obtained from the Far Detector data and the Beam Matrix extrapolation method.
	The values quoted are the average shifts for $\pm$ 1 standard deviation variations in each of the systematic parameters.  
        The last row of the table shows the expected statistical uncertainty on the measurement of \dmsq{} and \sintwo{} for an exposure of $\unit[1.27\times 10^{20}]{POT}$.
	The shifts on \dmsq{} and \sintwo{} are treated as uncorrelated, and correlations between the various systematic effects are not taken 
	into account.\label{tab:oscsyst}}
\end{center}  
\end{table}

\begin{figure}[h]
\begin{center}
\includegraphics[width=\columnwidth]{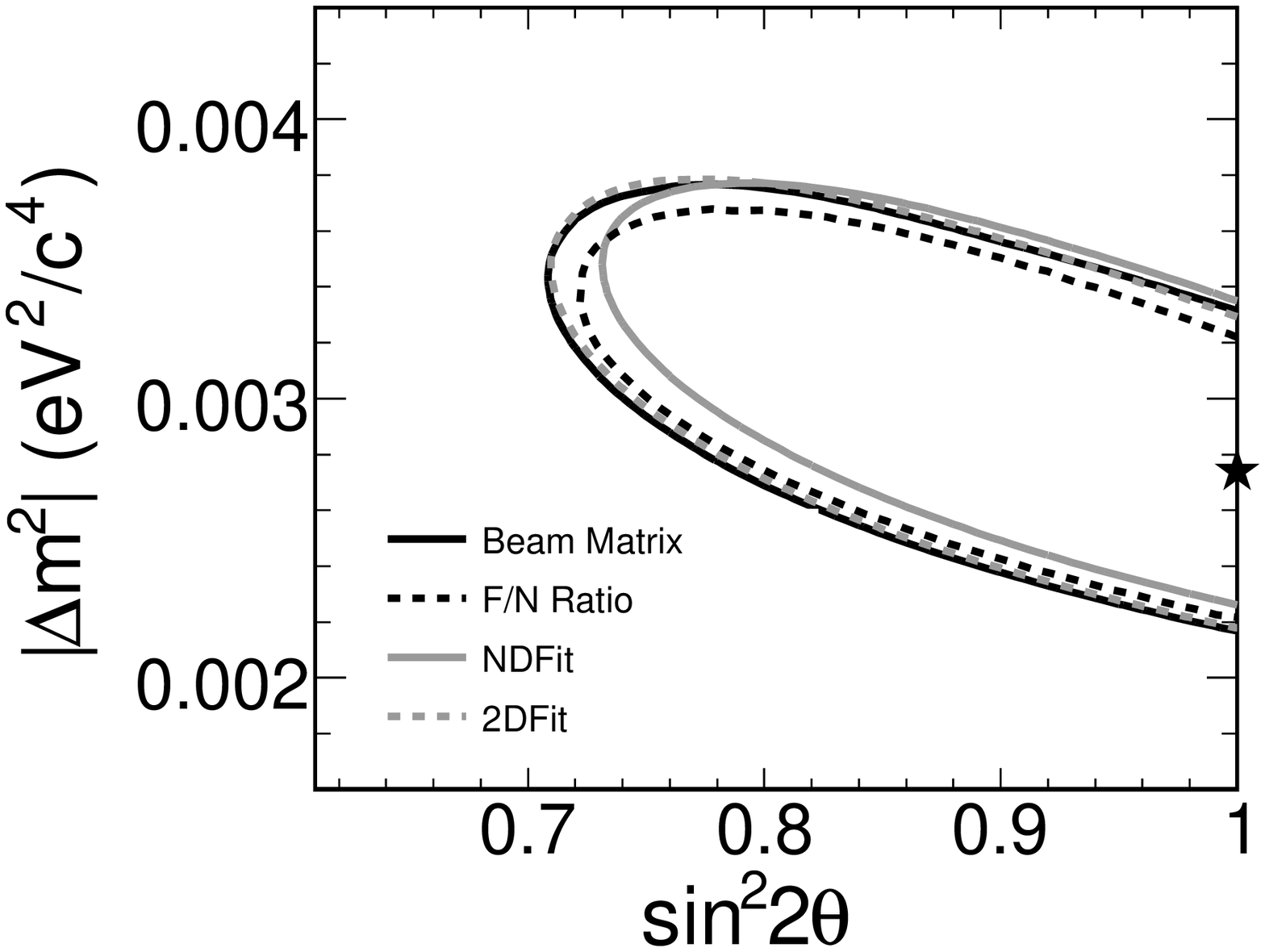}
\caption{\label{fig:region_4methods} Comparison of the 90\% C.L. regions from oscillation fits using the four extrapolation methods. The best-fit point from the Beam Matrix method is shown by the star. All of the contours include the three leading systematic errors described in the text.}
\end{center}
\end{figure}

\pstart
Figure~\ref{fig:region_4methods} shows the 90\% C.L. allowed regions obtained from fits to the 215 Far Detector data events using the Far Detector predictions obtained from the four extrapolation methods described in the previous section. Both the allowed regions, and the best-fit parameters, which are shown in Tab.~\ref{tab:fdfitres}, are in very good agreement between the methods. The spread in the allowed regions is small relative to the size of the regions --- this spread is due to the small differences in the predicted Far Detector spectra shown in the previous section. 
\pend



\begin{table*}[]
\centering

\begin{tabular}{lcccc}
\hline
\hline
Extrapolation method \hspace{0.2in} &\hspace{0.2in} $\dm$ (\evmass{}) \hspace{0.2in} &  $\st$ & \hspace{0.2in} $\chi^{2}/d.o.f$ \hspace{0.2in} & \hspace{0.2in}$\chi^{2}/d.o.f.$ (no osc.) \hspace{0.2in}\\
\hline
\hline
Beam Matrix & $2.74\times10^{-3}$ & 1.0 & 20.3/13 & 104/15\\
Far/Near ratio & $2.73\times10^{-3}$ & 1.0 & 52.8/58 & 132/60\\ 
NDFit & $2.82\times10^{-3}$ & 1.0 & 20.1/13 & 96/15\\ 
2DFit & $2.80\times10^{-3}$ & 0.98 & 34.2/28 & 107/30\\ 
\hline
\hline
\end{tabular}
\caption{Best-fit oscillation parameters and $\chi^2$ values returned from fits to the 215 Far Detector data events using four independent extrapolation methods. The right-hand column shows the values of $\chi^{2}$/d.o.f. obtained by each method for the null oscillation hypothesis. Fifteen reconstructed energy bins were used by the Beam Matrix and NDFit methods. The F/N ratio used 60 \unit[0.5]{GeV} bins and the 2DFit employed 30 \unit[1.0]{GeV} bins.\label{tab:fdfitres}}

\end{table*}

\vspace{0.1in}
\par





\begin{figure}[h]
\begin{center}
\includegraphics[width=\columnwidth]{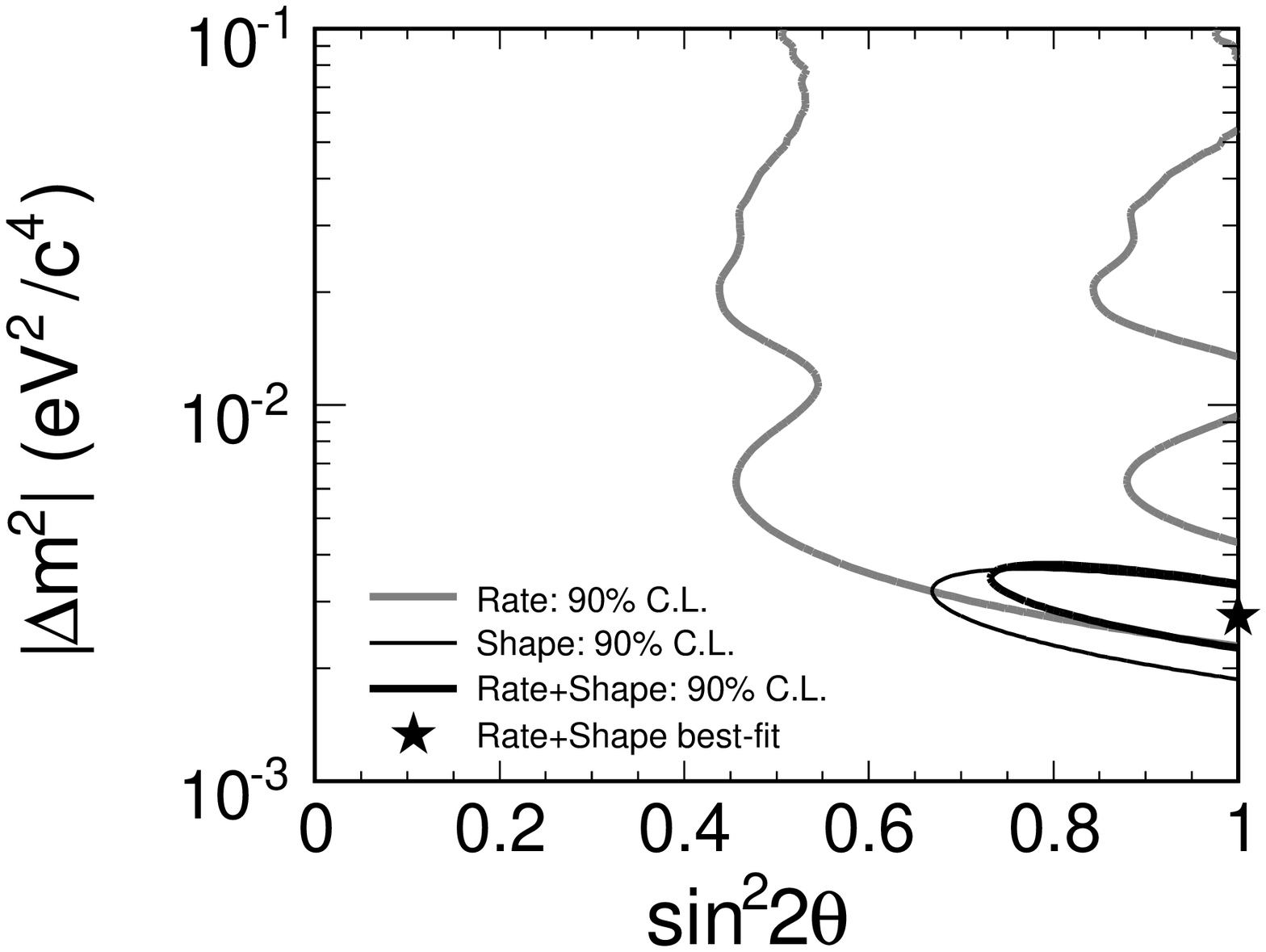}
\caption{\label{fig:region_normshape} Comparison of the 90\% C.L. regions from oscillation fits using shape and rate information only. The best-fit point and 90\% C.L. contour from the fit to shape and rate information is also shown. }
\end{center}
\end{figure}

\pstart
 The relative sensitivity of the oscillation fit to shape and rate information is illustrated in Fig.~\ref{fig:region_normshape}. This shows the 90\% C.L. regions that are obtained from fits to the total rate of Far Detector events, where no spectral information is used and fits to the shape of the spectrum only where data and MC distributions are normalized to the same number of events. The shape of the spectrum plays the most important role in defining the size of the allowed region; the rate information alone does not provide an upper bound on the value of $\dm{}$. 
\pend

\begin{figure}[h]
\begin{center}
\includegraphics[width=\columnwidth]{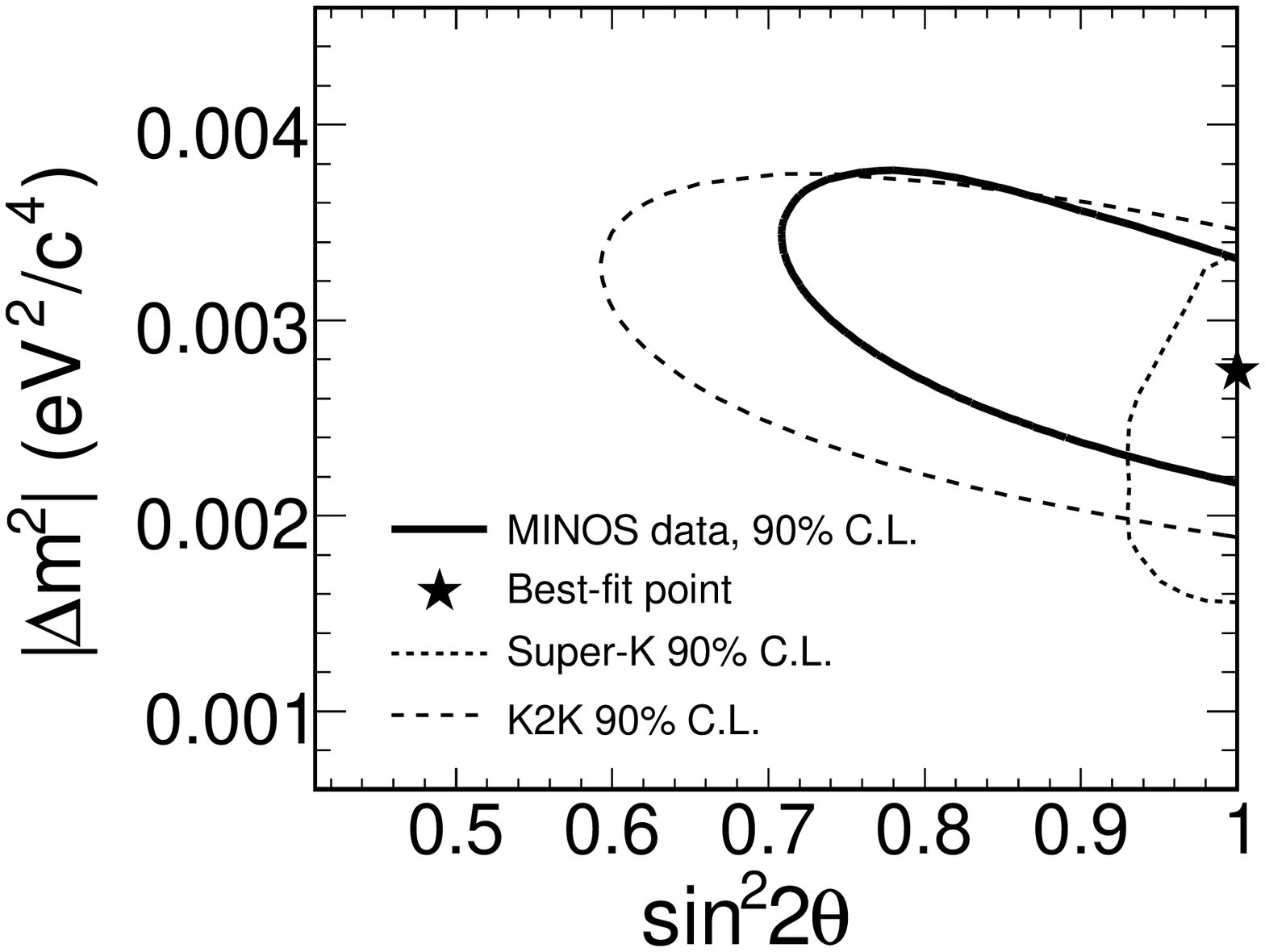}
\caption{\label{fig:region_k2ksk} Comparison of the 90\% C.L. region from MINOS with those previously reported by other experiments.}
\end{center}
\end{figure}

\pstart
Figure~\ref{fig:region_k2ksk} shows a comparison of the 90\% C.L. allowed region from MINOS with those previously reported from the K2K long-baseline~\cite{Ahn:2006zz} and the Super-Kamiokande atmospheric neutrino~\cite{Ashie:2004mr} oscillation analyses. Note that the MINOS results are for \numucc{} events, whereas the Super-Kamiokande results are for a combined $\nu_\mu+{\overline{\nu}}_\mu$ dataset. The allowed regions are in good agreement with each other.
\pend


\section{Summary}

\pstart
In this paper we report on an observation of \numu{} disappearance by the \minos{} long-baseline neutrino experiment using an accelerator based neutrino beam provided by the Fermilab Main Injector. The data were collected over a nine month period from May 2005 to February 2006, corresponding to a total of $1.27 \times 10^{20}$ protons on target. The experiment uses two detectors separated by \unit[734]{km}. The prediction of the unoscillated neutrino flux at the Far Detector site was obtained from the observed neutrino spectrum at the Near Detector location. This two-detector approach provides significant cancellation of systematic errors due to uncertainties in beam modeling and neutrino cross-sections. In addition, we used Near Detector data taken with several beam configurations in order to constrain some of these uncertainties.
\pend

\pstart
We have used four techniques, each with different sensitivities to systematic uncertainties, in order to obtain the unoscillated Far Detector flux from Near Detector data. All four methods predict very similar Far Detector spectra. The total number of \numucc{} events observed in the Far Detector is 215, compared to the expectation of $336 \pm 14$ for no oscillations. The deficit in the number of observed events shows a strong energy dependence, consistent with neutrino oscillations. 
\pend

\pstart
A fit to the observed energy spectrum, assuming two-flavor $\nu_\mu \leftrightarrow \nu_\tau$ mixing, yields best-fit parameters
\begin{displaymath}
\dm{}=\unit[2.74\times10^{-3}]{\evmass{}}, \qquad \sintwo{}=1
\end{displaymath}
\noindent with allowed ranges of $2.54\times10^{-3}<\dm<\unit[3.18\times 10^{-3}]{\evmass{}}$ and $\st>0.87$ (68\% C.L., 1 d.o.f.). All four analysis techniques give consistent results. These values are also consistent with those from existing experiments.
\pend


\pstart
The current estimate of our systematic uncertainty is approximately a factor of 2--3 smaller than our statistical error for $1.27 \times10^{20}$ protons on target. Continued data taking, together with refinements in the estimation of our systematic errors, will allow us to make significant improvements in our measurements of $\dm$ and $\st$ in future analyses.  Since the initial publication of these results in~\cite{prl}, \minos{} has accumulated a total of $\unit[3.5\times10^{20}]{POT}$  through July, 2007. Preliminary results from an exposure of $\unit[2.5\times10^{20}]{POT}$ recorded in the LE10/185kA beam configuration have been presented in~\cite{lp07} and analysis of the full $\unit[3.5\times10^{20}]{POT}$ dataset is in progress.

\pend

\begin{acknowledgments}
\pstart
This work was supported by the US DOE; the UK PPARC; the US NSF; the State and University of Minnesota; the University of Athens, Greece and Brazil's FAPESP and CNPq. We are grateful to the Minnesota Department of Natural Resources, the crew of the Soudan Underground Laboratory, and the staff of Fermilab for their contribution to this effort.
\pend


\end{acknowledgments}

\ifthenelse{\isundefined{\beginnumbering}}{}{\endnumbering}

\bibliography{cc_prd,beam}

\end{document}